\documentclass[superscriptaddress,aps,preprintnumbers,amsmath,amssymb,prd,nofootinbib,preprint]{revtex4-1}
\pdfoutput=1

\usepackage{graphicx}
\usepackage{epstopdf}
\usepackage{dcolumn}
\usepackage{bm}
\usepackage{bbm}
\usepackage{hyperref}
\usepackage[usenames, dvipsnames]{color}
\usepackage{amsmath,amssymb}
\usepackage[sort&compress]{natbib}
\usepackage{float}
\usepackage{multirow}
\usepackage{slashed}
\usepackage{cancel}
\usepackage[table]{xcolor}
\usepackage{booktabs}
\usepackage{makecell}
\usepackage{mathtools}
\usepackage{pifont}
\usepackage{enumitem}

\def\mpl{M_{\rm Pl}}

\numberwithin{equation}{section}

\makeatletter
\renewcommand{\p@subsection}{}
\renewcommand{\p@subsubsection}{}
\makeatother

\makeatletter
\def\l@subsubsection#1#2{}
\makeatother

\makeatletter
    \def\CT@@do@color{%
      \global\let\CT@do@color\relax
            \@tempdima\wd\z@
            \advance\@tempdima\@tempdimb
            \advance\@tempdima\@tempdimc
    \advance\@tempdimb\tabcolsep
    \advance\@tempdimc\tabcolsep
    \advance\@tempdima2\tabcolsep
            \kern-\@tempdimb
            \leaders\vrule
                    \hskip\@tempdima\@plus  1fill
            \kern-\@tempdimc
            \hskip-\wd\z@ \@plus -1fill }
    \makeatother
 
\makeatletter
\def\@ssect@ltx#1#2#3#4#5#6[#7]#8{%
  \def\H@svsec{\phantomsection}%
  \@tempskipa #5\relax
  \@ifdim{\@tempskipa>\z@}{%
    \begingroup
      \interlinepenalty \@M
      #6{%
       \@ifundefined{@hangfroms@#1}{\@hang@froms}{\csname @hangfroms@#1\endcsname}%
       {\hskip#3\relax\H@svsec}{#8}%
      }%
      \@@par
    \endgroup
    \@ifundefined{#1smark}{\@gobble}{\csname #1smark\endcsname}{#7}%
  }{%
    \def\@svsechd{%
      #6{%
       \@ifundefined{@runin@tos@#1}{\@runin@tos}{\csname @runin@tos@#1\endcsname}%
       {\hskip#3\relax\H@svsec}{#8}%
      }%
      \@ifundefined{#1smark}{\@gobble}{\csname #1smark\endcsname}{#7}%
      \addcontentsline{toc}{#1}{\protect\numberline{}#8}%
    }%
  }%
  \@xsect{#5}%
}%
\makeatother

\hypersetup{colorlinks,citecolor= blue,linkcolor= black}

\begin{document}


\newcommand{\meV}{ \ {\rm meV} }
\newcommand{\eV}{ \ {\rm eV} }
\newcommand{\keV}{ \ {\rm keV} }
\newcommand{\MeV}{\  {\rm MeV} }
\newcommand{\GeV}{\  {\rm GeV} }
\newcommand{\TeV}{\  {\rm TeV} }
\newcommand{\PeV}{\  {\rm PeV} }
\newcommand{\EeV}{\  {\rm EeV} }

\title{Gravitational Wave and CMB Probes of Axion Kination}

\preprint{UMN-TH-4023/21, FTPI-MINN-21-15, CERN-TH-2021-124}

\author{Raymond T.~Co}
\affiliation{\small William I. Fine Theoretical Physics Institute, School of Physics and Astronomy, University of Minnesota, Minneapolis, MN 55455, USA}
\author{David Dunsky}
\affiliation{\small Department of Physics, University of California, Berkeley, CA 94720, USA}
\affiliation{\small Theoretical Physics Group, Lawrence Berkeley National Laboratory, Berkeley, CA 94720, USA}
\author{Nicolas Fernandez}
\affiliation{\small Department of Physics, University of Illinois at Urbana-Champaign, Urbana, IL 61801, USA}
\affiliation{\small Illinois Center for Advanced Studies of the Universe, University of Illinois at Urbana-Champaign, Urbana, IL 61801, USA}
\author{Akshay~Ghalsasi}
\affiliation{\small Pittsburgh Particle Physics, Astrophysics, and Cosmology Center, 
Department of Physics and Astronomy,
University of Pittsburgh, Pittsburgh, PA 15260, USA}
\author{Lawrence J.~Hall}
\affiliation{\small Department of Physics, University of California, Berkeley, CA 94720, USA}
\affiliation{\small Theoretical Physics Group, Lawrence Berkeley National Laboratory, Berkeley, CA 94720, USA}
\author{Keisuke Harigaya}
\affiliation{\small Theoretical Physics Department, CERN, Geneva, Switzerland}
\affiliation{\small School of Natural Sciences, Institute for Advanced Study, Princeton, NJ 08540, USA}
\author{Jessie Shelton}
\affiliation{\small Department of Physics, University of Illinois at Urbana-Champaign, Urbana, IL 61801, USA}
\affiliation{\small Illinois Center for Advanced Studies of the Universe, University of Illinois at Urbana-Champaign, Urbana, IL 61801, USA}

\begin{abstract}
Rotations of an axion field in field space provide a natural origin for an era of kination domination, where the energy density is dominated by the kinetic term of the axion field, preceded by an early era of matter domination. Remarkably, no entropy is produced at the end of matter domination and hence these eras of matter and kination domination may occur even after Big Bang Nucleosynthesis. We derive constraints on these eras from both the cosmic microwave background and Big Bang Nucleosynthesis. We investigate how this cosmological scenario affects the spectrum of possible primordial gravitational waves and find that the spectrum features a triangular peak. We discuss how future observations of gravitational waves can probe the viable parameter space, including regions that produce axion dark matter by the kinetic misalignment mechanism or the baryon asymmetry by axiogenesis. For QCD axion dark matter produced by the kinetic misalignment mechanism, a modification to the inflationary gravitational wave spectrum occurs above 0.01 Hz and, for high values of the energy scale of inflation, the prospects for discovery are good. We briefly comment on implications for structure formation of the universe.
\end{abstract}

\maketitle

\begingroup
\hypersetup{linkcolor=black}
\renewcommand{\baselinestretch}{1.12}\normalsize
\tableofcontents
\renewcommand{\baselinestretch}{2}\normalsize
\endgroup

\section{Introduction}
\label{sec:intro}

The thermal history of the very early Universe remains uncertain. It involved a sequence of eras, where each era was characterized by  a certain expansion rate. The expansion rate is key to understanding the physical processes occurring during any era and is determined by $\rho(a)$, the dependence of the energy density on the Friedmann-Robertson-Walker scale factor $a$. From the precise observations of the cosmic microwave background (CMB), we know that as the temperature cooled through the eV region, the universe transitioned from being dominated by radiation, with $\rho(a) \propto 1/a^4$, to being dominated by matter,  with $\rho(a) \propto 1/a^3$.  This matter-dominated era lasted until relatively recently when the universe entered an era apparently dominated by vacuum energy, with $\rho$ independent of $a$.  Furthermore, at the time of Big Bang Nucleosynthesis (BBN), when the temperature was in the MeV region, the universe was radiation-dominated, and there was likely a very early era of vacuum domination known as inflation, when $\rho$ was independent of $a$.   Since this is the total observational evidence we have of the very early  evolution of our universe, there are clearly many possible cosmological histories, each having a different sequence of transitions between eras of differing $\rho(a)$.

It is remarkable that {\it if} the universe underwent an era of ``kination”, with $\rho$ dominated by the kinetic energy of a classical homogeneous scalar field, then $\rho(a)$ falls very rapidly as $1/a^6$,  which can lead to interesting physical phenomena.  Kination was first considered in the context of ending inflation~\cite{Spokoiny:1993kt}, and subsequently as a source for a strongly first-order electroweak phase transition that could enhance baryogenesis~\cite{Joyce:1996cp}.  Such a rapid evolution can also greatly affect the abundance of dark matter~\cite{Visinelli:2009kt, DEramo:2017gpl,Redmond:2017tja,DEramo:2017ecx,Visinelli:2017qga}, alter the spectrum of gravitational waves being emitted from cosmic strings~\cite{Cui:2017ufi, Cui:2018rwi, Bettoni:2018pbl, Ramberg:2019dgi, Auclair:2019wcv, Chang:2019mza, Gouttenoire:2019kij, Chang:2021afa} or originated from inflation~\cite{Giovannini:1998bp,Giovannini:1999bh,Giovannini:1999qj,Riazuelo:2000fc,Sahni:2001qp,Tashiro:2003qp,Boyle:2007zx,Giovannini:2008tm,Giovannini:2009kg,Kuroyanagi:2011fy,Li:2016mmc,Kuroyanagi:2018csn,Bernal:2019lpc,Figueroa:2019paj,Dalianis:2020cla, Hook:2020phx,Li:2021htg} during such an era, and boost the matter power spectrum, enhancing small-scale structure formation~\cite{Redmond:2018xty,Visinelli:2018wza}.

What is the underlying field theory and cosmology that leads to an era of kination?  Once kination starts, it is easy to end since the kination energy density dilutes under expansion much faster than radiation, so a transition to radiation domination will occur. But how does kination begin? Going to earlier times during the kination era, the kinetic energy density of the scalar field rapidly increases. This issue is particularly important for primordial gravitational waves, whether produced from quantum fluctuations during inflation and entering the horizon during a kination era or from emission from cosmic strings during a kination era, since the spectrum for both increases linearly with frequency.  This UV catastrophe must get cut off by the physics that initiates the kination era, and hence the peak of the gravitational wave distribution will have a shape determined by this physics.

Recently a field theory and cosmology for kination was proposed by two of us: axion kination~\cite{Co:2019wyp}.  An approximate $U(1)$ global symmetry is spontaneously broken by a complex field. Early on there are oscillations in both angular and radial modes, which we call  axion and saxion modes, in an approximately quadratic potential.  The radial oscillation results from a large initial field displacement, and the angular mode is excited by higher-dimensional operators that break the $U(1)$ symmetry.  At some point the radial mode is damped, but by this time the “angular momentum” in the field, that is the charge density, is conserved, except for cosmic dilution, and a period of circular evolution sets in.  This era is assumed to occur when the radial field is much larger than the vacuum value, $f_a$, the symmetry-breaking scale for the axion.  In fact, at first the trajectory is not quite circular, as the cosmic dilution of charge leads to a  slow inward spiral of the trajectory. If the energy density of the universe is dominated by the scalar field energy, this inspiral era is a matter-dominated era with $\rho(a) \propto 1/a^3$.  This era ends when the radial mode settles to $f_a$; the potential vanishes and the axion energy density is now entirely kinetic so that an era of kination ensues. Kination ends when the axion energy density falls below that of radiation.  

The rotation of an axion field was used in~\cite{Co:2019wyp} to generate a baryon asymmetry via Axiogenesis and in~\cite{Co:2019jts} to generate axion dark matter via the Kinetic Misalignment Mechanism.
In fact, such schemes did not rely on the axion field energy becoming larger than the radiation energy, so an era of kination was possible but not required.  In this paper we study the implications of a kination-dominated era from this mechanism, where the era is cut off in the UV by an early matter-dominated era.
We consider both the QCD axion~\cite{Peccei:1977hh,Peccei:1977ur,Weinberg:1977ma,Wilczek:1977pj} that solves the strong CP problem and generic axion-like particles (ALPs).

The early matter-dominated era regulates the spectrum of gravitational waves from inflation or cosmic strings. After the linear increase with frequency from the kination era, the magnitude of the gravitational wave spectrum decreases, producing a triangular peak in the spectrum.\footnote{This is in contrast to the scenarios previously considered in Refs.~\cite{Boyle:2007zx, Giovannini:2008tm, Li:2016mmc, Figueroa:2019paj, Li:2021htg}, where a kination era follows immediately after inflation and BBN limits the duration of the kination era through the dark radiation constraint.} Interestingly, the shape of the peak contains information about the shape of the potential of the complex field and could reveal the origin of the spontaneous $U(1)$ symmetry breaking resulting in an axion as a Nambu-Goldstone boson.

It is natural to expect this kination era to occur early in the cosmic history, ending well before BBN.  Remarkably, for low $f_a$ and large charge density, a late era of kination domination can occur {\it after} BBN, but before matter-radiation equality. This possibility arises because axion field rotations do not generate entropy. In such scenarios, the beginning of the matter-dominated era is constrained by BBN and the end of the kination-dominated era is constrained by the CMB. 

We also examine the implication of the NANOGrav signal~\cite{NANOGrav:2020bcs} to axion kination. Similar signals are also reported by PPTA~\cite{Goncharov:2021oub} and EPTA~\cite{Chen:2021rqp}. The signal may be explained by gravitational waves emitted from cosmic strings~\cite{Ellis:2020ena,Blasi:2020mfx}. We discuss how the gravitational wave spectrum from low to high frequency is modified by axion kination and the modification, including the peak, can be detected by future observations.

In Sec.~\ref{sec:rotation_kination}, we discuss the above mechanism of axion kination cosmology, including two theories for the potential of the radial mode.  We study the transition from the early matter-dominated era to the kination-dominated era, and also the thermalization of the radial mode, as this leads to important constraints on the signals. We derive the dependence of the matter and kination transition temperatures on the axion model parameters.
In Sec.~\ref{sec:constraints}, we analyze the constraints on axion kination from both BBN and the CMB. In Sec.~\ref{sec:DM_baryogenesis}, we review the kinetic misalignment mechanism and axiogenesis and derive predictions for the parameters of axion kination. In Sec.~\ref{sec:GW}, we compute the spectrum of gravitational waves produced from inflation and cosmic strings in the axion kination cosmology. The spectra depend on and can be used to infer the kination and matter transition temperatures and hence axion parameters. We discuss whether future observations of gravitational waves can detect the imprints of axion kination. We pay particular attention to the parameter space where the axion rotation also lead to dark matter or to the baryon asymmetry and also to the case of the QCD axion. Sec.~\ref{sec:summary} is dedicated to summary and discussion. 

\section{Axion rotations and kination}
\label{sec:rotation_kination}

\subsection{Axion rotations}
\label{subsec:rotations}
In field-theoretical realizations of an axion, the axion field $\phi_a$ is the angular direction $\theta \equiv \phi_a/f_a$ of a complex scalar field $P$,
\begin{align}
 P = \frac{1}{\sqrt{2}} S e^{i \theta},
\end{align}
where $S$ is the radial direction which we call the saxion. In the present universe, $S= f_a$, the decay constant, spontaneously breaking an approximate $U(1)$ symmetry.  The potential for $P$ contains small explicit $U(1)$-breaking terms that give a small mass to the axion. For simplicity, we take the domain wall number to be unity.

Given that the $U(1)$ symmetry is explicitly broken to yield a non-zero axion mass, from the effective field theory point of view, it is plausible that the symmetry is also explicitly broken by a higher dimensional operator, 
\begin{align}
    V = \frac{P^n}{M^{n-4}} + {\rm h.c.},
\end{align}
where $M$ is a dimensionful parameter.
Such a term is in fact expected in theories where the $U(1)$ symmetry arises as an accidental one~\cite{Holman:1992us,Barr:1992qq,Kamionkowski:1992mf,Dine:1992vx} and is broken by the effects of quantum gravity~\cite{Giddings:1988cx,Coleman:1988tj,Gilbert:1989nq,Harlow:2018jwu,Harlow:2018tng}.
In the early universe, $S$ may take on a field value much larger than $f_a$. The potential gradient to the angular direction given by the higher-dimensional operator then gives a kick to the angular direction and the complex field begins to rotate.
As the universe expands, the field value of $S$ decreases and the higher-dimensional operator becomes ineffective. The field $P$ continues to rotate while preserving its angular momentum $\dot{\theta} S^2$ up to the cosmic expansion.  
Such dynamics of complex scalar fields was proposed in the context of Affleck-Dine baryogenesis~\cite{Affleck:1984fy}.
The angular momentum $\dot{\theta} S^2$ is nothing but the conserved charge density associated with the $U(1)$ symmetry.
It is convenient to normalize this charge density by the entropy density of the universe $s$,
\begin{equation}
\label{eq:Y_theta_def}
    Y_\theta = \frac{\dot\theta S^2}{s},
\end{equation}
which remains constant as long as entropy is not produced. As we will see, the charge density must be large enough to obtain kination domination. This can be easily achieved in our scenario because of the large initial $S$.

Soon after the field rotation begins, the motion is generically a superposition of angular and radial motion and has non-zero ellipticity. Once $P$ is thermalized, the radial motion is dissipated, while the angular motion remains because of angular momentum conservation. One may think that the angular momentum is transferred into particle-antiparticle asymmetry in the thermal bath. It is, however, free-energetically favored to keep most of the charge in the form of axion rotations as long as $S\gg T$~\cite{Co:2019wyp}. The resultant motion after thermalizaion is a circular one without ellipticity. The parameter space that leads to successful thermalization is investigated in Sec.~\ref{subsec:thermalization}.

If the axion couples to a gauge field, the angular momentum can be also transferred into the helicity of the gauge field through tachyonic instability~\cite{Agrawal:2017eqm,Co:2021rhi}. The gauge field in the tachyonic instability band has a wavelength $\sim (\alpha \dot{\theta}/\pi)^{-1}\gg T^{-1}$ and the resultant gauge field cannot be treated as excitations in the thermal bath, and the above thermodynamical argument is not applicable. However, as is shown in Ref.~\cite{Domcke:2018eki}, the backreaction from charged fermions prevents efficient production of the gauge field. Using the upper bound on the magnitude of the gauge field derived in the reference, one can show that the energy density of the gauge field produced by the tachyonic instability is much smaller than the energy density of the axion rotation. Therefore, the axion rotation is not destroyed by the production of the gauge field. Note that Ref.~\cite{Domcke:2018eki} assumes no fermions besides those that are created from the gauge field. With the thermal bath in our setup, the production of the gauge fields will be even more ineffective.

\subsection{Axion kination}
\label{subsec:kination}

The evolution of the energy density of the axion rotations depends on the shape of the potential of the saxion. A very interesting evolution involving kination is predicted when the saxion potential is nearly quadratic at $S\gg f_a$~\cite{Co:2019wyp}. Such a potential arises naturally in supersymmetic theories, where the saxion is the scalar partner of the axion: the saxion potential may be flat in the supersymmetric limit and generated by supersymmetry breaking.

For example, the soft supersymmetry breaking coefficient of $P^*P=S^2/2$ may be positive at high scales but evolve under renormalization to negative values at low scales. This triggers spontaneous breaking of the $U(1)$ symmetry, which can be described by the potential~\cite{Moxhay:1984am},
\begin{align}
\label{eq:dim_trans}
V(P) = \frac{1}{2} m_S^2 |P|^2 \left( {\rm ln} \frac{2 |P|^2}{f_a^2} -1 \right),
\end{align}
which is nearly quadratic for $S \gg f_a$. Another example is a two-field model with soft masses,
\begin{align}
\label{eq:two_field}
W = X( P \bar{P} - v_{P}^2 ),~~V_{\rm soft} = m_P^2 |P|^2 +  m_{\bar{P}}^2 |\bar{P}|^2.
\end{align}
Here $X$ is a stabilizer field that fixes the symmetry breaking field $P$ and $\bar{P}$ on a moduli space $P\bar{P} = v_{P}^2$.
For $P \gg v_P$ or $\bar{P} \gg v_P$, the saxion potential is dominated by the soft mass $m_P$ and $m_{\bar{P}}$, respectively. Without loss of generality, we choose $P$ to be initially much larger than $v_P$ and identify the saxion with the radial direction of $P$. We neglect possible renormalization running of $m_P$ which modifies the saxion potential only by a small amount.

For these potentials, the axion rotations evolve as follows. When $S \gg f_a $, the potential of $S$ is nearly quadratic, and the equation of motion of the radial direction requires $\dot{\theta}^2 = V'(S)/S \simeq m_S^2$. The conservation of the angular momentum, $\dot{\theta}S^2 \propto a^{-3}$, then requires $S^2 \propto a^{-3}$. Here $a$ is the scale factor of the universe, not to be confused with the axion field which we denote as $\phi_{a}$. The potential energy $\sim m_S^2 S^2$ and the kinetic energy $\sim \dot{\theta}^2 S^2$ are comparable. Once $S$ decreases and $S \simeq f_a$, the conservation of the angular momentum requires $\dot{\theta}\propto a^{-3}$.
The kinetic energy dominates over the potential energy. The scaling of the energy density in these two regimes is
\begin{align}
    \rho_\theta \propto
    \begin{cases}
     a^{-3} & : S \gg f_a \\
     a^{-6} & : S \simeq f_a.
    \end{cases}
\end{align}

The scaling of the energy density naturally leads to kination domination~\cite{Co:2019wyp}. When $S \gg f_a$, the rotation behaves as matter and is  red-shifted slower than radiation is, so the universe may become matter-dominated by the axion rotation. Once $S$ approaches $f_a$, kination domination by the axion rotation begins. 

Throughout most of this paper, we adopt a piecewise approximation where $\rho_\theta \propto a^{-3}$ for $S>f_a$ and $\rho_\theta \propto a^{-6}$ for $S = f_a$. We will comment on how the actual evolution differs from this. Within this approximation, the Hubble expansion rate $H$ as a function of the temperature $T$ is given by
\begin{equation}
\label{eq:Hubble}
    H (T) = \frac{1}{M_{\rm Pl}} \sqrt{\frac{\pi^2}{90}g_*} \times
\begin{cases}
 T^2 & \text{~for~RD~:~} T \gg T_{\rm RM} \\
 T_{\rm RM}^2 \left( \frac{T}{T_{\rm RM}}\right)^{ \scalebox{1.01}{$\frac{3}{2}$} } & \text{~for~MD~:~}  T_{\rm RM} \gg T \gg T_{\rm MK}  \\
 T_{\rm KR}^2 \left( \frac{T}{T_{\rm KR}}\right)^3 & \text{~for~KD~:~}  T_{\rm MK} \gg T \gg T_{\rm KR} \\
 T^2 & \text{~for~RD~:~} T_{\rm KR} \gg T 
\end{cases} .
\end{equation}
Here $T_{\rm RM}$ is the temperature at which the matter domination (MD) by the axion rotation begins, $T_{\rm MK}$ is the temperature at which the kination domination (KD) begins, and $T_{\rm KR}$ is the temperature at which the KD ends and radiation domination (RD) begins.  

In Eq.~(\ref{eq:Hubble}), it is assumed that $P$ is thermalized when the rotation is still a subdominant component of the universe. It is also possible that the thermalization occurs after the rotation dominates the universe. In this case, the energy associated with the radial mode, which is comparable to or larger than the angular mode, is converted into radiation energy. Then the universe evolves as RD $\rightarrow$ MD $\rightarrow$ RD $\rightarrow$ MD $\rightarrow$ KD $\rightarrow$ RD, and the scaling in Eq.~(\ref{eq:Hubble}) is applicable to the last four eras. The second RD is very short when the radial and the angular component of the initial rotation is comparable, which naturally occurs in supersymmetric theories; see~\cite{Co:2020jtv} for details. $Y_\theta$ computed after the initiation of the rotation receives entropy production from the dissipation of the radial mode, and is conserved again afterward.  It is this final value of $Y_\theta$ that concerns us, and we do not investigate how it is related to the UV parameters of the theory.

The cosmological progression from RD $\rightarrow$ MD $\rightarrow$ KD $\rightarrow$ RD described above is determined by three parameters, $(f_a, m_S, Y_\theta)$, and the three temperatures $(T_{\rm RM}, T_{\rm MK}, T_{\rm KR})$ can be expressed in terms of these,
\begin{align}
\label{eq:TRM}
    T_{\rm RM} & = \frac{4}{3}m_S Y_{\theta} 
    \simeq  1.3 \times10^7 \GeV \left(\frac{m_S}{100 \TeV}\right) \left(\frac{Y_\theta}{100}\right),  \\
    \label{eq:TMK}
    T_{\rm MK} & = 
    \left(\frac{45}{2\pi^2 g_*} \frac{m_S f_a^2}{Y_{\theta}}\right)^{ \scalebox{1.01}{$\frac{1}{3}$} } \\
    & \simeq  2.8\times 10^6 \GeV \left(\frac{m_S}{100 \TeV}\right)^{ \scalebox{1.01}{$\frac{1}{3}$} } 
    \left(\frac{f_a}{10^9 \GeV}\right)^{ \scalebox{1.01}{$\frac{2}{3}$} }
    \left(\frac{100}{Y_\theta} \right)^{ \scalebox{1.01}{$\frac{1}{3}$} }
    \left(\frac{g_{*,{\rm SM}}}{g_*}\right)^{ \scalebox{1.01}{$\frac{1}{3}$} },  \\
    \label{eq:TKR}
    T_{\rm KR} & = \frac{3 \sqrt{15}}{2 \sqrt{g_*}\pi } \frac{f_a}{Y_{\theta}} 
    \simeq   1.8 \times 10^6 \GeV \left( \frac{f_a}{10^9 \GeV} \right) \left(\frac{100}{Y_\theta} \right) \left(\frac{g_{*,{\rm SM}}}{g_*}\right)^{ \scalebox{1.01}{$\frac{1}{2}$} }.
\end{align}
The kination-dominated era exists if $T_{\rm KR} < T_{\rm RM}$,
\begin{align}
    Y_\theta \gtrsim 37 
    \left(\frac{100 \TeV}{m_S}\right)^{ \scalebox{1.01}{$\frac{1}{2}$} } 
    \left(\frac{f_a}{10^9 \GeV}\right)^{ \scalebox{1.01}{$\frac{1}{2}$} } 
    \left(\frac{g_{*,{\rm SM}}}{g_*}\right)^{ \scalebox{1.01}{$\frac{1}{4}$} }.
\end{align}
The scalings of (\ref{eq:Hubble}) imply that these three temperatures are not independent, but are related by $T_{\rm MK}^3 \simeq T_{\rm RM} T_{\rm KR}^2$. The expansion history of the universe is therefore determined by two combinations of $(f_a, m_S, Y_\theta)$ such as $(m_Sf_a, Y_\theta)$, but other phenomenology depends also on the third combination. 
In what follows, according to convenience, we use a variety of ways of spanning the 3-dimensional parameter space. For discussion of axion physics we must include the axion mass $m_a$ as a fourth parameter. The axion mass is determined by $f_a$ for the QCD axion~\cite{Peccei:1977hh,Peccei:1977ur,Weinberg:1977ma,Wilczek:1977pj}.

Note that $T_{\rm RM} T_{\rm KR} \propto m_S f_a$, so one can in principle determine the product $m_S f_a$ by a measurement of $T_{\rm RM}$ and $T_{\rm KR}$ through gravitational wave spectra discussed in Sec.~\ref{sec:GW}. Moreover, additional theoretical considerations such as the baryon asymmetry and/or axion dark matter abundance from the axion rotation, as discussed in Sec.~\ref{sec:DM_baryogenesis}, will help determine $m_S$ and $f_a$ individually. If $f_a$ and $m_a$ are also measured by axion experiments, then the theory parameters are over-constrained so that the theory can be confirmed or ruled out.

\begin{figure}
    \centering
    \includegraphics[width=\columnwidth]{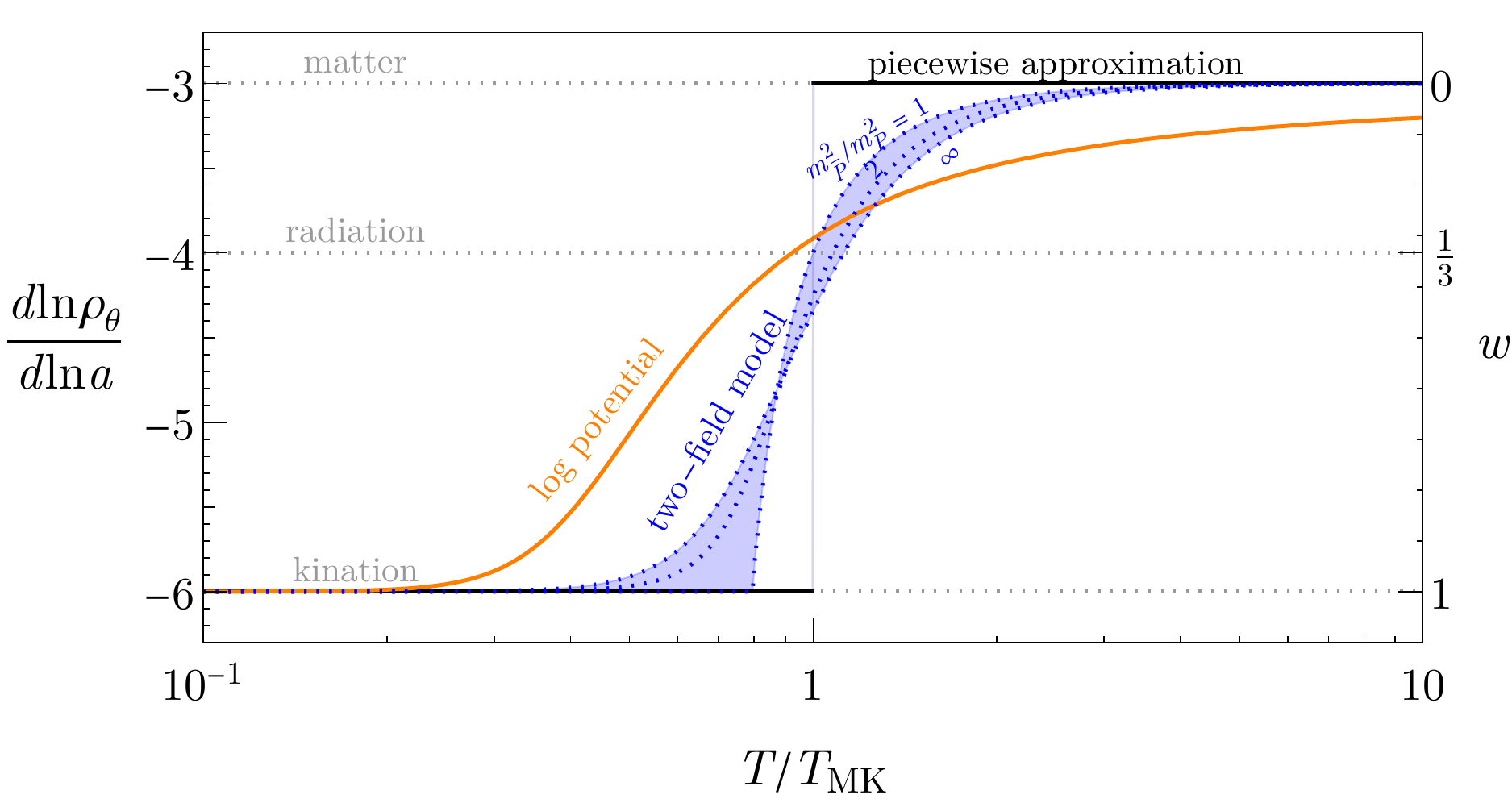}
    \caption{Scaling evolution of the energy density $\rho$ with scale factor $a$ (left axis) as well as the equation of state $w$ (right axis) as a function of temperature in units of $T_{\rm MK}$, the transition temperature from matter to kination. The colored curves are for the two-field model (blue) and the logarithmic potential (orange), whereas the step function (black) is the piecewise approximation we employ in the remainder of the paper. For the two-field model, we show the blue dotted curves for different ratios of the soft masses of the two fields $\bar{P}$ and $P$, and the blue shading indicates the entire possible range of the model.}
    \label{fig:EoS}
\end{figure}

A unique feature of our kination scenario is that matter domination preceding kination domination ends without creating entropy. This is quite different from usual matter domination, where matter domination ends by dissipation of matter into radiation creating a huge amount of entropy.
Because of the absence of entropy production in our scenario, matter and kination domination can occur even after BBN and before recombination.

The precise evolution of the universe differs from the piecewise approximation and depends on the saxion potential.  
The evolution for the one-field model of Eq.~(\ref{eq:dim_trans}) is derived in~\cite{Co:2019wyp} and reviewed in Appendix~\ref{app:evol_axion}, and is shown by the orange solid line in Fig.~\ref{fig:EoS}.
Beyond the piecewise approximation, there is no sharply defined $T_{\rm MK}$, so we first define $T_{\rm RM}$ and $T_{\rm KR}$ by the equality of the axion energy density with the radiation energy density and then define $T_{\rm MK} \equiv T_{\rm RM}^{1/3} T_{\rm KR}^{2/3}$.
The transition from matter to kination domination is not sharp, but still occurs within a temperature change of $\mathcal{O}(10)$. The evolution for the two-field model of Eq.~(\ref{eq:two_field}) is derived in Appendix~\ref{app:evol_axion} and is shown by the blue-dashed lines. The evolution depends on the ratio $m_P/ m_{\bar{P}}$, but the transition is sharper than the one-field model and occurs within a temperature change of $\mathcal{O}(1)$.
As we will see, this difference shows up in the spectrum of primordial gravitational waves and allows the determination of the saxion potential.

\subsection{Thermalization}
\label{subsec:thermalization}

As discussed in the previous subsection, the motion of the field $P$ is initiated in both angular and radial components, and the energy density associated with the radial mode is comparable to or more than the rotational energy. Since the radial mode also evolves as matter, if it is thermalized after $S$ reaches $f_a$, no kination-dominated era is present. Thus earlier thermalization is required. 
In the simplest case, we consider a Yukawa coupling between the saxion and fermions $\psi$ and $\bar\psi$ that couple with the thermal bath,
\begin{equation}
    \mathcal{L} \supset y_\psi S \psi  \bar\psi .
\end{equation}
The simplest possibility is a Standard Model gauge charged fermion, but we may also consider a dark sector fermion.
The thermalization rate is given by~\cite{Mukaida:2012qn},
\begin{equation}
\label{eq:Gamma_S_psi_psi}
    \Gamma_{S\psi\bar\psi} = b y_\psi^2 T, 
\end{equation}
where $b$ is a constant and is $\mathcal{O}(0.1)$ when the coupling of the fermion with the thermal bath is $\mathcal{O}(1)$. The fermion is heavy in the early universe because of a large saxion field value, $m_\psi = y_\psi S$, while the fermions themselves need to be populated in the thermal bath in order to thermalize the saxion at the temperature $T_{\rm th}$. Such a requirement, $y_\psi S_{\rm th} \le T_{\rm th}$, leads to an upper bound on the Yukawa coupling as well as on the rate
\begin{equation}
\label{eq:Gamma_S_psi_psi_max}
    \Gamma_{S\psi\bar\psi} \le \frac{b T_{\rm th}^3}{S_{\rm th}^2} .
\end{equation}
We obtain the same bound for a saxion-scalar coupling. For gauge boson couplings, which arise after integrating out charged fermions or scalars, the thermalization rate $\simeq 10^{-5} T^3/S^2$~\cite{Mukaida:2012qn}, so the constraints on the parameter space for this case can be obtained by putting $b= 10^{-5}$ in the following equations.

For a fixed $Y_\theta = m_S S_{\rm th}^2 / (2\pi^2 g_* T_{\rm th}^3/45 )$,
one can now use Eq.~(\ref{eq:Gamma_S_psi_psi_max}) to derive the maximal thermalization temperature as well as the saxion field value at the time,
\begin{align}
    T_{\rm th}^{\rm max} & \simeq 4 \times 10^7 \GeV 
    \left( \frac{b}{0.1} \right)^{ \scalebox{1.01}{$\frac{1}{2}$} }
    \left( \frac{m_S}{\rm TeV} \right)^{ \scalebox{1.01}{$\frac{1}{2}$} }
    \left( \frac{10^3}{Y_\theta} \right)^{ \scalebox{1.01}{$\frac{1}{2}$} }
    \left( \frac{g_{*,{\rm SM}}}{g_*(T_{\rm th})} \right)^{ \frac{3}{4} }  \\
     & \simeq 3 \times 10^7 \GeV  
    \left( \frac{b}{0.1} \right)^{ \scalebox{1.01}{$\frac{1}{2}$} }
    \left( \frac{m_S}{\rm TeV} \right)^{ \scalebox{1.01}{$\frac{1}{2}$} }
    \left( \frac{T_{\rm KR}}{10^5 \GeV} \right)^{ \scalebox{1.01}{$\frac{1}{2}$} }
    \left( \frac{10^9 \GeV}{f_a} \right)^{ \scalebox{1.01}{$\frac{1}{2}$}}
    \left( \frac{g_{*,{\rm SM}}}{g_*(T_{\rm th})} \right)^{ \frac{3}{4} }
    \left( \frac{g_*(T_{\rm KR})}{g_{*,{\rm SM}}} \right)^{ \frac{1}{4} },
    \nonumber \\
    S_{\rm th}^{\rm max} & \simeq 2 \times 10^{12} \GeV
    \left( \frac{b}{0.1} \right)^{ \scalebox{1.01}{$\frac{3}{4}$} }
    \left( \frac{m_S}{\rm TeV} \right)^{ \scalebox{1.01}{$\frac{1}{4}$} }
    \left( \frac{10^3}{Y_\theta} \right)^{ \scalebox{1.01}{$\frac{1}{4}$} }
    \left( \frac{g_{*,{\rm SM}}}{g_*(T_{\rm th})} \right)^{ \frac{5}{8} }  \\
     & \simeq 1 \times 10^{12} \GeV 
     \left( \frac{b}{0.1} \right)^{ \scalebox{1.01}{$\frac{3}{4}$} }
     \left( \frac{m_S}{\rm TeV} \right)^{ \scalebox{1.01}{$\frac{1}{4}$} }
     \left( \frac{T_{\rm KR}}{10^5 \GeV} \right)^{ \scalebox{1.01}{$\frac{1}{4}$}}
     \left( \frac{10^9 \GeV}{f_a} \right)^{ \scalebox{1.01}{$\frac{1}{4}$}}
     \left( \frac{g_{*,{\rm SM}}}{g_*(T_{\rm th})} \right)^{ \frac{5}{8} }
     \left( \frac{g_*(T_{\rm KR})}{g_{*,{\rm SM}}} \right)^{ \frac{1}{8} }. \nonumber
\end{align}
Here $Y_\theta$ is determined from a fixed $T_{\rm KR}$ using Eq.~(\ref{eq:TKR}). In this case, the thermalization constraints can be imposed by the consistency conditions: 1) the saxion field value at thermalization must be larger than $f_a$, i.e., $S_{\rm th}^{\rm max} \ge f_a$; otherwise, thermalization would not occur or the Universe would not be kination-dominated, 2) the radiation energy density after thermalization is at least that of the saxion, i.e., $\pi^2 g_*(T_{\rm th}) T_{\rm th}^4 /30 \ge m_S^2 (S_{\rm th}^{\rm max})^2$, where the inequality is saturated when the saxion dominates and reheats the universe. In fact, upon assuming the existence of kination domination, $T_{\rm RM} > T_{\rm KR}$, condition (2) automatically guarantees that (1) is satisfied. Therefore, only condition (2) is relevant and leads to the constraint on the saxion mass
\begin{align}
    \label{eq:therm_constraint}
    m_S & \lesssim  1.5 \times 10^5 \GeV  
    \left( \frac{b}{0.1} \right)
    \left( \frac{10^9 \GeV}{f_a}\right)^3 
    \left( \frac{T_{\rm KR}}{10^5 \GeV}\right)^3 
    \left( \frac{g_*(T_{\rm KR})}{g_*(T_{\rm th})} \right)^{ \frac{3}{2} }.
\end{align}
This thermalization constraint is shown as the green regions in the figures we will show in the following sections. For consistency with the assumption of the rotation in the vacuum potential, the thermal mass of $m_S$ must be subdominant to the vacuum one, $y_\psi T_{\rm th} < m_S$. However, using condition (2), one finds that this constraint becomes $y_\psi < (\pi^2 g_*(T_{\rm th})/30)^{1/2} T_{\rm th}/S_{\rm th}$, which is always weaker than the earlier constraint from requiring $\psi$ in thermal equilibrium, $y_\psi < T_{\rm th}/S_{\rm th}$.

When the thermalization temperature is much lower than the QCD scale, additional constraints may become important~\cite{Co:2021rhi}.  For example, the energy density deposited into the bath (or dark radiation) at late thermalization may contribute to excessive $\Delta N_{\rm eff}$ since this energy deposit cannot be absorbed by the Standard Model bath nor diluted by the change of $g_*$ in the Standard Model across the QCD phase transition. Effectively, the constraint from condition (2) above is strengthened by replacing $g_*(T_{\rm th})$ by $7 \Delta N_{\rm eff}/4$. We impose the limit $\Delta N_{\rm eff} > 0.17$ from the CMB and BBN~\cite{Fields:2019pfx}. 

In the case when $T_{\rm RM} \ll \mathcal{O}({\rm GeV})$, the saxion has to be very light and is thus subject to quantum corrections, which require the coupling $y_\psi < 4\pi m_S/m_{\rm soft}$ where $m_{\rm soft}$ is the soft mass of $\psi$'s scalar partner, $\tilde\psi$. We do not impose this constraint since one can simply assume that $m_{\rm soft}$ is generated in the same way as and is of the same order of $m_S$, in which case the constraint is trivially satisfied.   

One may expect additional constraints from the relic density and warmness of $\psi$ especially when $f_a$ is very small and $\psi$ may be very light. However, one may consider a model where $\psi$ has a sufficiently large vector-like mass and freezes out non-relativistically much before BBN (see Ref.~\cite{Co:2021rhi} for details), or $\psi$ is dark gauge-charged and effectively annihilates into massless dark gauge bosons.

\section{Cosmological constraints}
\label{sec:constraints}

The axion rotations lead to matter and kination-dominated eras. If these happen close to BBN or recombination, the modified expansion rate alters primordial light element abundances or the spectrum of the CMB. Constraints from BBN divide kination into two paradigms - ``early kination" for $T_{\rm KR} > \mathcal{O}({\rm MeV})$ and ``late kination" with $T_{\rm RM} < \mathcal{O}(10 \keV)$. In this section, we discuss the constraints on the axion rotation on both early and late  kination from BBN and the late kination from CMB.

\subsection{BBN}

\begin{figure}
    \centering
    \includegraphics[width=0.495\columnwidth]{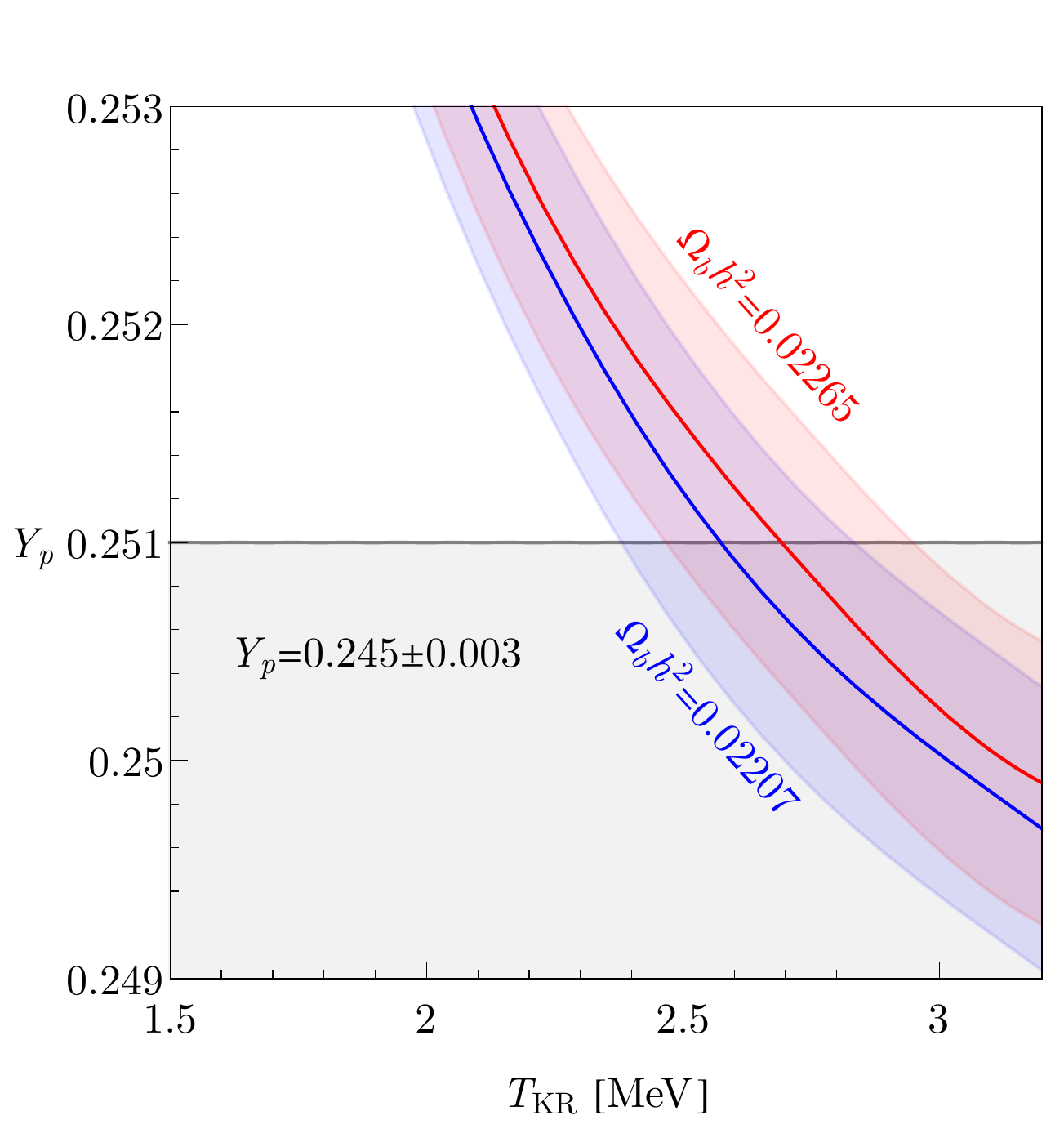}
     \includegraphics[width=0.495\columnwidth]{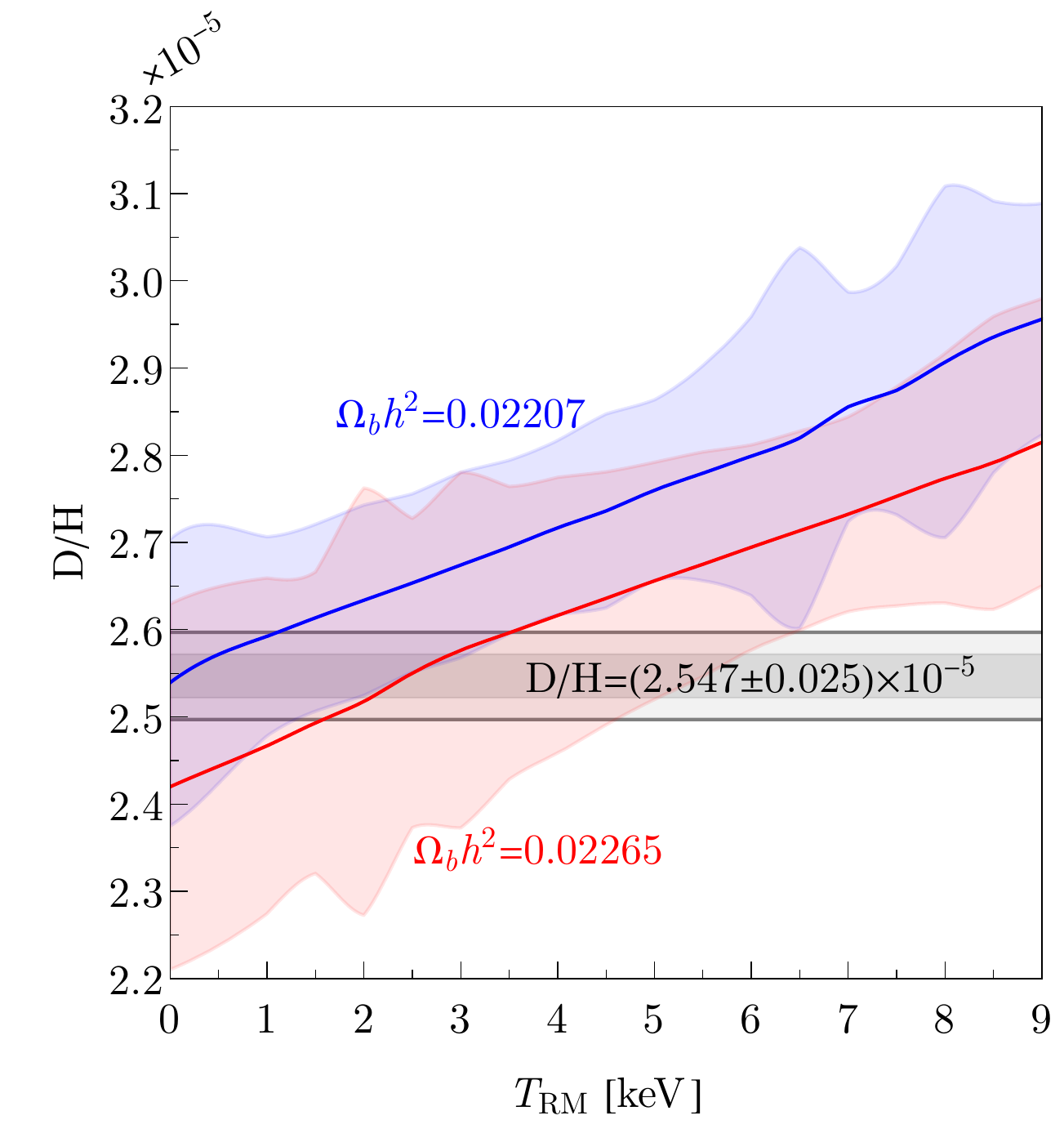}
    \caption{Primordial helium (left panel) and deuterium (right panel) abundances as a function of $T_{\rm KR}$ and $T_{\rm RM}$, respectively. The gray bands show the experimental constraints.}
    \label{fig:BBN}
\end{figure}

When kination domination occurs before BBN, the strongest constraint comes from the helium abundance, since it is sensitive to the freeze-out of neutron-proton conversions, which occurs at an early stage of BBN. Using AlterBBN~\cite{Arbey:2011nf,Arbey:2018zfh}, we show the prediction on the primordial helium abundance as a function of $T_{\rm KR}$ with varying the baryon abundance within values allowed by \textit{Planck} 2018 (TT,TE,EE+lowE)~\cite{Planck:2018vyg},
together with the constraint on the abundance~\cite{ParticleDataGroup:2020ssz}, in the left panel of Fig.~\ref{fig:BBN}. The width of the prediction originates from the uncertainty in nuclear reaction rates and the neutron lifetime. From this, we obtain the constraint $T_{\rm KR} \gtrsim 2.5 \MeV$. This is stronger than the bounds obtained from other more simplified approaches in Refs.~\cite{Li:2013nal, DEramo:2017gpl}.

Here we use \textit{Planck}'s allowed range for the baryon abundance with the BBN consistency condition imposed on the helium abundance for the following reason: the BBN prediction for the helium abundance with kination does not deviate from the standard prediction as much to make the helium abundance a free parameter (see Fig.~40 of~\cite{Planck:2018vyg}).  Since BBN and CMB results for baryon abundance are in excellent agreement, relaxing the consistency condition and allowing the baryon abundance to range more freely gives very similar results for the allowed parameter space.

When kination domination occurs after BBN, but before recombination, the strongest constraint comes from the abundance of deuterium whose destruction freezes out at a late stage of BBN. We show the prediction on the primordial deuterium abundance as a function of $T_{\rm RM}$ in the right panel of Fig.~\ref{fig:BBN} from which we obtain $T_{\rm RM} \lesssim 6 \keV$.

\subsection{CMB}
\label{subsec:CMB}

The case of an early kination-dominated era with $T_{\rm KR} > 2.5 \MeV$ has no observable impact on the CMB. On the other hand, in the case with $T_{\rm RM} < 6 \keV$ the modified expansion rate of the universe can potentially lead to significant deviations in the evolution of modes on scales probed by the CMB.

The angular size of the sound horizon at the surface of last scattering, which is precisely measured, is one quantity that can be altered by a modified cosmic expansion history. We will develop some intuition for how the sound horizon is changed by assuming a kination-dominated era with $T_{\rm RM} < 6\keV$ and $T_{\rm KR} > T_{eq}$, where $T_{eq} \simeq 0.8 \eV$ is the approximate temperature at matter-radiation equality. We will use the piece-wise approximation to develop our intuition.

The comoving sound horizon can be written as
\begin{align}
    \label{eq:rs}
    r_{s}(\eta) = \int^{\eta}_{0} d\eta^{'} c_{s}(\eta^{'}),
\end{align}
where $\eta$ is the comoving horizon and $c_{s} = \sqrt{\frac{1}{3\left(1+\left(\frac{3 \rho_{b}}{4 \rho_{\gamma}}\right)\right)}}$. Here for simplicity we will assume $c_{s} = \sqrt{\frac{1}{3}}$. We can then rewrite the integral for the comoving sound horizon at matter-radiation equality as
\begin{align}
    \label{eq:rsa}
    r_{s}(\eta_{eq}) = \frac{\eta_{eq}}{\sqrt{3}} =\frac{1}{\sqrt{3}} \int_{0}^{a_{eq}} da \frac{1}{a^{2} H(a)},
\end{align}
where $a_{eq}$ is the scale factor at matter-radiation equality.
The $\rm{\Lambda CDM}$ comoving sound horizon at matter-radiation equality can then be written as
\begin{align}
    \label{eq:rslcdm}
    r_{s}(\eta_{eq},\mathrm{\Lambda CDM}) =\frac{1}{\sqrt{3}} \int_{a_{i}}^{a_{eq}} da \frac{1}{a^{2}H_{i} \left(\frac{a_{i}}{a}\right)^{2}} \simeq \frac{1}{\sqrt{3}}\frac{a_{eq}}{a^{2}_{i} H_{i}},
\end{align}
where $a_{i}\rightarrow 0$ and $H_{i}$ denote the scale factor and the Hubble scale deep in radiation domination and in the last equality we have used $a_{i} \ll a_{eq}$.
For the comoving sound horizon in the case of kination cosmology, we can divide the universe into two successive eras, $a < a_{\rm MK}$ and $a_{\rm MK}\leq a \leq a_{eq}$, where we assume that $a_{eq} \gg a_{\rm KR} \gg a_{\rm RM}$. We consider the energy density of a scalar field that behaves like matter for $a < a_{\rm MK}$ and kination for $a> a_{\rm MK}$ in addition to the standard radiation density. The sound horizon in our kination cosmology up to matter-radiation equality can then be written as
\begin{align}
    \label{eq:rskin}
    r_{s}(\eta_{eq},{\rm kination}) &= \frac{1}{\sqrt{3}}\int_{a_{i}}^{a_{\rm MK}} da \frac{1}{a^{2}H_{\rm RM} \left(\left(\frac{a_{\rm RM}}{a}\right)^{3}+\left(\frac{a_{\rm RM}}{a}\right)^{4}\right)^{1/2}} \nonumber \\
    &+\frac{1}{\sqrt{3}}\int_{a_{\rm MK}}^{a_{eq}} da \frac{1}{a^{2}H_{\rm RM} \left(\left(\frac{a_{\rm RM}}{a_{\rm MK}}\right)^{3}\left(\frac{a_{\rm MK}}{a}\right)^{6}+\left(\frac{a_{\rm RM}}{a_{\rm MK}}\right)^{4}\left(\frac{a_{\rm MK}}{a}\right)^{4}\right)^{1/2}}\nonumber \\ 
     &\simeq \frac{1}{\sqrt{3}}\frac{a_{eq}\left(1- \frac{a_{\rm KR}}{a_{eq}}\left(1+\mathcal{O}\left(\frac{a^{2/3}_{\rm RM}}{a^{2/3}_{\rm KR}},\frac{a_{\rm KR}}{a_{eq}}\right)\right)\right)}{a^{2}_{\rm RM} H_{\rm RM}},
\end{align}
where in the last approximation we have used $a_{i} \ll a_{eq}$ and the identity $a_{\rm RM} = \frac{a^{3}_{\rm MK}}{a^{2}_{\rm KR}}$. Here $\sqrt{2} H_{\rm RM}$ is the Hubble at the early matter radiation equality. Assuming $a_{eq} \gg a_{\rm KR}$ and $a^{2}_{i} H_{i} = a^{2}_{\rm RM} H_{\rm RM}$ in Eq.~(\ref{eq:rslcdm}), the relative difference between the sound horizons in $\Lambda \rm{CDM}$ and kination cosmology at last scattering is approximated by
\begin{align}
    \label{eq:error}
    \frac{\Delta r_{s}(\eta_{ls})}{r_{s}(\eta_{ls},\Lambda \rm{CDM})} \simeq \frac{a_{\rm KR}}{2.4 a_{eq}} = 3 \times 10^{-3}\left(\frac{100\eV}{T_{\rm KR}}\right),
\end{align}
where $\eta_{ls}$ is the comoving horizon at the surface of last scattering and we have assumed $a_{ls} \simeq 3 a_{eq}$, $T_{eq} = 0.8\eV$ in the last equality. Thus for large enough $T_{\rm KR}$, the deviation in the angular scale of the sound horizon at last scattering can be minimal.

\begin{figure}[t!]
    \centering
    \includegraphics[width=0.6\columnwidth]{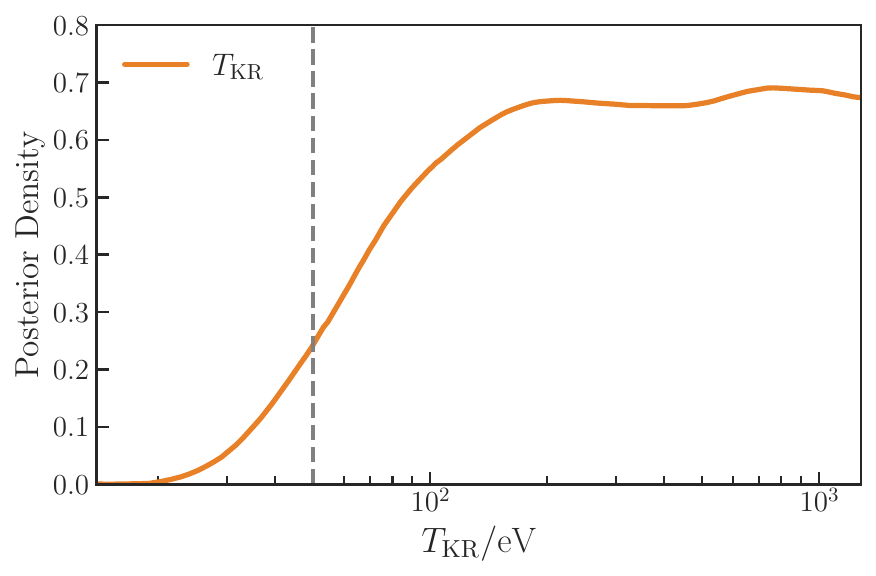}
    \caption{Posterior distribution for $T_{\rm KR}$ for a late era of kination. We use \textit{Planck} temperature and polarization data (highTTTEEE+lowEE+lowTT) to constrain $T_{\rm{KR}} > 50\, \rm{eV}$ at 95\% (vertical dashed line). See Fig.~\ref{fig:cornerplot} for the complete 2{-}dimensional posterior distributions for $\Lambda\mathrm{CDM} + T_{\mathrm{KR}}$ parameters.}
    \label{fig:TKR_CMB}
\end{figure}

The above gives some intuition for how, at fixed values of the other cosmological parameters, a kination cosmology changes the sound horizon at last scattering and by implication the angular scale of the sound horizon $\theta_{s}$. However, allowing the remaining cosmological parameters to vary---in particular $H_{0}$ or the baryon fraction $\Omega_{b}$, which enters the speed of sound---can also alter the angular scale of the sound horizon at last scattering and possibly compensate.  Moreover, the enhanced Hubble rate during the early matter-dominated era, as well as kination, changes the time-temperature relationship and modifies the evolution of perturbations, which in turn substantially impacts the oscillations in the CMB power spectrum in detail.
To assess the impact of low-scale kination on the CMB in full, we thus
need to solve the coupled Boltzmann equations governing the evolution of gravitational and matter perturbations in the modified cosmology.

In order to quantify the bounds on the kination parameters $T_{\rm KR}$ and $T_{\rm RM}$, we modify the publicly available CMB code, CLASS \cite{Blas:2011rf}, to include kination cosmology. We also use Monte Python \cite{Brinckmann:2018cvx}, a Markov Chain Monte Carlo (MCMC) code, along with CLASS to derive the posterior probability distribution on the cosmological parameters. We use a log potential discussed in the first subsection of Appendix~\ref{app:evol_axion} to describe the background cosmology as well as to derive our perturbation equations of the kination field. We choose our parameters such that $T_{\rm RM} \simeq \mathcal{O} ({\rm keV})$, consistent with the BBN bounds. For further details, see Appendix~\ref{app:parameters}.

We find that the lower bound on $T_{\rm KR}$ from the CMB is insensitive to $T_{\rm RM}$, as the CMB is mainly probing modes that enter the horizon at lower temperatures.
We consider the following cosmological parameters $\left(\Omega_{b},\, \Omega_{c}, \, \Omega_{\Lambda}, \, Y_{He},\, \theta_{s}, \, A_{s}, \, n_{s}, \,\tau, \, T_{\rm KR}\right)$. We use the \textit{Planck} 2018 CMB data (TT,TE,EE+lowE) to derive our constraints.
The posterior distribution of $T_{\rm KR}$ is shown in Fig.~\ref{fig:TKR_CMB}, from which we obtain the constraint $T_{\rm KR} > 50 \,\rm{eV}$ at 95\%.
The posterior distributions of other parameters are shown in Figs.~\ref{fig:cornerplot} and \ref{fig:cornerplot_d} in Appendix~\ref{app:parameters}.

\begin{figure}[t!]
    \centering
    \includegraphics[width=0.7\columnwidth]{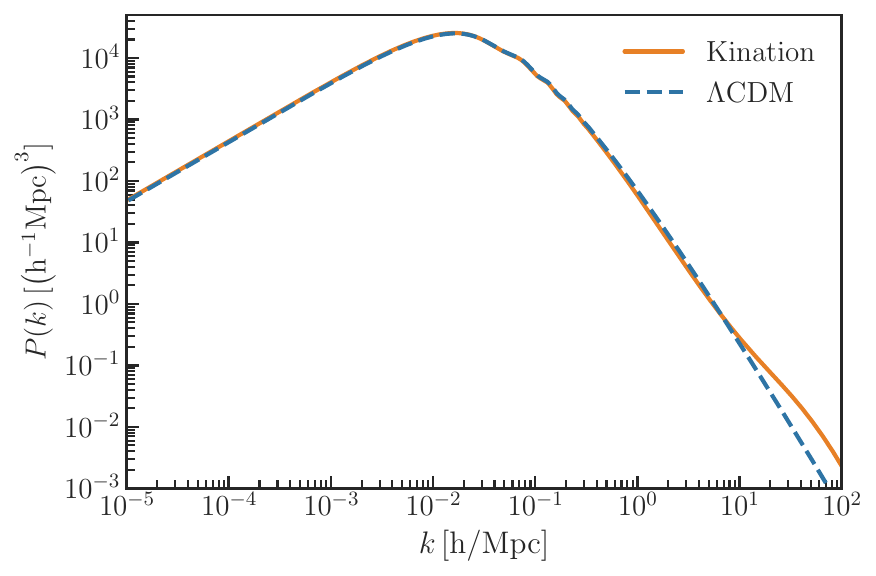}
    \caption{The  \emph{linear} matter power spectrum for $\Lambda$CDM and kination cosmology at $z = 0$. For kination cosmology we use $T_{\rm{KR}} = 50 \, \rm{\eV}$ and $T_{\rm{RM}} = 5 \,\rm{\keV}$. Kination leads to an enhanced linear power spectrum above $k \approx \mathcal{O}(1) \, \rm{h/Mpc}$.}
    \label{fig:ps}
\end{figure}

We find that matter perturbation modes that enter the horizon around $a_{\rm MK}$ grow linearly.  This rapid growth can result in an enhancement of the matter power spectrum on scales that were inside the horizon during the epochs of modified expansion.  The excellent constraints provided by the CMB require that substantial modifications to the matter power spectrum must occur on scales
$k \gg \mathcal{O}(0.1) \, \mathrm{h} \, \mathrm{Mpc^{-1}}$ (see Fig.~\ref{fig:ps}).   
Probes of the matter power spectrum at low redshifts (such as Lyman-$\alpha$) can be used to constrain the non-linear power spectrum at $k \simeq \mathcal{O}( 10) \mathrm{h}\, \mathrm{Mpc^{-1}}$ and hence raise the lower bound on $T_{\rm KR}$. However, in order to accurately derive the constraint one needs to evolve the kination matter power spectrum into the non-linear regime and then use hydrodynamical simulations to derive the Lyman-$\alpha$ flux power spectrum to compare to experiments.  This is beyond the scope of the present publication, but we will return to this in future work.

\section{Dark matter and baryogenesis from axion rotations}
\label{sec:DM_baryogenesis}
In this section, we discuss the production of axion dark matter and baryon asymmetry from axion rotations by the kinetic misalignment and axiogenesis mechanisms in the following subsections, respectively. We show the implications of these mechanisms for the parameter space $(f_a, m_S, Y_\theta,m_a)$.

\subsection{Axion dark matter from kinetic misalignment}
\label{subsec:KMM}

Axion rotations can lead to a larger axion abundance today via the kinetic misalignment mechanism~\cite{Co:2019jts} than that from the conventional misalignment mechanism~\cite{Preskill:1982cy,Abbott:1982af,Dine:1982ah}. As long as the axion field velocity is much larger than its mass $\dot\theta \gg m_a$, the axion continues to run over the potential barriers. If this motion continues past the time when the mass is equal to Hubble, then the axion kinetic energy $\dot\theta^2 f_a^2$ is much larger than the maximum possible potential energy $\theta_i^2 m_a^2 f_a^2$ in the conventional case and thus the abundance is enhanced.

\begin{figure}[t!]
    \centering
    \includegraphics[width=0.49\columnwidth]{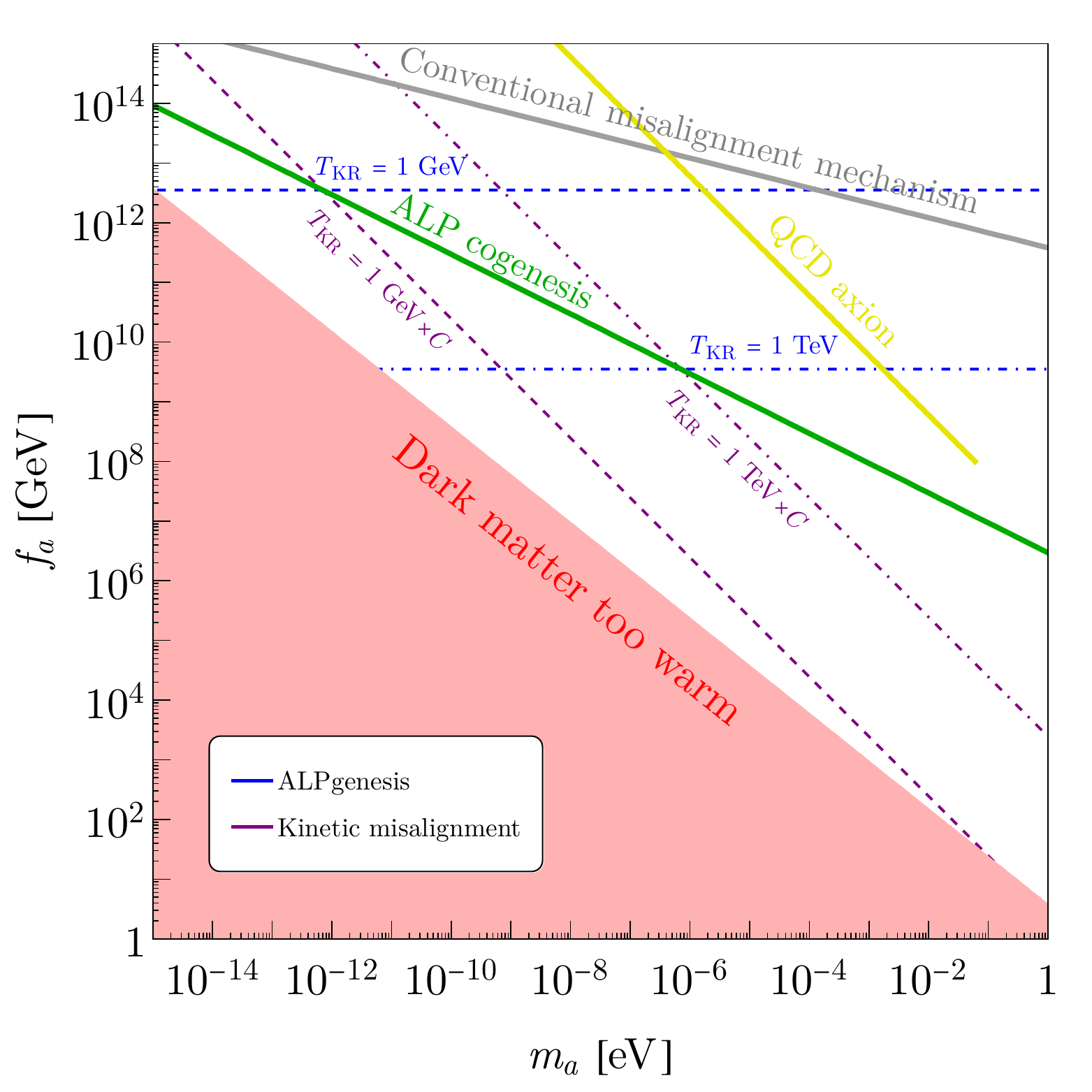}
    \includegraphics[width=0.49\columnwidth]{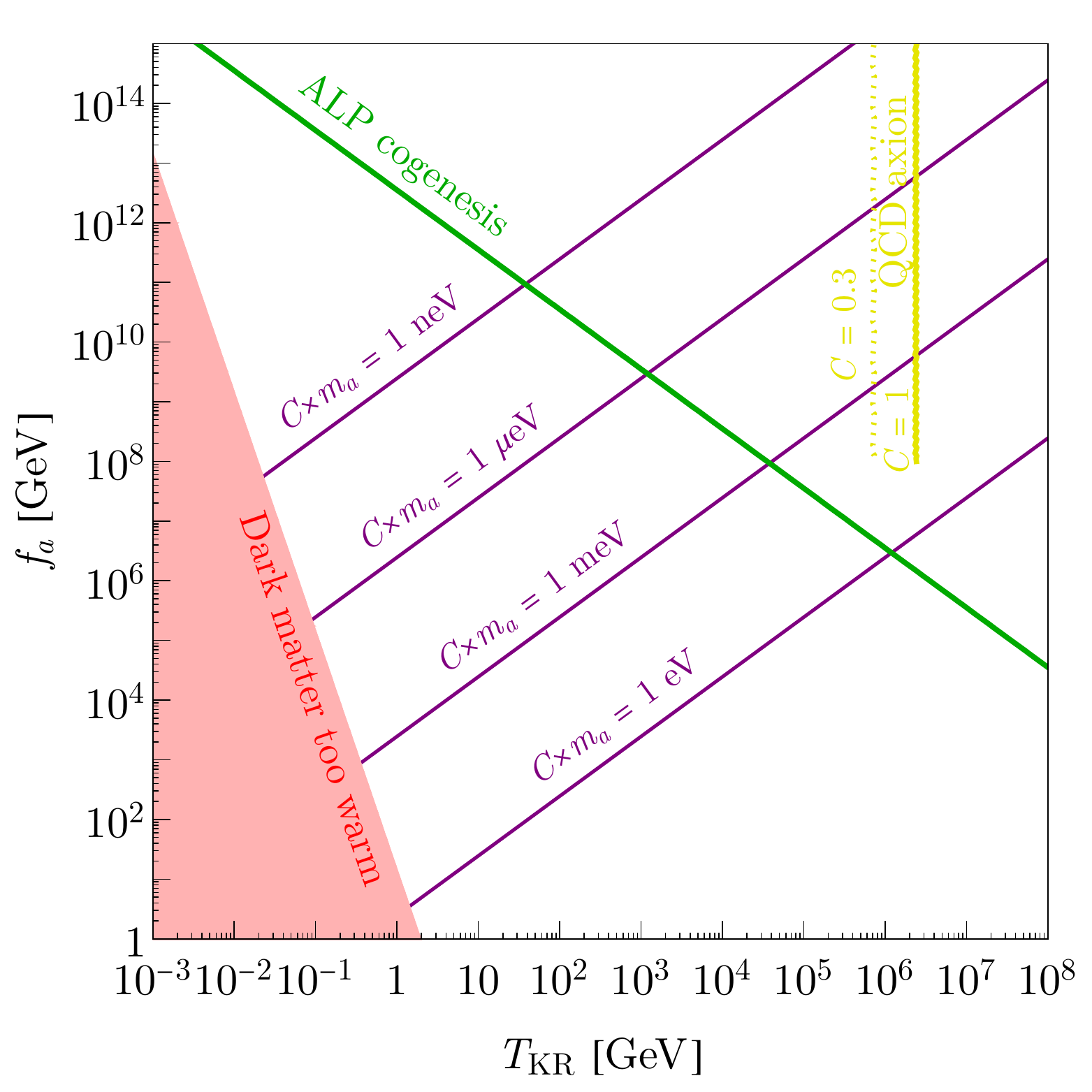}
    \caption{Axion dark matter and the baryon asymmetry from axion rotation. Left panel: in the axion parameter space, contours of $T_{\rm KR} = 1 \GeV$ ($1 \TeV$) are shown in dashed (dot-dashed) lines as predicted by dark matter from kinetic misalignment (purple) and for the baryon asymmetry from minimal ALPgenesis (blue). The contours intersect along the green line where dark matter and the baryon asymmetry are simultaneously explained as in ALP cogenesis. Right panel: the purple lines are the contours of the mass of axion dark matter predicted by kinetic misalignment as a function of $f_a$ and $T_{\rm KR}$. In both panels, the red region is excluded by the warmness of axion dark matter from kinetic misalignment. The yellow line in either plot shows the prediction assuming a QCD axion which terminates at $f_{a} = 10^{8} \GeV $ since lower $f_{a}$ is disfavored by astrophysical constraints.}
    \label{fig:axioKD}
\end{figure}

Even when the axion field velocity is larger than the mass, the axion self-interactions can cause parametric resonance (PR)~\cite{Dolgov:1989us, Traschen:1990sw, Kofman:1994rk, Shtanov:1994ce, Kofman:1997yn}, which fragments the axion rotation into axion fluctuations~\cite{Jaeckel:2016qjp, Berges:2019dgr, Fonseca:2019ypl}.
The production of fluctuations by PR occurs at an effective rate%
\footnote{The effective rate is much smaller than the PR rate at the center of the first resonance band $\sim m_a^2/\dot{\theta}$ because of the narrow width of the band, the reduction of the axion velocity by the PR production~\cite{Fonseca:2019ypl}, and the reduction of the axion momentum by cosmic expansion.}
given by~\cite{Fonseca:2019ypl}
\begin{equation}
\label{eq:Gamma_PR}
    \Gamma_{\rm PR} = \frac{m_a^4}{\dot\theta^3}.
\end{equation}
In order for kinetic misalignment to be effective, $\dot{\theta}$ must be larger than $m_a$ when $H\sim m_a$. Before $\dot{\theta}$ would become as small as $m_a$ so that the axion field would be trapped by the potential barrier, $\Gamma_{\rm PR}$ already becomes larger than $H$. Therefore, unless the angular momentum is close to the critical value for kinetic misalignment to occur, parametric resonance always becomes effective before the trapping by the potential occurs. 
On the other hand, the axion momentum $k_{\rm PR}$ generated at the time of PR is of order $\dot\theta/2$ due to the resonance condition. Therefore, the abundance of the axion is estimated as
\begin{equation}
\label{eq:rho_KMM}
    \frac{\rho_{a}}{s} =  m_a Y_a = C m_a \frac{\rho_\theta/s}{k_{\rm PR}} = C m_a Y_\theta,
\end{equation}
where the axion yield $Y_a$ is approximately conserved after PR.
Here $C$ is a factor that should be determined by numerical computation. In Ref.~\cite{Co:2019jts}, $C\simeq 2$ was derived assuming the coherence of the axion rotation throughout the evolution. As noted in Ref.~\cite{Co:2021rhi}, the axion abundance is reduced by an $\mathcal{O}(1)$ factor in comparison with the estimation in Ref.~\cite{Co:2019jts} because of the extra energy of axions from non-zero momenta sourced by PR.
Just after PR effectively occurs, the number-changing scattering rate of axion fluctuations is comparable to the Hubble expansion rate while axion fluctuations are over-occupied, so the number density may be further reduced by an $\mathcal{O}(1)$ factor, which should be determined by lattice computation; see also the discussion in~\cite{Micha:2002ey,Micha:2004bv,Co:2017mop}. We use the reference value $C=1$ in this paper and demonstrate the impact of $C<1$ on observations by showing results for $C = 0.3$.
Requiring axion dark matter from the kinetic misalignment mechanism, we obtain a prediction on $T_{\rm KR}$,
\begin{align}
\label{eq:TKR_KMM}
    T_{\rm KR} 
    & \simeq  2.4\times10^6 \GeV \times C 
    \left(\frac{f_a}{10^9 \GeV} \right)
    \left( \frac{m_a}{6~{\rm meV}} \right)
    \left( \frac{g_{*,{\rm SM}}}{g_*(T_{\rm KR})}\right)^{ \scalebox{1.01}{$\frac{1}{2}$} } 
    &  & \text{for ALPs}  \nonumber\\
    & \simeq 2.4\times 10^6 \GeV \times C 
    \left( \frac{g_{*,{\rm SM}}}{g_*(T_{\rm KR})}\right)^{ \scalebox{1.01}{$\frac{1}{2}$} } 
    &  & \text{for the QCD axion} .
\end{align}
The prediction for an ALP is shown by the purple dashed lines in the left panel of Fig.~\ref{fig:axioKD}.
To avoid overproduction of axion dark matter by the kinetic misalignment mechanism, this prediction is also a lower bound on $T_{\rm KR}$. The bound can be avoided if the rotation is washed out at $T<T_{\rm KR}$. This is difficult for the QCD axion with the Standard Model because of the suppression of the washout rate by the small up Yukawa coupling~\cite{Co:2019wyp,McLerran:1990de}, but is possible in extensions of the Standard Model. For example, squark mixing in the minimal supersymmetric standard model can indeed wash out the rotation~\cite{Co:2021qgl}. We do not pursue this possibility further. The prediction on $m_a$ as a function on $f_a$ and $T_{\rm KR}$ is shown in the right panel of Fig.~\ref{fig:axioKD}; here the prediction is also an upper bound on $m_a$.

Parametric resonance becomes effective when $\Gamma_{\rm PR} \simeq H$. One can obtain $\dot\theta(T)$, which is relevant for $\Gamma_{\rm PR}$, by using Eq.~(\ref{eq:Y_theta_def}) and requiring the axion abundance in Eq.~(\ref{eq:rho_KMM}) to reproduce the observed dark matter abundance $\rho_{\rm DM}/s \simeq 0.44$ eV. The temperature $T_{\rm PR}$ when PR occurs is given by
\begin{equation}
    T_{\rm PR} \simeq 100 \MeV 
    \left( \frac{f_a}{10^9 \GeV} \right)^{ \scalebox{1.01}{$\frac{6}{11}$} } 
    \left( \frac{m_a}{10^{-6} \eV} \right)^{ \scalebox{1.01}{$\frac{7}{11}$} } 
    \left( \frac{g_{*,{\rm SM}}}{g_*(T_{\rm PR})} \right)^{ \scalebox{1.01}{$\frac{7}{22}$} } C^{3/11} ,
\end{equation}
where we assume that the axion mass at $T = T_{\rm PR}$, is the same as the one in vacuum, $m_a(T_{\rm PR}) = m_a$, and the saxion is at the minimum of the potential, $S(T_{\rm PR}) = f_a$.
As with PR production from radial motion of the symmetry breaking field~\cite{Co:2017mop,Co:2020dya}, the produced axions are initially relativistic. They may become cold enough to be dark matter by red-shifting, and will have residual warmness~\cite{Berges:2019dgr,Co:2021rhi}.
They become non-relativistic at temperature
\begin{equation}
\label{eq:T_NR}
    T_{\rm NR} \simeq 10 \MeV 
    \left( \frac{f_a}{10^9 \GeV} \right)^{ \scalebox{1.01}{$\frac{10}{11}$} } 
    \left( \frac{m_a}{10^{-6} \eV} \right)^{ \scalebox{1.01}{$\frac{8}{11}$} } 
    \left( \frac{g_{*,{\rm SM}}}{g_*(T_{\rm PR})} \right)^{ \frac{1}{33} } C^{5/11}.
\end{equation}
For sufficiently small $m_a$ and/or $f_a$, these axions are too warm to be dark matter based on the current warmness constraint from the Lyman-$\alpha$ measurements~\cite{Irsic:2017ixq}, $T_{\rm NR} > 5 \keV$. This constraint is shown by the red regions of Fig.~\ref{fig:axioKD}.

\subsection{Baryon asymmetry from axiogenesis}
\label{subsec:baryogenesis}

The observed cosmological excess of matter over antimatter can also originate from the axion rotation. The $U(1)$ charge associated with the rotation defined in Eq.~(\ref{eq:Y_theta_def}) can be transferred to the baryon asymmetry as shown in~\cite{Co:2019wyp,Co:2020xlh,Co:2020jtv}. In the case of the QCD axion, the strong anomaly necessarily transfers the rotation into the quark chiral asymmetry, which is distributed into other particle-antiparticle asymmetry.
More generically, the couplings of the QCD axion or an ALP with the thermal bath can transfer the rotation into particle-antiparticle asymmetry.
The particle-antiparticle asymmetry can be further transferred to baryon asymmetry
via processes that violate the baryon number.
We call this scheme, applicable to the QCD axion and ALPs, axiogenesis. To specifically refer to the QCD axion and ALPs, we use QCD axiogenesis and ALPgenesis, respectively. 

\subsubsection{Minimal axiogenesis}
In the minimal scenario, which we call minimal axiogenesis, the particle-antiparticle asymmetry is reprocessed into the baryon asymmetry via  the electroweak sphaleron processes.
If the QCD axion or an ALP has an electroweak anomaly, then the rotation can directly produce the baryon asymmetry by the electroweak sphaleron processes.
The contribution to the yield of the baryon asymmetry is given by~\cite{Co:2019wyp}
\begin{equation}
\label{eq:YB_minaxio}
    Y_B = \frac{n_B}{s} = \left. \frac{45 c_B}{2 g_* \pi^2} \frac{\dot{\theta}}{T} \right|_{T = T_{\rm ws}} \hspace{-0.2 in} \simeq 8.2 \times 10^{-11} \left( \frac{c_B}{0.1} \right) \left( \frac{\dot{\theta}(T_{\rm ws})}{5 \keV} \right) \left( \frac{130 \GeV}{T_{\rm ws}} \right),
\end{equation}
where $T_{\rm ws}$ is the temperature at which the electroweak sphaleron processes go out of equilibrium and is approximately $130 \GeV$ in the Standard Model~\cite{DOnofrio:2014rug}, and $c_B$ is a model-dependent coefficient given in Ref.~\cite{Co:2020xlh} that parameterizes the anomaly coefficients and the axion-fermion couplings. When the transfer is dominated by axion-gauge boson couplings, $c_B$ is typically $\mathcal{O}(0.1)$, while if dominated by axion-fermion couplings, it can be much smaller.

For the QCD axion, to produce sufficient $Y_B$ and to avoid overproduction of dark matter by kinetic misalignment requires that $f_a \lesssim 10^7 \GeV c_B/C $, which is disfavored by astrophysical constraints~\cite{Ellis:1987pk,Raffelt:1987yt,Turner:1987by,Mayle:1987as,Raffelt:2006cw,Payez:2014xsa,Bar:2019ifz} unless $c_B/C > 10$. The baryon asymmetry can be enhanced if the electroweak phase transition occurs at a higher temperature,
\begin{align}
 T_{\rm ws} \geq 1~{\rm TeV} 
 \left( \frac{f_a}{10^8 \GeV}\right)^{ \scalebox{1.01}{$\frac{1}{2}$} } 
 \left( \frac{0.1}{c_B}\right)C^{1/2},
\end{align}
with both dark matter and baryon asymmetry of the universe explained by the rotation of the QCD axion when the inequality is saturated.

For an ALP, we may choose sufficiently small $m_a$ to avoid the over production without modifying the electroweak phase transition temperature.
Requiring that the baryon yield of Eq.~(\ref{eq:YB_minaxio}) match the observed baryon asymmetry gives a constraint on $\dot\theta(T_{ws})$; using Eq.~(\ref{eq:TKR}) this can be converted to a prediction for $T_{\rm KR}$
\begin{align}
\label{eq:TKRALPgen}
    T_{\rm KR} = 3.5~{\rm TeV} 
    \left(\frac{10^9 \GeV}{f_a}\right) 
    \left(\frac{f_a}{S(T_{\rm ws})}\right)^2 
    \frac{c_B}{0.1} \left(\frac{g_{*,{\rm SM}}}{g_*}\right)^{ \scalebox{1.01}{$\frac{1}{2}$} } ,
\end{align}
which is shown by the blue dot-dashed line in the left panel of Fig.~\ref{fig:axioKD}, assuming $S(T_{\rm ws}) = f_a$. Note that this is necessarily the case when $T_{\rm MK} >T_{\rm ws}$. For lower $T_{\rm MK}$, $S(T_{\rm ws}) > f_a$ is possible.
Since $\dot\theta \simeq m_S$ when $S > f_a$ and $\dot\theta \propto T^3$ after $S = f_a$, we have $\dot{\theta}(T_{\rm ws}) \lesssim m_S$ and therefore the saxion mass is predicted to be
\begin{align}
    m_S > 5 \keV \left( \frac{0.1}{c_B} \right).
\end{align}
The bound is saturated when $S(T_{\rm ws}) > f_a$, which is the case if $T_{\rm MK} < T_{\rm ws}$.

Both dark matter and baryon asymmetry of the universe is explained by the axion rotation, which is called ALP cogenesis~\cite{Co:2020xlh}, when
\begin{align}
\label{eq:ALPcogenesis}
    m_a = 8.5~{\rm \mu eV} \left(\frac{10^9 \GeV}{f_a}\right)^2 \left( \frac{f_a}{S(T_{\rm ws})}\right)^2 
    \left( \frac{c_B}{0.1} \right) 
    \left( \frac{1}{C} \right).
\end{align}
This prediction is shown by the green lines in  Fig.~\ref{fig:axioKD}.

\subsubsection{$B-L$ number violation by new physics}

In the presence of an operator that violates lepton number and generates Majorana neutrino masses, the transfer of asymmetries can be more efficient. The operator creates
a non-zero $B-L$ number, which is preserved by Standard Model electroweak sphaleron processes. Since the production of $B-L$ at high temperatures depends on whether the lepton number violating interactions are in equilibrium, the determination of the final baryon number in this scenario is sensitive to the full cosmological evolution. As an example, for the models studied in Ref.~\cite{Co:2020jtv}, the baryon asymmetry is given by
\begin{equation}
\label{eq:YB_lepto-ALPgenesis}
    Y_{B} \simeq 8.7 \times 10^{-11} N_{\rm DW}  \left( \frac{c_B}{0.1} \right) \left(\frac{g_{\rm MSSM}}{g_*} \right)^{ \scalebox{1.01}{$\frac{3}{2}$} } \left( \frac{\bar{m}^2}{0.03~{\rm eV}^2} \right) \left( \frac{m_S}{30 \TeV} \right) \left( \frac{D}{23} \right) ,
\end{equation}
where $\bar{m}^2 = \Sigma_{i}m^{2}_{i}$ is the sum of the square of the neutrino masses $m_i$, $N_{\rm DW}$ is the domain wall number (which is assumed to be unity in other parts of the paper), and the function $D$ parameterizes the different cosmological scenarios. In particular, $D = \mathcal{O}(20)$ for the case where no entropy is produced after the production of $B-L$ and is logarithmically dependent on $f_a$ as well as the saxion field values at various temperatures. Alternatively, if the saxion dominates before decaying and reheating the Universe $D = 1$, and (\ref{eq:YB_lepto-ALPgenesis}) yields a sharp prediction for $m_S$, valid when the saxion is thermalized before settling to $f_a$. For details of the evaluation of $D$, one can refer to Ref.~\cite{Co:2020jtv}. Regardless, the saxion mass is generically predicted to be $\mathcal{O}(30-10^4) \TeV \times (0.1/c_B)$ by this baryogenesis mechanism, named lepto-axiogenesis generically, QCD lepto-axiogenesis for the QCD axion, and lepto-ALPgenesis for the ALP. While $m_S \ll 30 \TeV$ appears difficult based on Eq.~(\ref{eq:YB_lepto-ALPgenesis}), the case of TeV scale supersymmetry is possible in some special cases presented in Ref.~\cite{Co:2020jtv}, using a thermal potential.

Other axiogenesis scenarios are also considered in the literature~\cite{Harigaya:2021txz,Chakraborty:2021fkp}. Baryon asymmetry may be dominantly produced at a temperature $T_{\rm dec}$ with $Y_B\sim \dot{\theta}/T \times {\rm min}(1,\Gamma_B/H)$, where $\Gamma_B$ is the transfer rate of the axion rotation into baryon asymmetry. For models with $T_{\rm dec} > T_{\rm ws}$, the lower bound on $m_S$ is generically stronger than that for ALPgenesis.

\begin{figure}
    \centering
    \includegraphics[width=0.495\columnwidth]{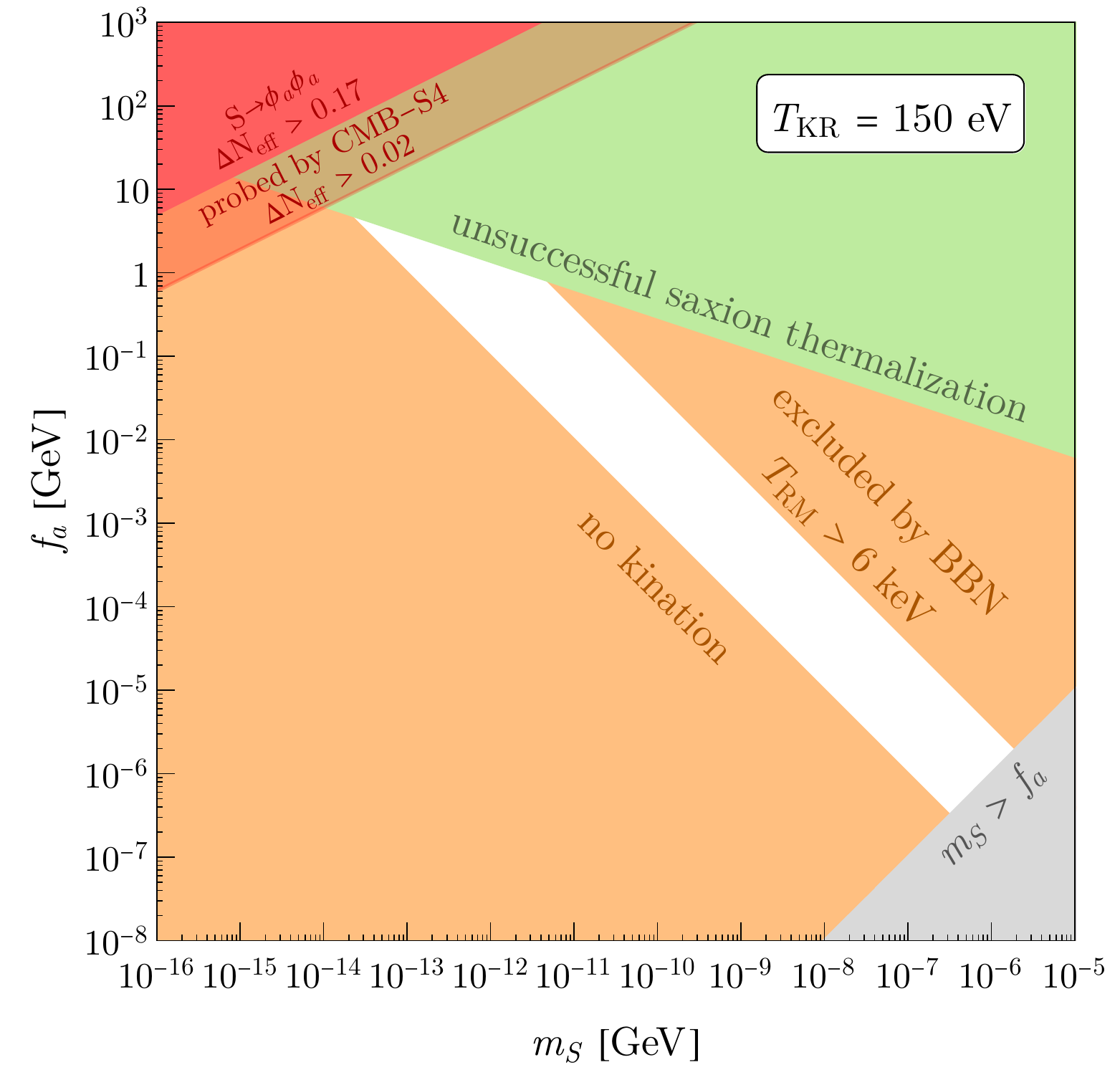}
    \includegraphics[width=0.495\columnwidth]{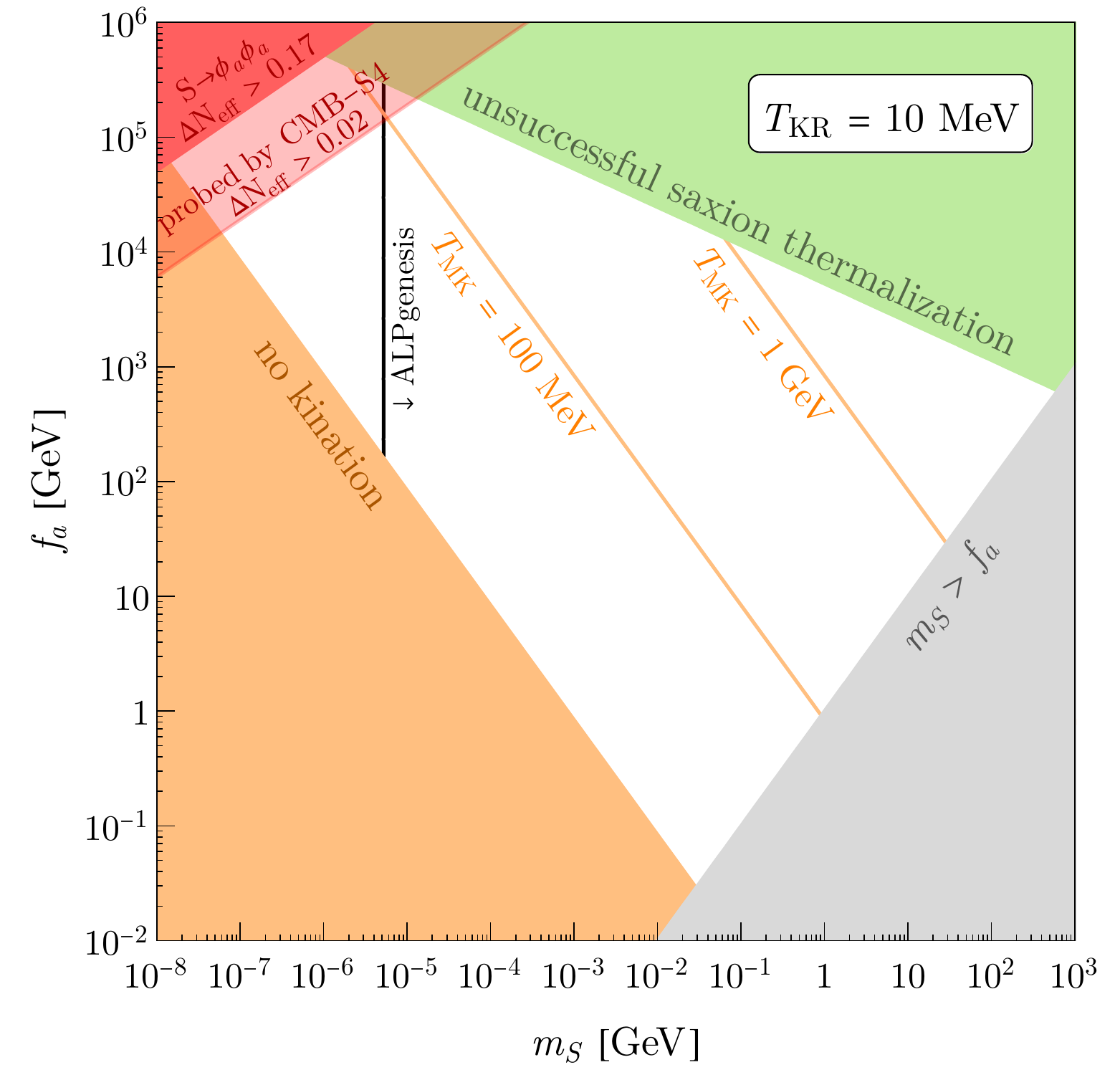}
    \includegraphics[width=0.495\columnwidth]{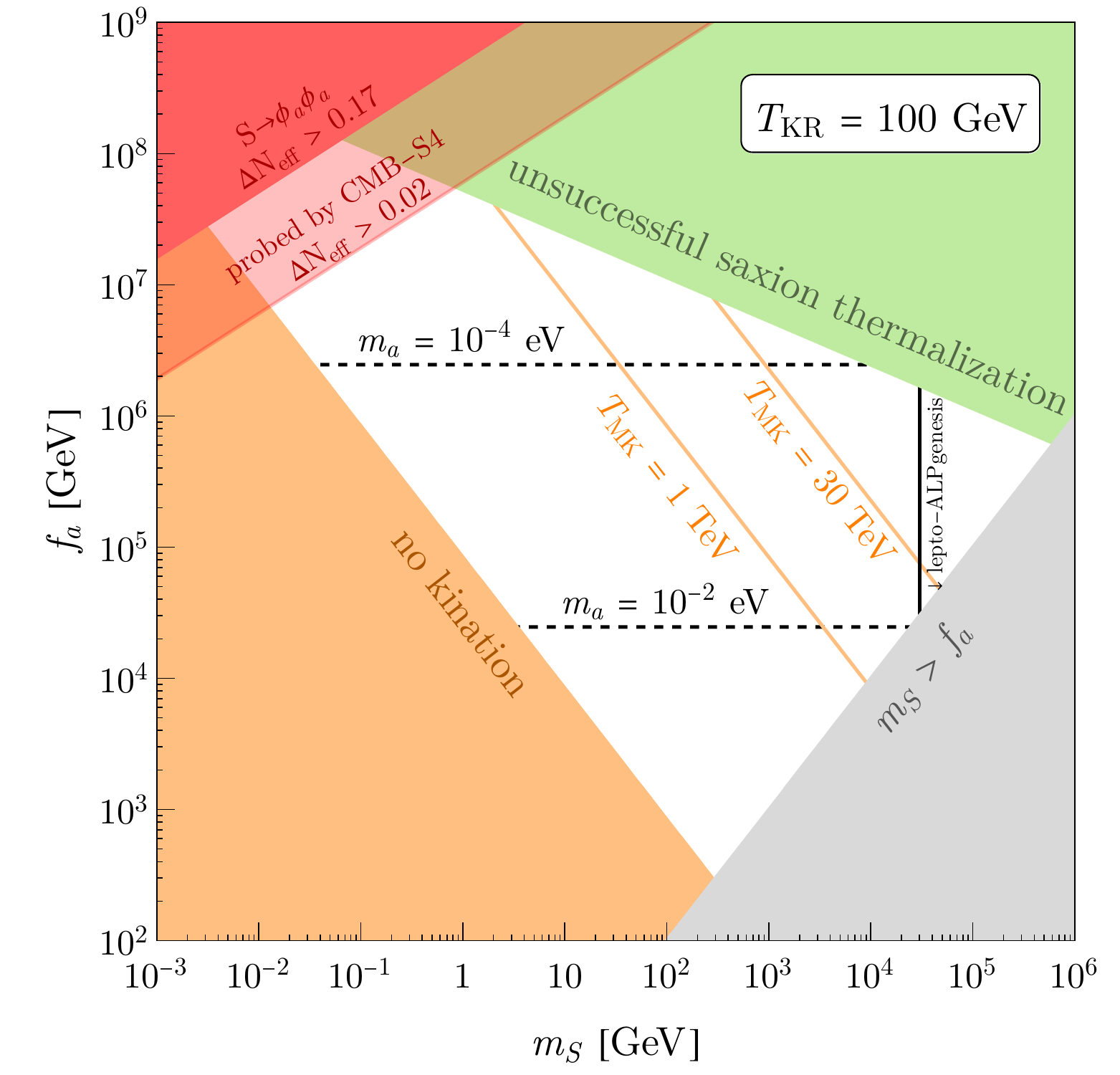}
    \includegraphics[width=0.495\columnwidth]{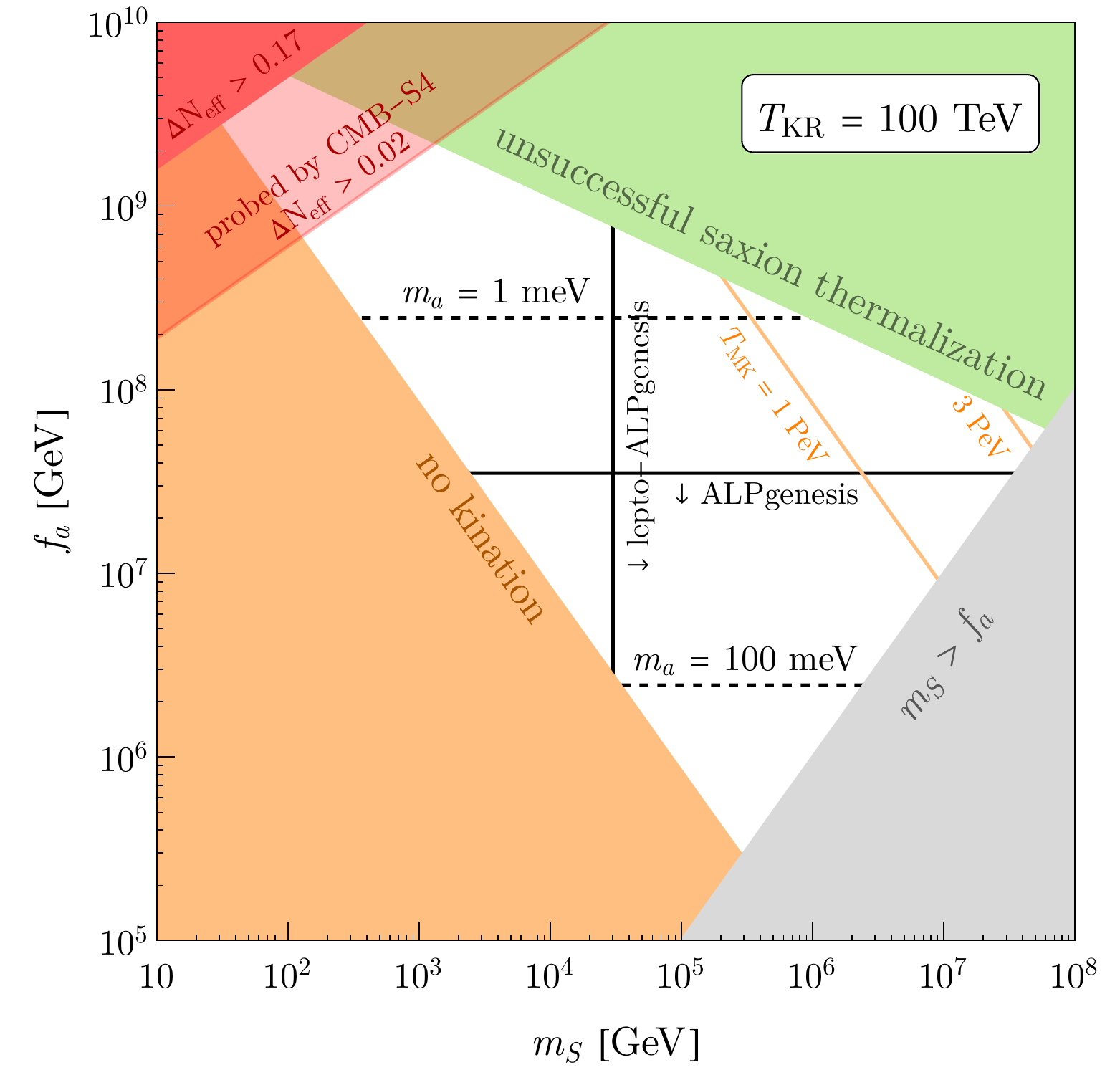}
    \caption{The unshaded regions show the allowed parameter space for axion kination for the fixed values of $T_{\rm KR}$ labeled in each panel.  Contours of $T_{\rm MK}$ are shown in these regions with kination. The excluded shaded regions are discussed in the text. To achieve minimal ALPgenesis, the parameter space collapses into $m_S \simeq 5 \keV (0.1/c_B)$ as shown by the black solid line in the upper-right panel, or into $f_a$ given by Eq.~(\ref{eq:TKRALPgen}) with $S(T_{\rm ws}) = f_a$ as shown by the black solid line in the lower-right panel, where we take $c_B=0.1$. On the other hand, lepto-ALPgenesis restricts the parameter space to $m_S \gtrsim 30 \TeV$. The axion cannot constitute dark matter via kinetic misalignment in the upper panels due to the warmness constraint in Eq.~(\ref{eq:T_NR}).}
    \label{fig:kination_fixTKR}
\end{figure}

\subsubsection{Implications to axion kination parameters}

In Fig.~\ref{fig:kination_fixTKR}, we show constraints on the parameter space for several reference values of $T_{\rm KR}$. The green-shaded regions are excluded because of the failure of thermalization of the initial radial oscillation, as described in Sec.~\ref{subsec:thermalization}. No kination-dominated era arises in the lower orange-shaded regions. The upper orange-shaded region in the upper-left panel is excluded by BBN. The orange lines are the contours of $T_{\rm MK}$. In the gray-shaded region, $m_S$ is above $f_a$ and the perturbativity of the potential of the $U(1)$ symmetry breaking field breaks down.
The red-shaded region is excluded by dark radiation produced by the decay of thermalized saxions into axions.  The remaining unshaded regions give the allowed parameter space where axion rotation leads to realistic cosmologies with early eras of matter and kination domination. Contours of $T_{\rm MK}$ are shown in these kination regions.

In the lower two panels, the horizontal black dashed lines  show the prediction for the axion mass from requiring that the observed dark matter result from the kinetic misalignment mechanism. In the upper two panels, axion dark matter from kinetic misalignment is excluded as it is too warm.

In the upper-left panel, no parameter region is consistent with the lower bound on $m_S$ from axiogenesis above the electroweak scale. In the upper-right panel, $m_S$ can be above the keV scale. $T_{\rm MK}$ is below the electroweak scale, so $S(T_{\rm ws}) > f_a$ and $\dot{\theta}_{\rm ws} = m_S = 5 $ keV $(0.1/c_B)$ is required.
In the lower-left panel, $T_{\rm MK} > 100$ GeV, so that the condition for successful minimal ALPgenesis is given by (\ref{eq:TKRALPgen}) with $S(T_{\rm ws}) = f_a$, 
requiring $c_B \ll 1$. Lepto-ALPgenesis is possible to the right of the black solid line.
In the lower-right panel, where $T_{\rm MK} > 100$ GeV, minimal ALPgenesis requires $f_a$ shown by the vertical black solid line according to Eq.~(\ref{eq:TKRALPgen}) with $c_B=0.1$. Lower $f_a$ is possible if $c_B < 0.1$.

\section{Gravitational waves}
\label{sec:GW}

In this section, we discuss how the spectrum of primordial gravitational waves is modified by eras of matter and kination domination generated from axion rotation, as discussed in Sec.~\ref{sec:rotation_kination}. We consider gravitational waves created by quantum fluctuations during inflation and by local cosmic strings. In both production mechanisms, the spectrum is nearly flat in the standard cosmology with radiation domination.
As we will see, the evolution of a universe with successive eras dominated by radiation, matter, kination, and back to radiation induces a triangular peak in the gravitational wave spectrum that can provide a unique signal for axion rotation and kination cosmology.

\subsection{From inflation}
\label{subsec:GW_inflation}

We first discuss the primordial gravitational waves produced from quantum fluctuations during inflation~\cite{Starobinsky:1985ww}. In the standard cosmology with radiation domination, the spectrum is nearly flat for the following reason. After inflation, a given mode $k$ is frozen outside the horizon, $k< H$. As the mode reenters the horizon, $k >H$, it begins to oscillate and behaves as radiation. The energy density of the mode at that point $\sim k^2 h^2(k) \mpl^2 $, where $h$ is the metric perturbation, whose spectrum is almost flat for slow-roll inflation. Since $k\sim H$ at the beginning of the oscillation, the energy density of the gravitational waves normalized by the radiation energy density $\sim H^2 \mpl^2$ is nearly independent of $k$ up to a correction by the degree of freedom of the thermal bath%
\footnote{Free-streaming neutrinos damp the amplitude of the gravitational waves for $f\lesssim 0.1 {\rm nHz}$~\cite{Weinberg:2003ur}.} 

During matter or kination domination in our scenario, the energy density at the horizon crossing is still $H^2 h^2 \mpl^2$, but the radiation energy density is now much smaller than $H^2 \mpl^2$. As a result, the energy density of gravitational waves with a mode $k$ is inversely proportional to the fraction of the radiation energy density to the total energy density when the mode enters the horizon. This means that the spectrum should feature a triangular peak in axion kination. The modes that enter the horizon at $T > T_{\rm RM}$ are not affected and remain flat. (See, however, a comment below). Therefore, as the horizon-crossing temperature decreases below $T_{\rm RM}$, the gravitational waves are enhanced and reach the maximal value at $T_{\rm MK}$, where the fraction of the radiation energy is minimized. The gravitational wave strength decreases again for the horizon crossing temperatures below $T_{\rm MK}$, and returns to a flat spectrum below $T_{\rm KR}$. Note that gravitational waves that enter the horizon during matter domination are also enhanced because of the absence of entropy production after matter domination.
We use an analytical approximation where each mode begins oscillations suddenly at the horizon crossing. 
Also approximating the evolution of $H$ by a piecewise function with kinks at the three transition, the resultant spectrum of gravitational waves is given by
\begin{align}
    \Omega_{\rm GW} h^2 & \simeq 1.4\times 10^{-17} \left( \frac{V_{\rm inf}^{1/4}}{10^{16} \GeV} \right)^4 \left( \frac{g_{*,{\rm SM}}}{g_*(T_{\rm hc})} \right)^{ \scalebox{1.01}{$\frac{1}{3}$} } 
\begin{cases} 
 1  
    & \text{RD~:~} f_{\rm RM} < f \\
 \left(\frac{f_{\rm RM}}{f}\right)^2 
    & \text{MD~:~} f_{\rm MK} < f < f_{\rm RM} \\
 \frac{f}{f_{\rm KR}} 
    & \text{KD~:~} f_{\rm KR} < f < f_{\rm MK} \\
 1  
    & \text{RD~:~} f < f_{\rm KR} 
\end{cases},  \\
f_{\rm RM,KR} & \simeq 27~\mu{\rm Hz} 
\left( \frac{T_{\rm RM,KR}}{{\rm TeV}} \right) 
\left( \frac{g_*(T_{\rm RM,KR})}{g_{*,{\rm SM}}} \right)^{ \scalebox{1.01}{$\frac{1}{6}$} },  \\
f_{\rm MK} & = (f_{\rm RM}^2 f_{\rm KR})^{1/3} \simeq 27~\mu{\rm Hz} 
\left( \frac{T_{\rm RM}}{{\rm TeV}} \right)^{ \scalebox{1.01}{$\frac{2}{3}$} }
\left( \frac{T_{\rm KR}}{{\rm TeV}} \right)^{ \scalebox{1.01}{$\frac{1}{3}$} }
\left( \frac{g_*(T_{\rm RM})}{g_{*,{\rm SM}}} \right)^{ \scalebox{1.01}{$\frac{1}{9}$} }
\left( \frac{g_*(T_{\rm KR})}{g_{*,{\rm SM}}} \right)^{ \scalebox{1.01}{$\frac{1}{18}$} },
\end{align}
where $V_{\rm inf}$ is the potential energy during inflation.
We normalize the spectrum to match a full numerical result at $f< f_{\rm KR}$ and $f> f_{\rm RM}$, which is found to be consistent with the result of the numerical computation in Ref.~\cite{Saikawa:2018rcs}.

\begin{figure}
    \centering
    \includegraphics[width=\columnwidth]{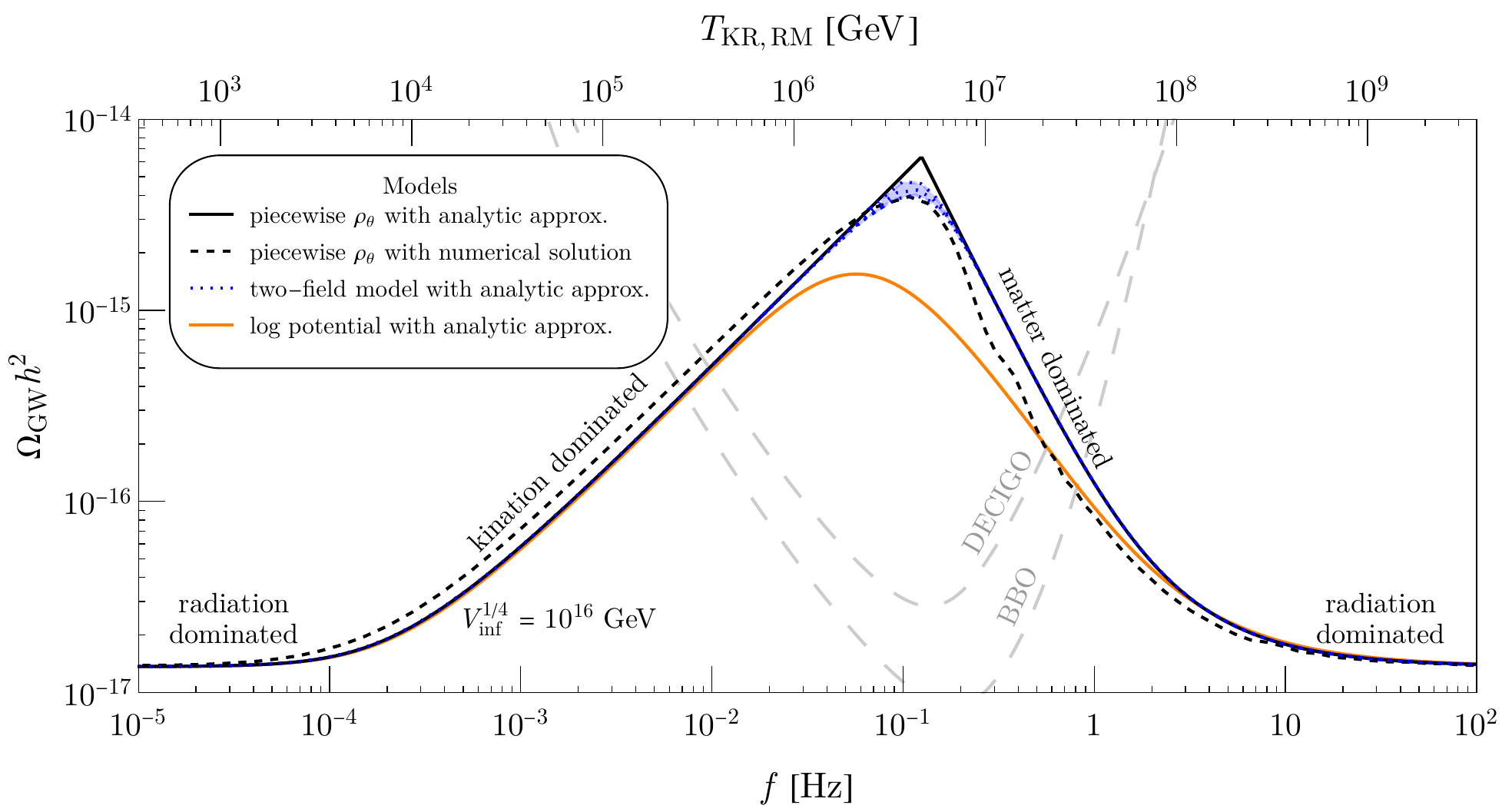}
    \caption{An illustration of the model dependence in the primordial gravitational wave spectrum. Here we fix $T_{\rm KR} = 10^4 \GeV$, $T_{\rm RM} = 10^8 \GeV$ (and accordingly $T_{\rm MK} \simeq 2 \times 10^5 \GeV$), and the inflationary energy scale $V_{\rm inf}^{1/4} = 10^{16} \GeV$. The black lines are for the case where the rotation energy density $\rho_\theta$ follows a piecewise scaling when $T \lessgtr T_{\rm MK}$ as shown in Fig.~\ref{fig:EoS}. The solid (dashed) black lines are obtained from an analytic (numerical) derivation of the evolution of the metric perturbations. The colored curves are for the two-field model (blue) and the logarithmic potential (orange) with evolution demonstrated in Fig.~\ref{fig:EoS}. For the two-field model, we show the blue dotted curves for different ratios of the soft masses of the two fields $\bar{P}$ and $P$, $m_{\bar P}^2/m_P^2 = 1, 2, \infty$ from top to bottom.}
    \label{fig:PGW_models}
\end{figure}

In Fig.~\ref{fig:PGW_models}, we illustrate the spectrum of gravitational waves in axion kination.
Throughout this paper, we use the sensitivity curves derived in Ref.~\cite{Schmitz:2020syl} for NANOGrav~\cite{McLaughlin:2013ira,NANOGRAV:2018hou,Aggarwal:2018mgp,Brazier:2019mmu}, PPTA~\cite{Manchester:2012za,Shannon:2015ect}, EPTA~\cite{Kramer:2013kea,Lentati:2015qwp,Babak:2015lua}, IPTA~\cite{Hobbs:2009yy,Manchester:2013ndt,Verbiest:2016vem,Hazboun:2018wpv}, SKA~\cite{Carilli:2004nx,Janssen:2014dka,Weltman:2018zrl}, LISA~\cite{LISA:2017pwj, Baker:2019nia}, BBO~\cite{Crowder:2005nr,Corbin:2005ny,Harry:2006fi}, DECIGO~\cite{Seto:2001qf,Kawamura:2006up,Yagi:2011wg}, CE~\cite{LIGOScientific:2016wof,Reitze:2019iox} and ET~\cite{Punturo:2010zz, Hild:2010id,Sathyaprakash:2012jk, Maggiore:2019uih}, and aLIGO and aVirgo~\cite{Harry:2010zz,LIGOScientific:2014pky,VIRGO:2014yos,LIGOScientific:2019lzm}.
The black solid and dashed curves are both based on the piecewise approximation of the $\rho_\theta$ contribution to the Hubble rate, whereas the black solid (dashed) curve is with the analytical approximation (numerical solution) of the horizon crossing. Here $H$ is computed by the addition of $\rho_\theta$ and the radiation energy density, so smoothly changes around $T_{\rm RM}$ and $T_{\rm KR}$. As the analytic approximation reproduces the numerical result very well, we use the analytic approximation of the horizon crossing in the remainder of this paper. The blue dotted and orange solid lines show the spectrum for the two-field model and the log potential, respectively. The spectrum for the two-field model is close to that for the piecewise approximation, while that for the log potential deviates from them. Remarkably, the measurement of the gravitational wave spectrum around the peak can reveal the shape of the potential that spontaneously breaks the $U(1)$ symmetry.

We comment on possible further modification of the spectrum. Axion kination relies on a nearly quadratic saxion potential, which is natural in supersymmetric theories.
The degrees of freedom of the thermal bath change by about a factor of two across the superpartner mass threshold, suppressing the gravitational wave signals by a few tens of a percent at high frequency~\cite{Watanabe:2006qe}. This depends on the superpartner masses, and we do not include this effect for simplicity.
We also assume that the radial mode of $P$ does not dominate the energy density of the universe. If it does, entropy is created by the thermalization of the radial mode and gravitational waves at $f > f_{\rm RM}$ can be suppressed. If the initial rotation before thermalization is highly elliptical, after the thermalization the universe is radiation dominated for a long time because of the radial mode energy much larger than the angular mode energy, so the suppression occurs at $f \gg f_{\rm RM}$. If the initial rotation is close to a circular one, the universe is radiation dominated only for a short period, so the suppression occurs right above $f_{\rm RM}$. In principle, we can learn about the very UV dynamics of axion rotations through the observations of gravitational waves.

\begin{figure}
    \centering
    \includegraphics[width=0.495\columnwidth]{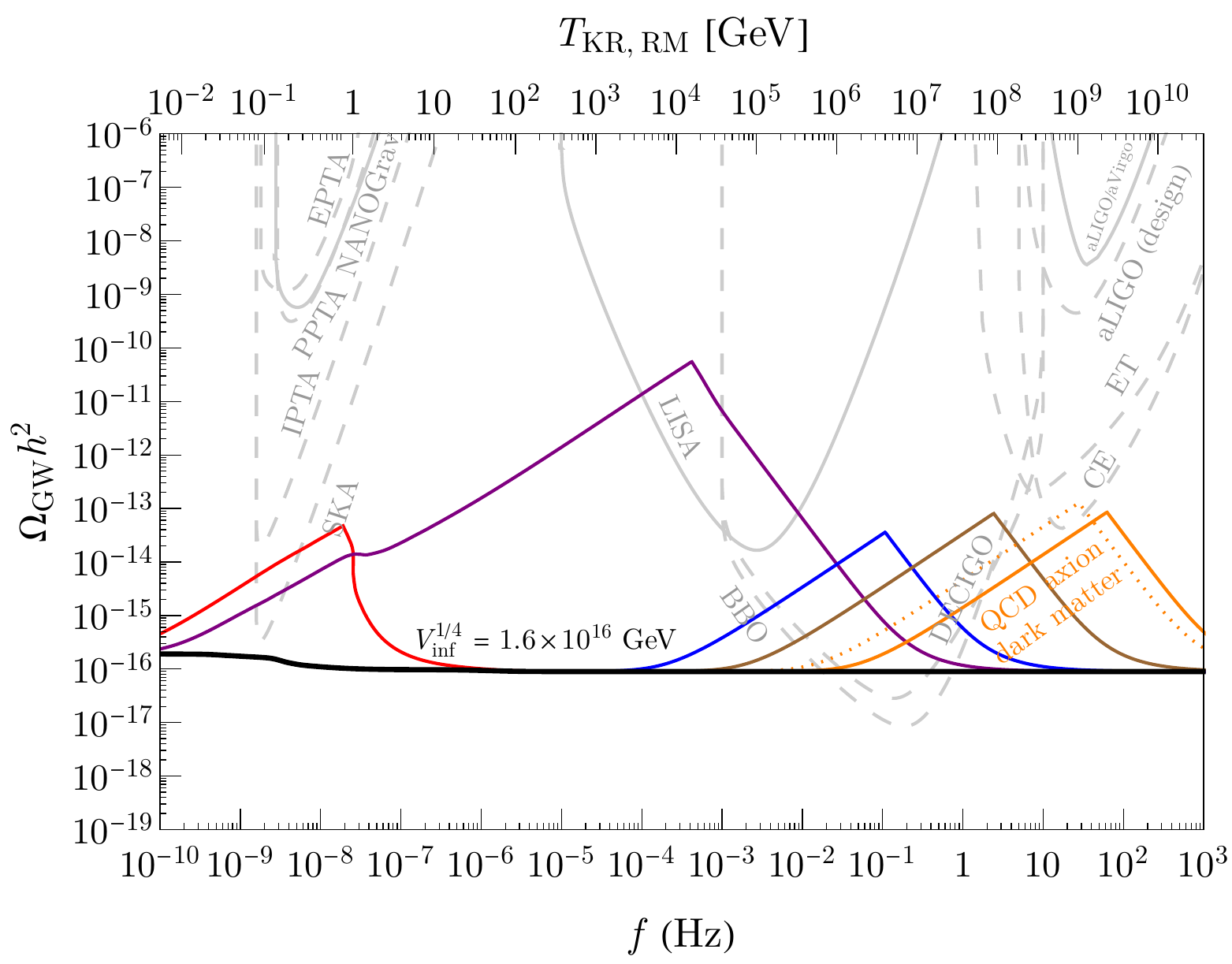}
    \includegraphics[width=0.495\columnwidth]{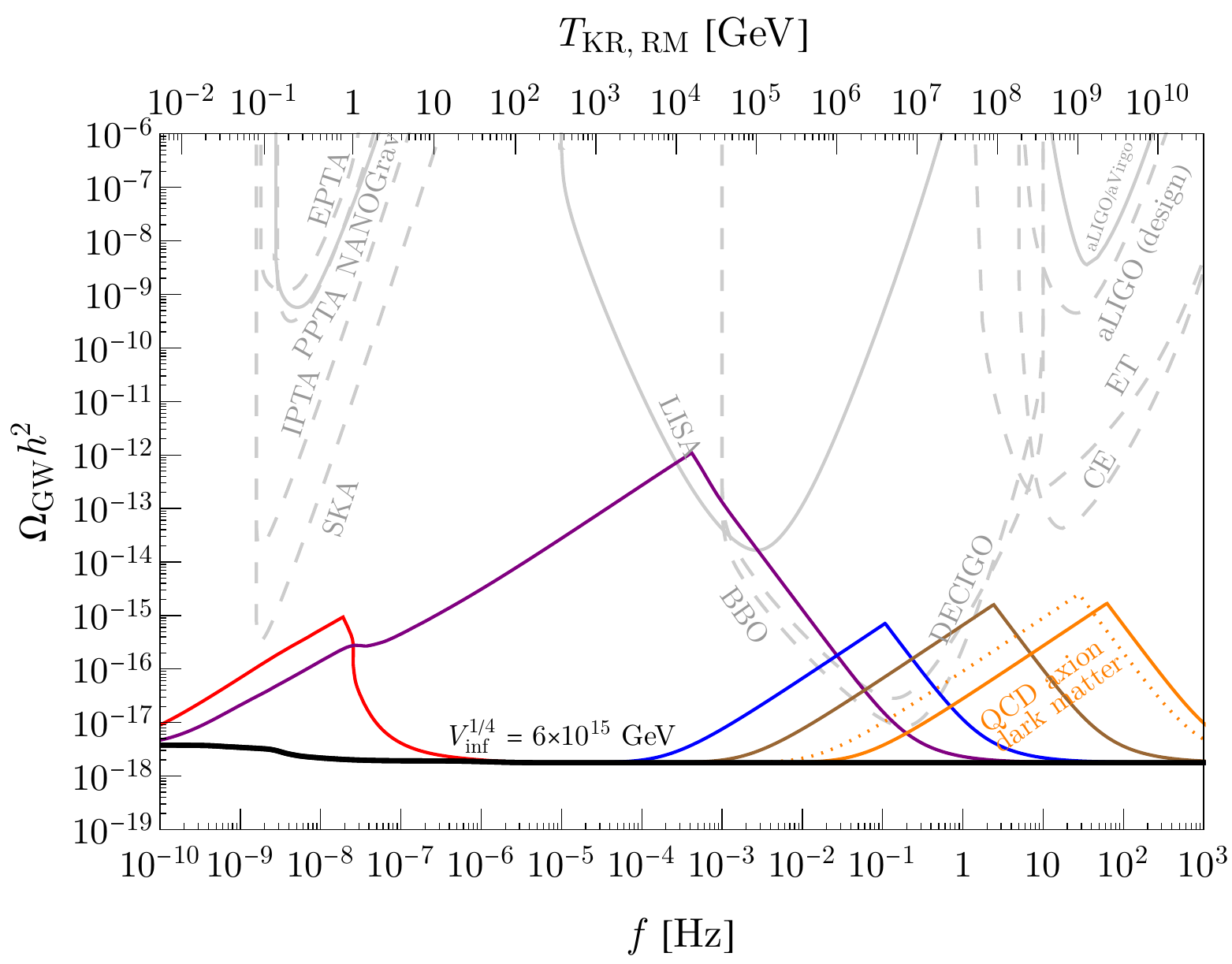}
    \caption{GW spectra from inflation for inflationary energy scale $V_{\rm inf}^{1/4}$ of $1.6 \times 10^{16}$ GeV (left panel) and $6 \times 10^{15}$ GeV (right panel). Each panel contains various choices of $(T_{\rm KR}, T_{\rm RM})$. The left (right) vertex of each triangle approximately indicates the choice of $T_{\rm KR}$ ($T_{\rm RM}$) labeled at the top axis, while $T_{\rm MK}^3 = T_{\rm RM}  T_{\rm KR}^2$. The $(T_{\rm KR}, T_{\rm RM})$  choices are $(3 \MeV, 3 \GeV)$ for red, $(10^{-2}, 10^7) \GeV$ for purple, $(10^4, 8 \times 10^7) \GeV$ for blue, and $(10^5, 3 \times 10^9) \GeV$ for brown. Finally, for QCD axion dark matter to be produced by kinetic misalignment with $C = 1$ and $0.3$,
    $T_{\rm KR}$ is predicted to be $2 \times 10^6$ and $7 \times 10^5$ GeV as shown in the solid and dotted orange curves with the maximal $T_{\rm RM}$ of $7 \times 10^{10}$ and $4 \times 10^{10}$ GeV allowed by the constraints shown in Fig.~\ref{fig:kination_QCD_axion}. 
    These curves assume $g_*(T)$ for the Standard Model and $H$ with individual energy density contributions including a piecewise $\rho_\theta$.}
    \label{fig:PGW}
\end{figure}

In Fig.~\ref{fig:PGW}, we show the gravitational wave spectra for two choices of the inflaton potential energy scale $V_{\rm inf}$. The inflationary energy scales of $V_{\rm inf}^{1/4} = 1.6\times 10^{16}$ GeV and $6 \times 10^{15}$ GeV correspond to the tensor fractions of $r=0.056$ near the upper bound from the CMB~\cite{Planck:2018vyg} and $r=0.001$ near the sensitivity limit of future CMB observations~\cite{CMB-S4:2016ple}, respectively. We show the spectrum for several sets of $(T_{\rm KR},T_{\rm RM})$ in different colored curves.

\begin{figure}
    \centering
    \includegraphics[width=0.495\columnwidth]{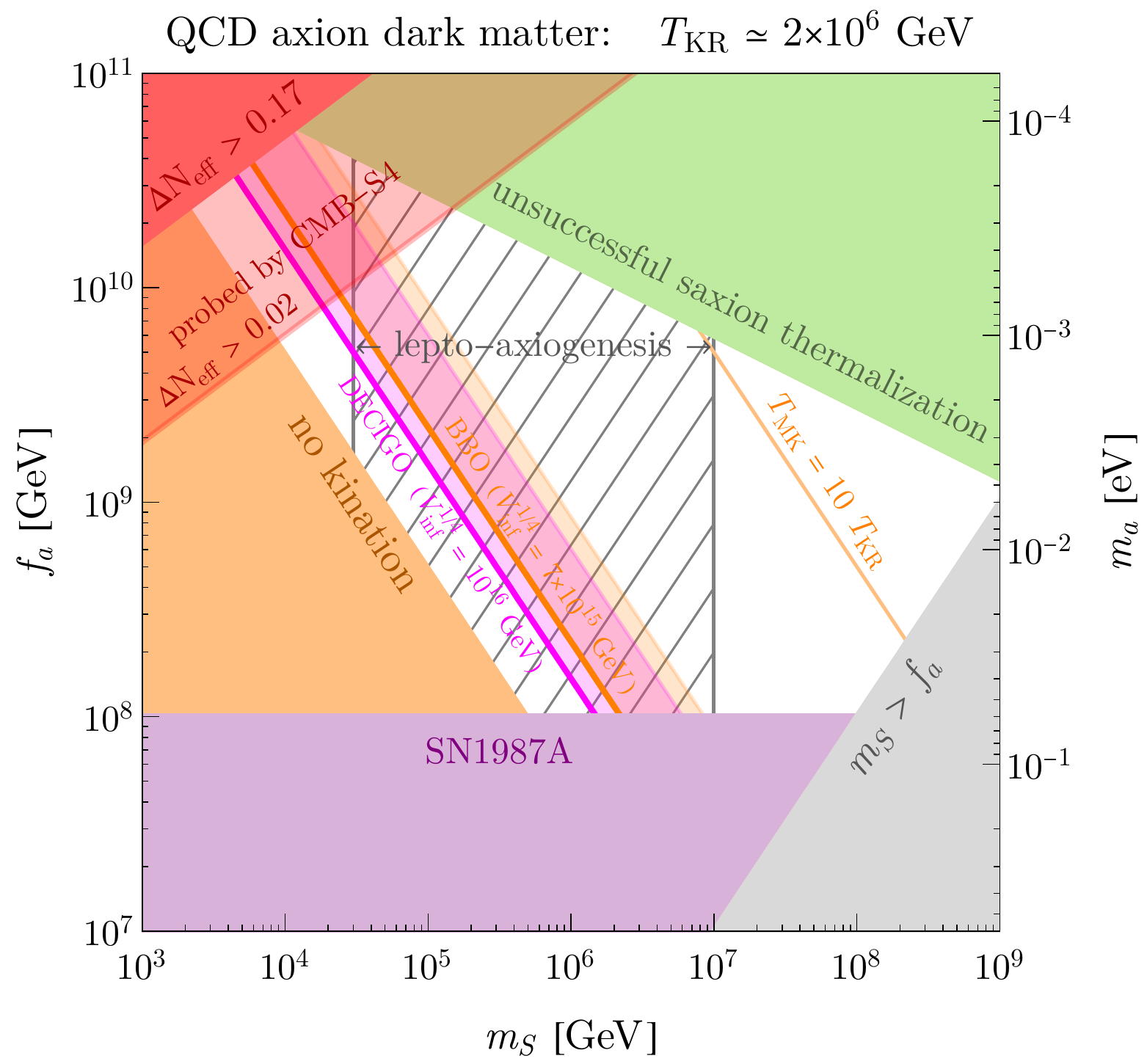}
    \includegraphics[width=0.495\columnwidth]{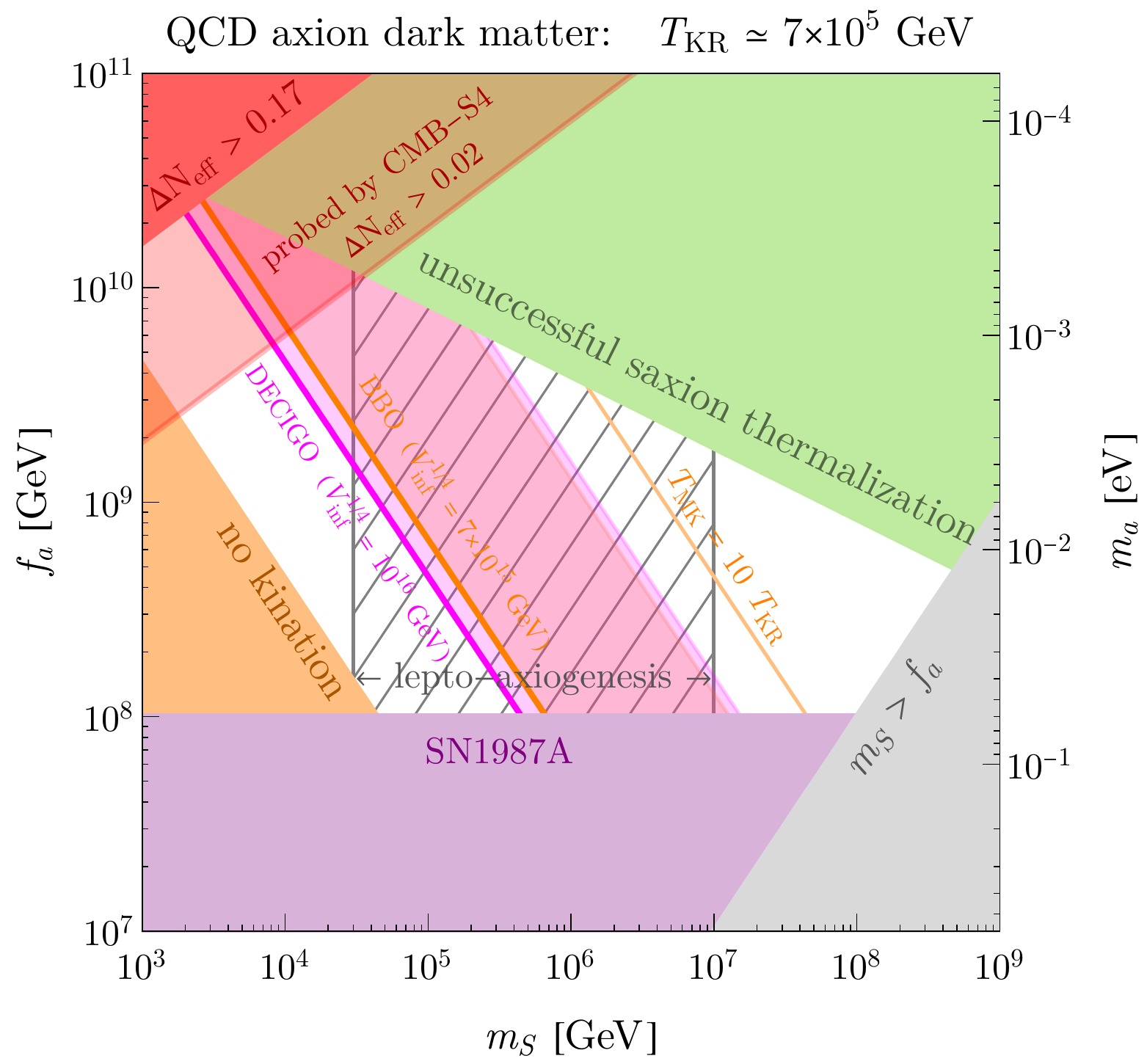}
    \caption{Parameter space for the QCD axion dark matter produced by kinetic misalignment, which predicts $T_{\rm KR} \simeq C \times 2 \times 10^{6} \GeV $ as can be seen in Fig.~\ref{fig:axioKD}. The left (right) panel assumes $C = 1~(0.3)$. The regions above the thick magenta and orange lines lead to a primordial gravitation wave signal that can be probed by DECIGO and BBO for the labeled choices of $V_{\rm inf}^{1/4}$, while within the adjacent transparent shadings, the peak of the spectrum can be detected by each observatory. The signal is made possible by the kination era; otherwise, $V_{\rm inf}^{1/4} > 1.2 \times 10^{16} \GeV$ is required for DECIGO.}
    \label{fig:kination_QCD_axion}
\end{figure}

The spectrum shown in the solid (dotted) orange curve corresponds to the value of $T_{\rm KR}$ predicted from QCD axion dark matter via the kinetic misalignment mechanism with $C=1$ ($C=0.3$) according to Eq.~(\ref{eq:TKR_KMM}), with the maximal $T_{\rm RM}$ allowed by the constraints shown in Fig.~\ref{fig:kination_QCD_axion}. In Fig.~\ref{fig:kination_QCD_axion}, we explore the parameter space for the QCD axion for $C = 1$ (left panel) and $C = 0.3$ (right panel). Most features of Fig.~\ref{fig:kination_QCD_axion} are analogous to those in Fig.~\ref{fig:kination_fixTKR}, whereas the gray hatched region indicates the range of $m_S$ compatible with lepto-axiogenesis based on Eq.~(\ref{eq:YB_lepto-ALPgenesis}).
If the inflation scale is not much below the present upper bound, DECIGO and BBO can detect the modification of the spectrum arising from the QCD axion kination era if the parameters of the theory lie anywhere above the thick magenta and orange lines in Fig.~\ref{fig:kination_QCD_axion}. Inside the transparent shaded regions with $T_{\rm RM}$ lower than the maximum allowed, BBO and DECIGO can also observe the peak of the spectrum peculiar to axion kination and identify how the Peccei-Quinn symmetry is spontaneously broken. If $C<1$ and $T_{\rm RM}$ is close to the maximum, CE can also observe the signal, but the inflation scale must be almost at the present upper bound as seen in Fig.~\ref{fig:PGW}.

\begin{figure}
    \centering
    \includegraphics[width=0.495\columnwidth]{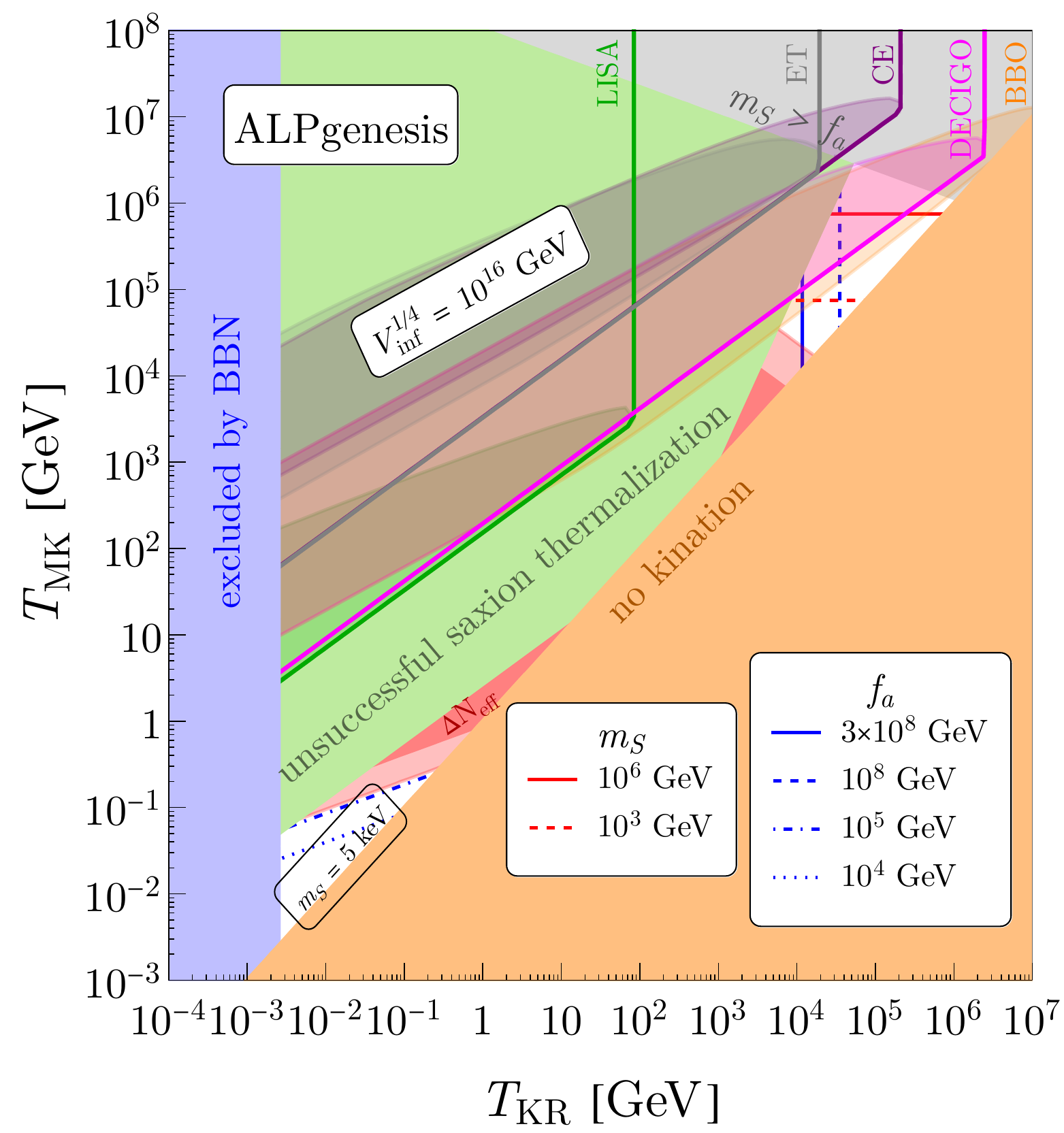}
    \includegraphics[width=0.495\columnwidth]{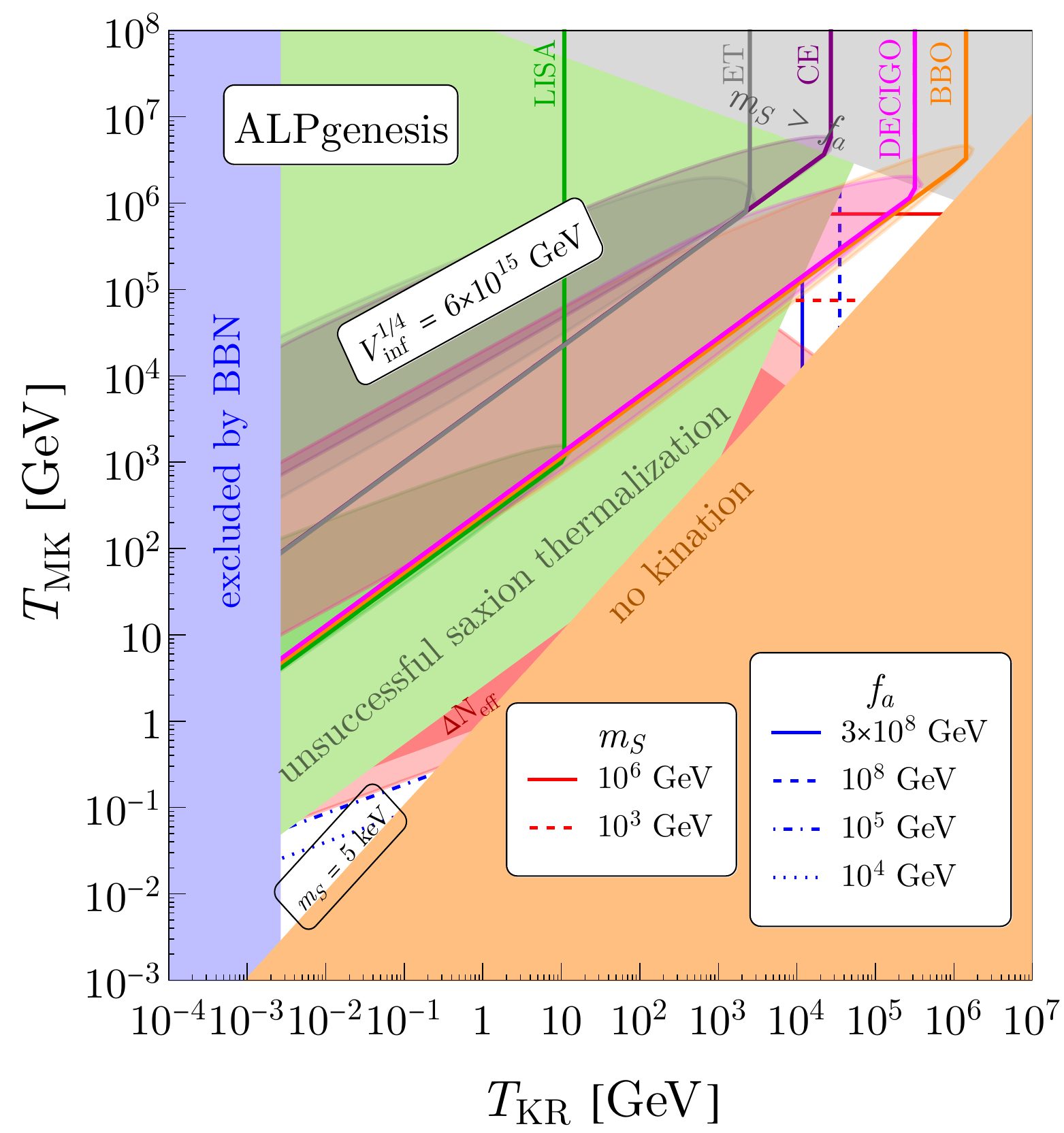}
    \caption{{Possible ranges of temperatures are shown for ALPgenesis assuming $c_B=0.1$. Contours of required $f_a$ and $m_S$ are shown by the blue and red lines respectively. White and transparent regions are allowed. Thanks to a kination era, the primordial gravitational waves for $V_{\rm inf}^{1/4} = 10^{16} \GeV$ (left panel) and $6 \times 10^{15} \GeV$ (right panel) become detectable by the experiments specified next to the colored sensitivity curves. The transparent colored shading for each gravitational wave observatory indicates the regions where the peak in the gravitational wave spectrum falls within the experimental sensitivity.}}
    \label{fig:ALPgenesis}
\end{figure}

The parameter space for ALPgenesis is shown in Fig.~\ref{fig:ALPgenesis}.
In the allowed parameter region at the bottom-left of the figure, $T_{\rm MK}$ is below the electroweak scale and $\dot{\theta}_{\rm ws} \simeq m_S = 5$ keV $(0.1/c_B)$. $f_a$ is determined according to Eqs.~(\ref{eq:TMK}) and (\ref{eq:TKR}). In the allowed parameter region in the upper-right corner, $T_{\rm MK}$ is above the electroweak scale and $S(T_{\rm ws}) = f_a$, so $f_a$ is determined by Eq.~(\ref{eq:TKRALPgen}). $m_S$ is determined by Eqs.~(\ref{eq:TMK}) and (\ref{eq:TKR}). Above each colored line, each experiment can detect the gravitational wave spectrum enhanced by axion kination. In the transparent shaded region, the triangular peak can be detected. Here we take $c_B = 0.1$. For smaller $c_B$, the prediction on $f_a$ and $m_S$ becomes larger, and the allowed range of $(T_{\rm KR},T_{\rm MK})$ expands, as can be seen from the black solid lines in Fig.~\ref{fig:kination_fixTKR}.

\begin{figure}
    \centering
    \includegraphics[width=0.32\columnwidth]{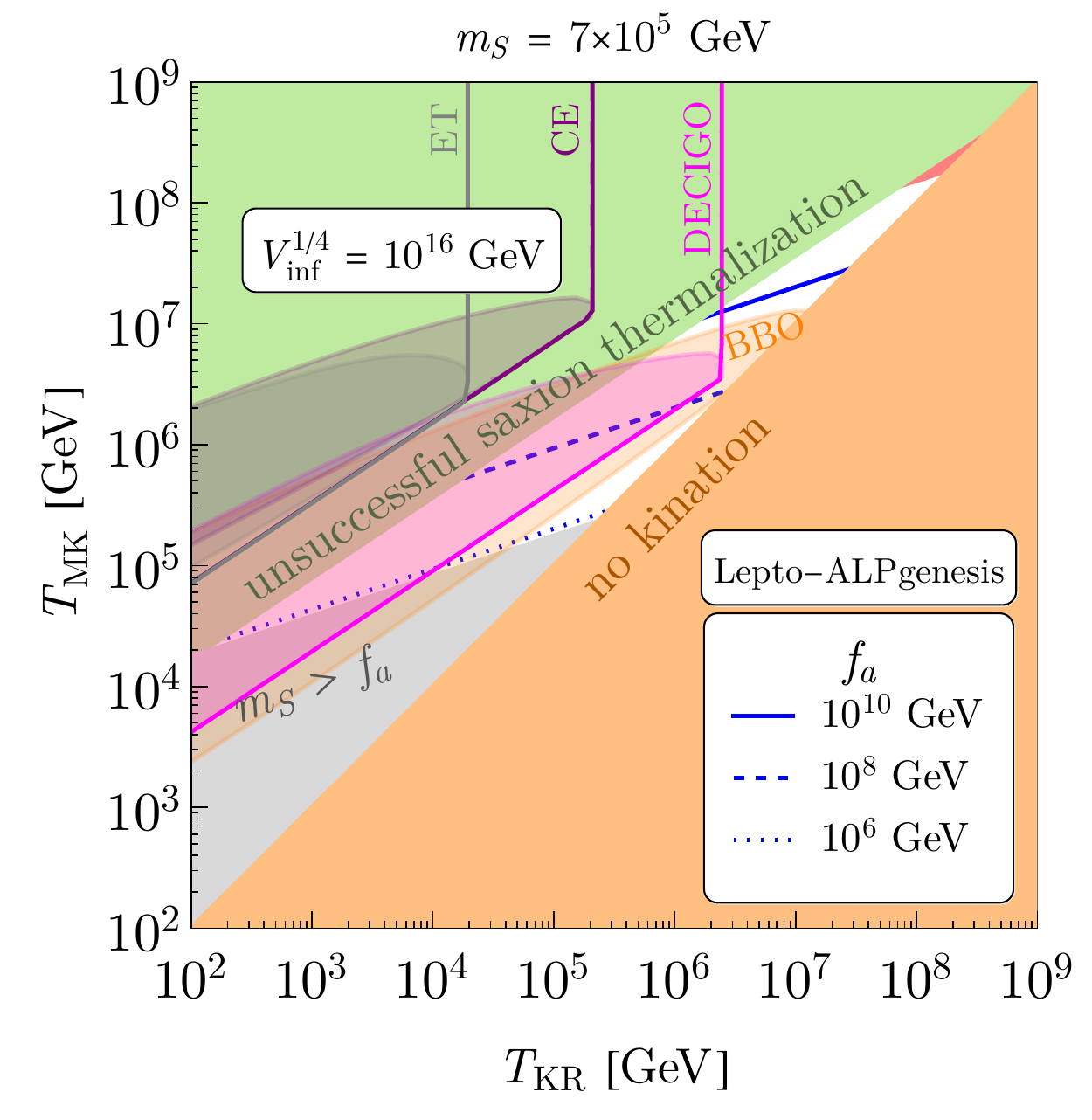}
    \includegraphics[width=0.32\columnwidth]{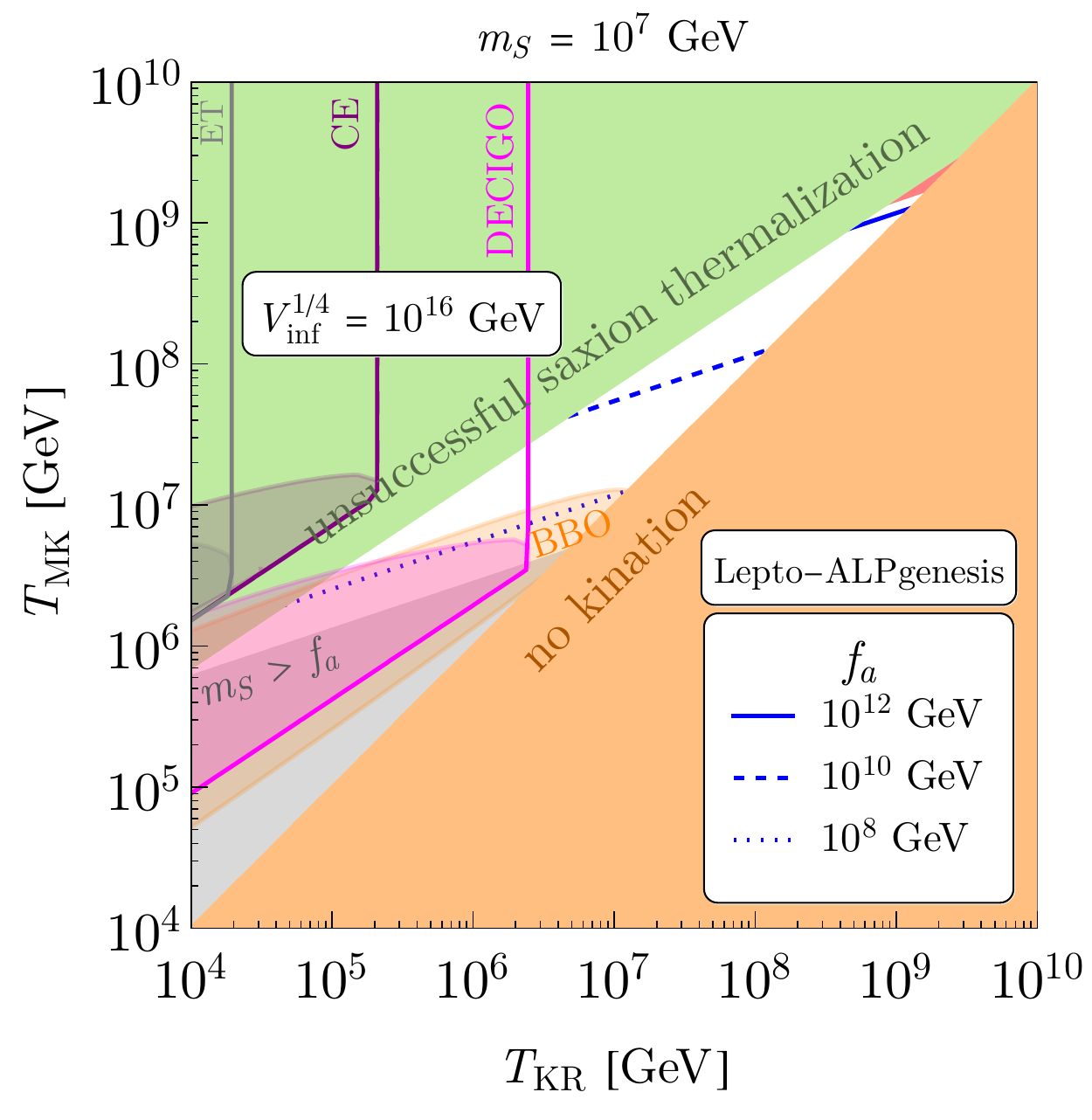}
    \includegraphics[width=0.32\columnwidth]{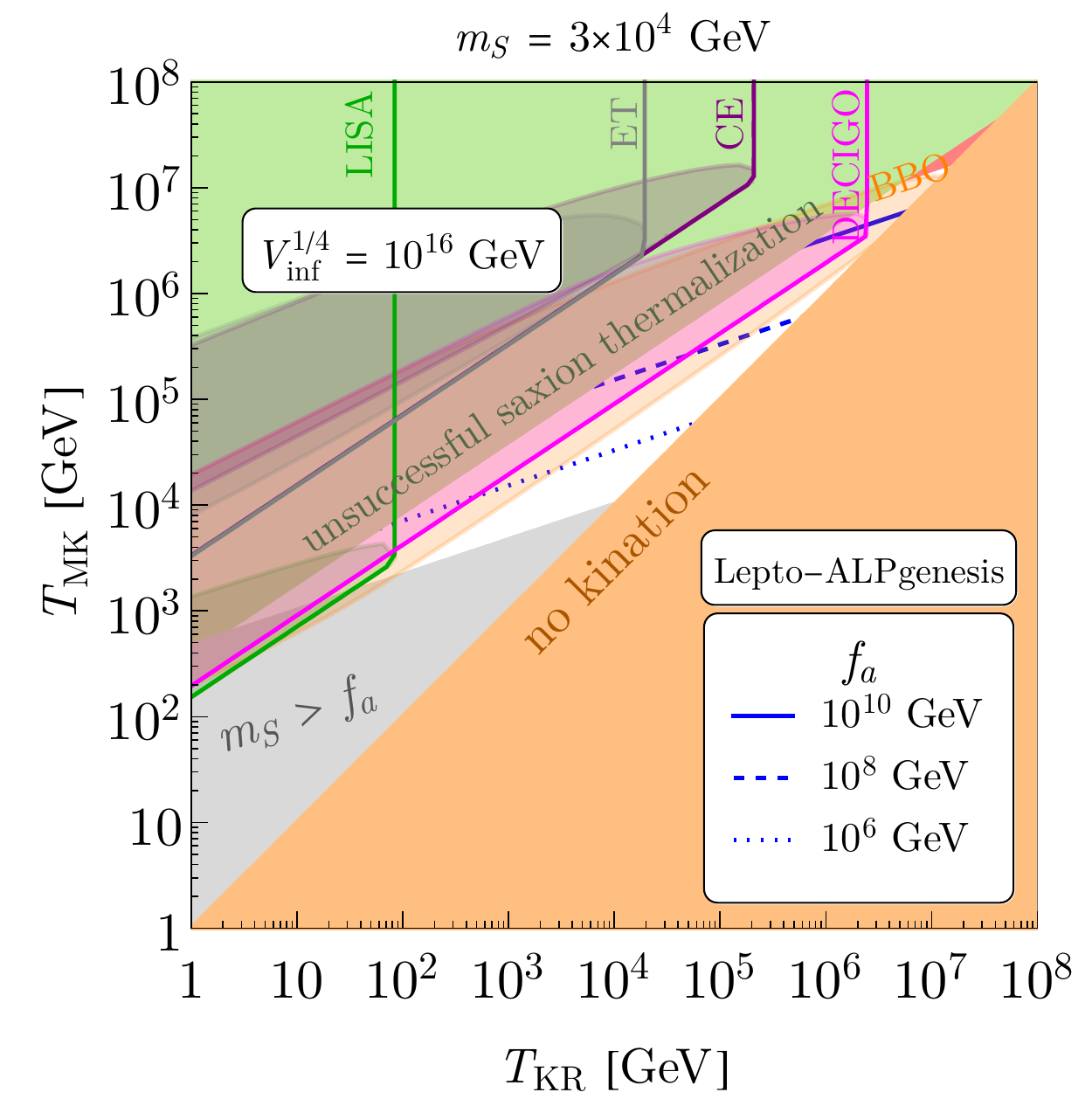}
    \includegraphics[width=0.32\columnwidth]{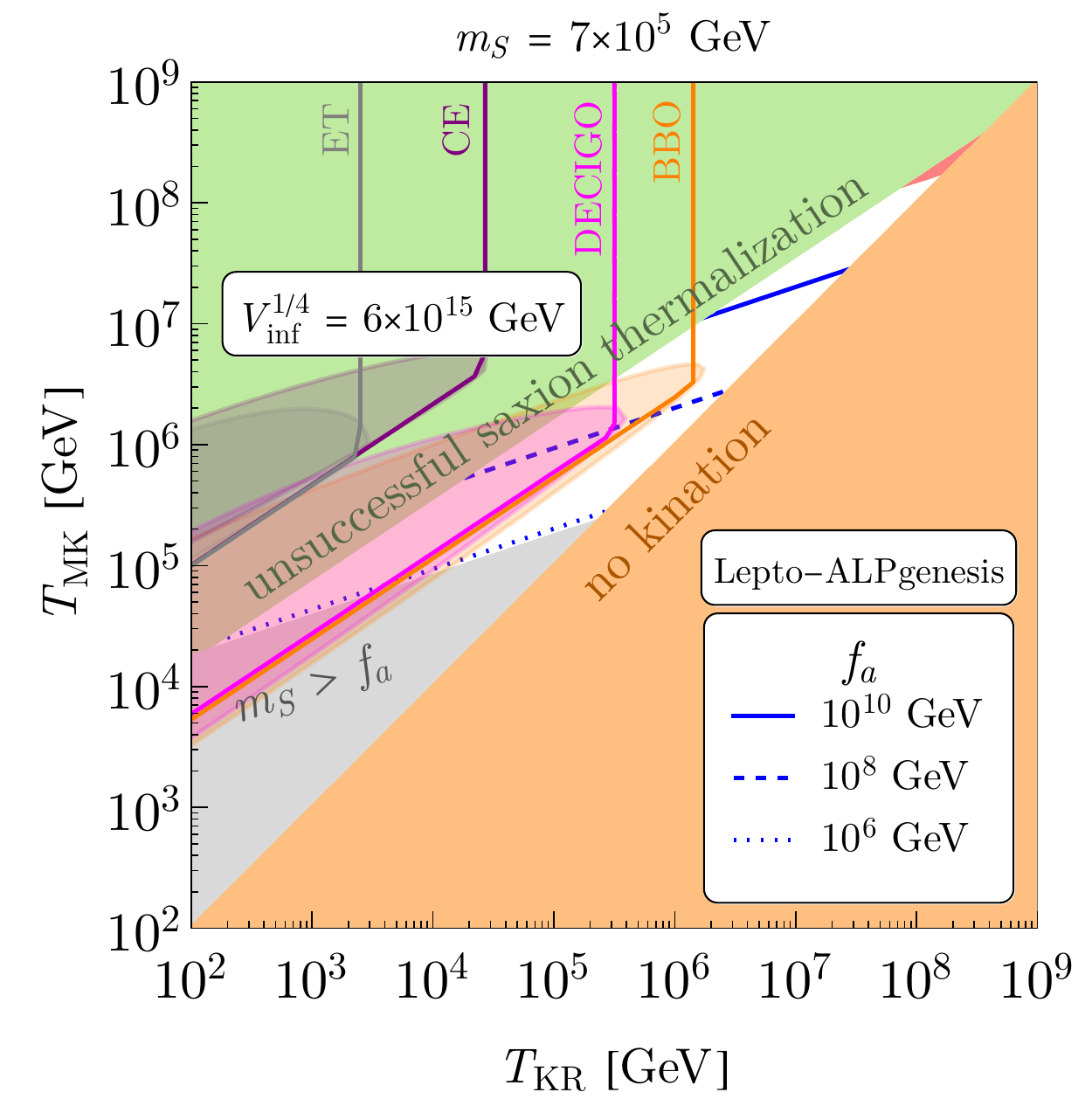}
    \includegraphics[width=0.32\columnwidth]{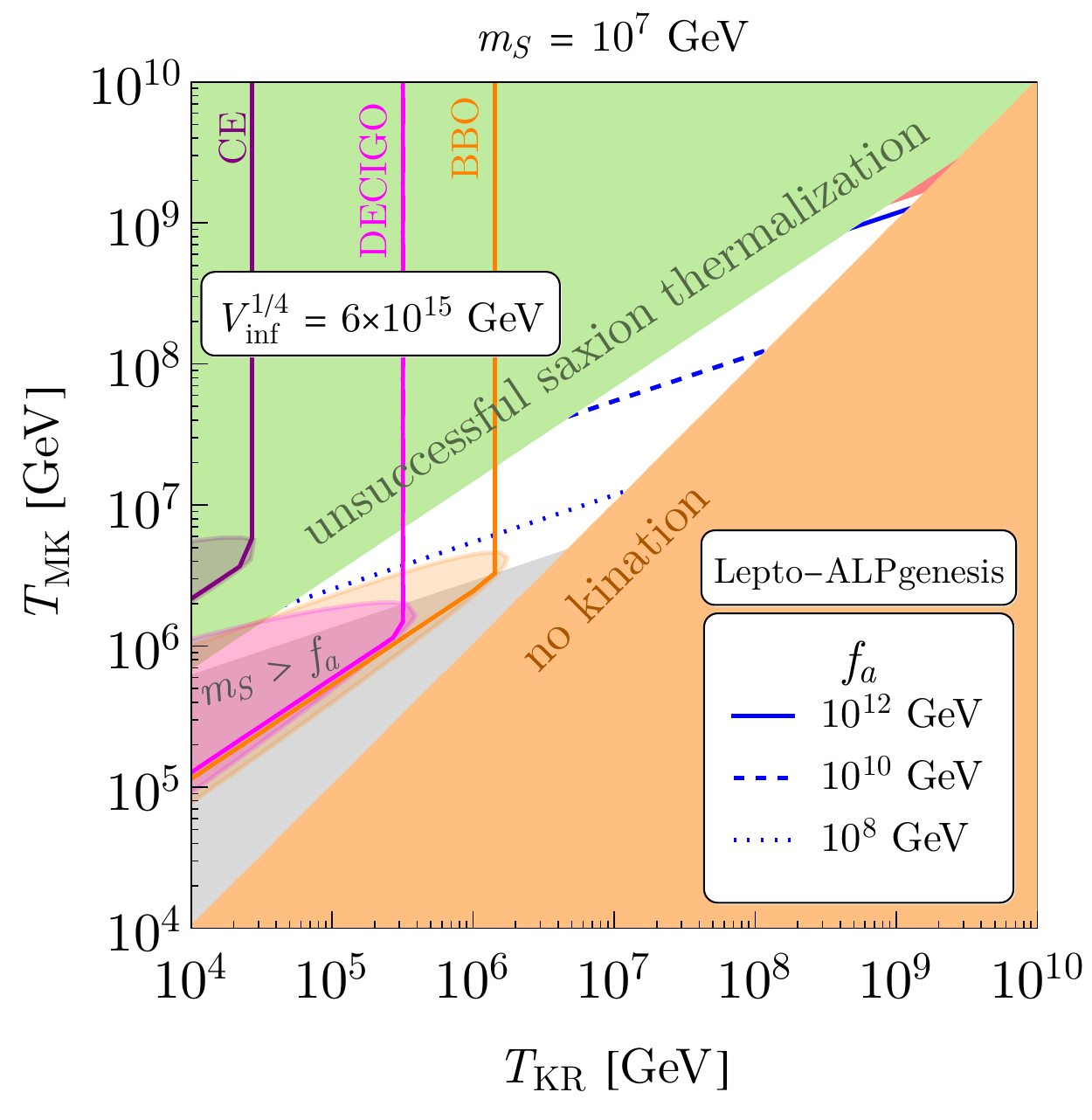}
    \includegraphics[width=0.32\columnwidth]{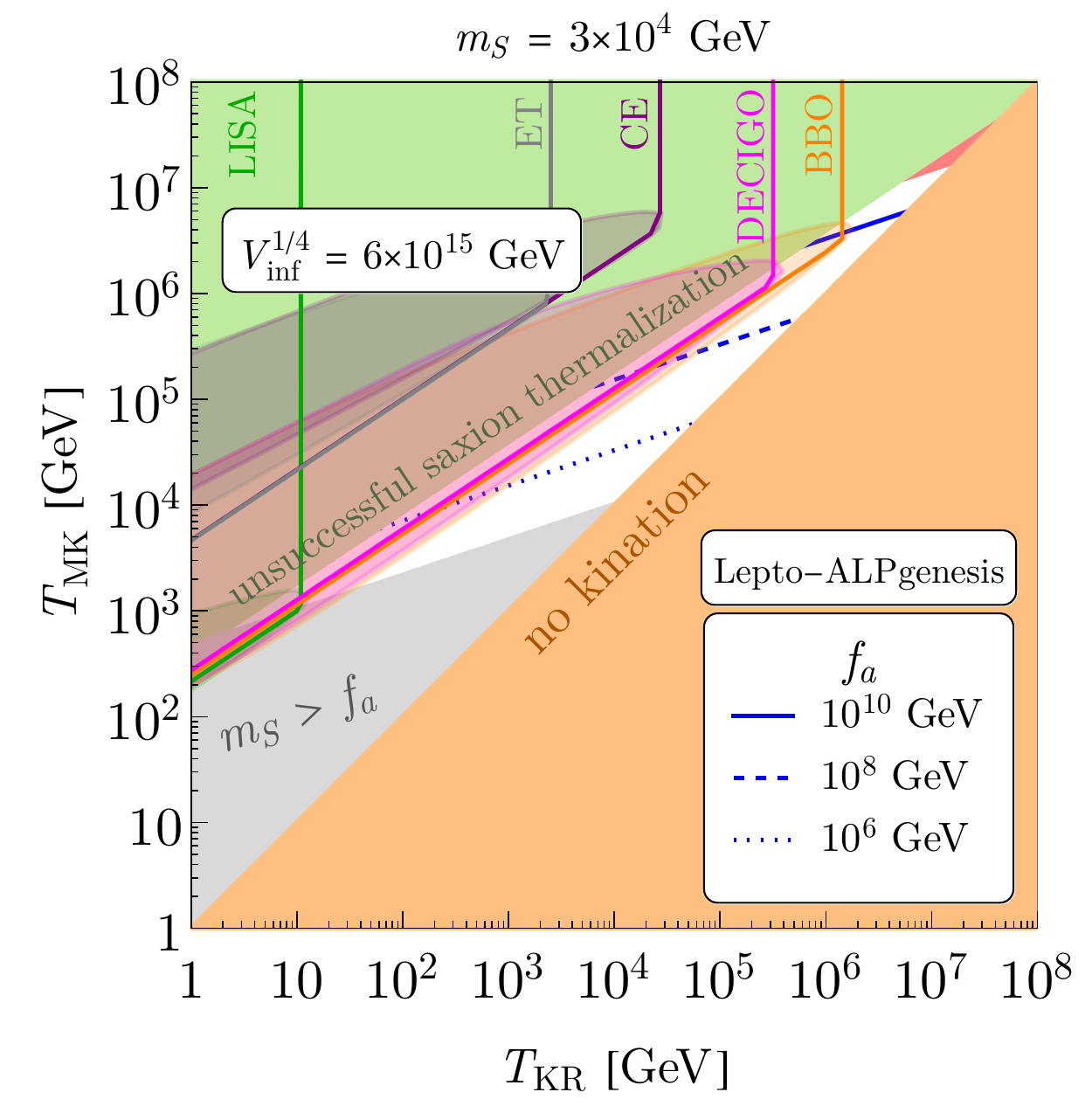}
    \caption{Possible ranges of temperatures are shown for lepto-ALPgenesis. The left two columns are for the case with entropy production from saxion domination ($D = 1$), while the right column assumes radiation domination ($D = \mathcal{O}(20)$) with degenerated neutrinos. These different cases are explained in Sec.~\ref{subsec:baryogenesis} and Ref.~\cite{Co:2020jtv}. The dark matter abundance is explained by an appropriate ALP mass determined by $f_a$ and $T_{\rm KR}$ using Fig.~\ref{fig:axioKD}. Thanks to a kination era, the primordial gravitational waves for $V_{\rm inf}^{1/4} = 10^{16} \GeV$ ($6 \times 10^{15} \GeV$) in the upper (lower) panels become detectable by the experiments labeled next to the colored sensitivity curves. The transparent colored shadings indicate that the peak of the gravitational wave spectrum due to kination lies inside the corresponding experimental reach.}
    \label{fig:lepto-ALP}
\end{figure}

The parameter space for lepto-ALPgenesis is shown in Fig.~\ref{fig:lepto-ALP}. Here $m_S$ is fixed so that the observed baryon asymmetry is explained by lepto-ALPgenesis; see Eq.~(\ref{eq:YB_lepto-ALPgenesis}). $f_a$ is then fixed by Eqs.~(\ref{eq:TMK}) and (\ref{eq:TKR}). The meaning of shaded regions and contours are the same as in Fig.~\ref{fig:ALPgenesis}.

Lastly, we comment on a potential constraint from high scale inflation. During inflation, light scalar fields receive quantum fluctuations with a magnitude $\sim H_{\rm inf}/2\pi$ with $H_{\rm inf}$ the Hubble scale during inflation~\cite{Mukhanov:1981xt,Hawking:1982cz,Starobinsky:1982ee,Guth:1982ec,Bardeen:1983qw}. If the axion rotation is responsible for the dark matter or baryon density, this leads to the matter isocurvature perturbation of order $P_{\rm iso} \simeq H_{\rm inf}^2/(2\pi S_{\rm inf})^2$ with $S_{\rm inf}$ the saxion field value during inflation. For the high inflation scale considered in this section, this is in conflict with the CMB observation~\cite{Planck:2018jri} unless $S_{\rm inf}$ is close to the Planck scale, placing constraints in the parameter space. However, since the PQ symmetry is explicitly broken by higher dimensional operators, the axion field does not necessarily stay light during inflation.
If the axion mass during inflation exceeds $H_{\rm inf}$, the quantum fluctuations are exponentially damped instead~\cite{Jeong:2013xta}. Therefore, we do not impose the model-dependent constraints from isocurvature perturbations.
(The higher dimensional operator that gives the axion mass during inflation should be different from the one that initiates the axion rotation, since otherwise the angular kick is suppressed.)

\subsection{From cosmic strings}
\label{subsec:GW_string}

We next discuss gravitational waves emitted from local cosmic strings~\cite{Vachaspati:1984gt}. Local cosmic strings are topological defects produced upon gauge symmetry breaking in the early universe, such as $U(1)$ symmetry breaking~\cite{Kibble:1976sj}. 
The breaking of a local $U(1)$ symmetry, and hence formation of a cosmic string network, arises in many theories beyond the Standard Model. For example, one of the best motivated cases is $U(1)_{\rm B-L}$, which is the unique flavor universal $U(1)$ symmetry that does not have a mixed anomaly with the Standard Model gauge symmetry. Moreover, $U(1)_{\rm B-L}$ can be embedded into $SO(10)$ together with the Standard Model gauge group, and whose spontaneous symmetry breaking can provide the right-handed neutrino masses in the see-saw mechanism~\cite{Minkowski:1977sc,Yanagida:1979as,Gell-Mann:1979vob,Mohapatra:1979ia}. 

After production, the cosmic string network follows a scaling law with approximately ${\mathcal O}(1)$ long strings per Hubble volume which is maintained from the balance between conformal expansion with the universe and losses from self-intercommutation. The self-intercommutation byproducts of the long string network lead to the formation of a network of string loops with a new loop forming nearly every Hubble time and with a loop size proportional to the horizon \cite{vilenkin2000cosmic}. These subhorizon loops oscillate and redshift like matter before decaying from the emission of gravitational waves. Because of the specific scaling law of the string network, the energy density fraction of the cosmic strings is nearly independent of temperature, and the spectrum of gravitational waves emitted from the local cosmic strings during radiation domination is nearly flat.

During kination or matter domination by axion rotation, the size of the horizon for a given temperature is smaller than it would be in a radiation dominated universe. This enhances the energy density of strings relative to the radiation density, and the spectrum of gravitational waves feature a triangular peak in our axion kination cosmology. Since the production of gravitational waves involves two steps that occur at widely separated times--the production of string loops and their later decay--the computation is more involved than the inflation case of Sec.~\ref{subsec:GW_inflation}.

The present day gravitational wave spectrum from a stochastic background of cosmic string loops is \cite{Vachaspati:1984gt,Auclair:2019wcv}
\begin{align}
    \label{eq:omegaGWSpec}
    \Omega_{\rm GW}(f) \equiv \frac{1}{\rho_c}\frac{d \rho_{\rm GW}}{d \ln f} = \frac{8\pi}{3 H_0^2} f \sum_{m = 1}^\infty  G \mu^2 P_m C_m.
\end{align}
Here $G \mu^2 P_m = \Gamma G \mu^2  m^{-q}/\zeta(q)$ is the power radiated by the $m$th mode of an oscillating string loop with $\Gamma \simeq 50$ being a constant determined from the average power over many types of string loop configurations~\cite{Blanco-Pillado:2017oxo,Auclair:2019wcv,vilenkin2000cosmic}.
The power index $q$ is $\frac{4}{3}$, $\frac{5}{3}$, or $2$, if the gravitational power is dominated by cusps, kinks, or kink-kink collisions, respectively~\cite{Vachaspati:1984gt,Binetruy:2009vt, Auclair:2019wcv}. We will take $q=4/3$, but for now we keep it as a free parameter. The present day critical density is $\rho_c$, and the factor $C_m$ is given by
\begin{align}
    \label{eq:CmInt}
    C_m &= \int_{t \rm scl}^t dt' \left(\frac{a(t')}{a(t)}\right)^3 \frac{dn}{df'}(t',f') 
    \\
    \frac{dn}{df'}(t',f') &= \frac{dn}{d t_k}\frac{d t_k}{dl}\frac{dl}{df'} = \left(\frac{\mathcal{F} C_{\rm eff}(t_k)}{\alpha t_k^4} \frac{a(t_k)^3}{a(t')^3}\right) \left(\frac{1}{\alpha + \Gamma G \mu}\right)\left(\frac{2m}{f^2}\frac{a(t')^2}{a(t)^2}\right).
\end{align}
Here $t_k(t',f) = \left(\frac{2m}{f}\frac{a(t')}{a(t)} + \Gamma G \mu t' \right) \left(\alpha + \Gamma G \mu \right)^{-1}$ denotes the formation time of a string loop of length $l$ that emits gravitational waves at frequency $f' = 2m/l$ at time $t'$. The lower integration time, $t_{\rm scl}$, is the time the infinite string network reaches scaling. $\mathcal{F} \approx 0.1$ \cite{Blanco-Pillado:2013qja} characterizes the fraction of energy that is transferred by the infinite string network into loops of size $l_k = \alpha t_k$
\footnote{Simulations suggest that roughly $90 \%$ of the energy transferred by the infinite string network into loops goes into loops smaller than $l_k$ which are short lived and subdominantly contribute to $\Omega_{\rm GW}$, or into translational kinetic energy which redshifts away \cite{Blanco-Pillado:2013qja,Dubath:2007mf} }
, and $C_{\rm eff}$ characterizes the loop formation efficiency which depends on the equation of state of the universe at loop formation time $t_k$. $C_{\rm eff}$ can be estimated from the velocity-one-scale model of the infinite string network and is found to be \cite{Blasi:2020wpy,Cui:2017ufi}
\begin{align}
    C_{\rm eff}(t_k) \approx 
    \begin{dcases}
        5.4 \quad &\text{$t_k$ during RD} 
        \\
        0.39 \quad &\text{$t_k$ during MD}
        \\
        30 \quad &\text{$t_k$ during KD}.
    \end{dcases}
\end{align}

The effect of the equation of state of the universe on the frequency dependence of $\Omega_{\rm GW}(f)$ can be seen by piecewise integrating $C_m$ in two regions: one where $t_k \approx (2/f)(\alpha + \Gamma G \mu)^{-1} a(t')/a(t)$ and the other where $t_k \approx (\Gamma G \mu t')(\alpha + \Gamma G \mu)^{-1}$ \cite{Cui:2018rwi}. The split occurs at time $t_\Gamma$ when the length of string lost to gravitational radiation, $\Gamma G \mu t_\Gamma$, equals the original loop formation length, $(2m/f) a(t_\Gamma)/a(t)$. The integral over $C_m$ is easily computed in either integration region by considering string loops that form when the equation of state of the universe is $w_1$, ($a(t_k) \propto t_k^{-3(1+w_1)}$), and emit gravitational radiation when the equation of state is $w_2$ ($a(t') \propto t'^{-3(1+w_2)}$). The spectral frequency dependence of the $m=1$ mode of oscillation is then \cite{Cui:2018rwi}
\begin{align}
    \label{eq:freqDependence}
    \Omega_{\rm GW}^{(1)}(f) \propto f^\lambda \qquad \lambda =  
    \begin{dcases}
        \frac{w_1 (6 w_2 +4)-2}{(w_1+1) (3w_2+1)}   &\quad t_\Gamma < t, \; p < 0 \\
        -1                                          &\quad t_\Gamma < t, \; p \geq 0 \\
        3 - \frac{2}{w_1 + 1}                       &\quad t_\Gamma > t
    \end{dcases}
\end{align}
where $p = -3 + 2/(1 + w_1) + 4/(3 (1 + w_2))$ characterizes whether the integral \eqref{eq:CmInt} is dominated at $t_{\Gamma}$ ($p < 0$) or the latest possible emission time $t'$ in that cosmological era ($p \geq 0$)~\cite{Cui:2018rwi}.%
\footnote{In a standard radiation dominated era, $\Omega_{\rm GW}$ is dominantly sourced by the smallest loops in the horizon due to their greater population and the independence of gravitational wave power, $\Gamma G \mu^2 $, on loop size. The smallest loops are those about to decay and hence for the standard cosmology, $p < 0$. However, this is not the case in more general cosmologies.}
The frequency dependence of $ \Omega_{\rm GW}^{(1)}(f)$ according to Eq.~(\ref{eq:freqDependence}) is shown in Table \ref{tab:frequencyDependence}.
\begin{table}
\[
  \begin{array}{cc|ccc}
    &\multicolumn{1}{c}{} & \multicolumn{3}{c}{\text{Loop Formation Era}} \\
    && \text{Radiation} & \text{Matter} & \text{Kination} \\
    \cline{2-5}
    & \text{Radiation} & f^0 & f^{-1} & f  \\[1.75ex]
    \smash{\rotatebox[origin=c]{90}{\text{Loop Decay Era}}} & \text{Matter} & f^{-1/2} & f^{-1} & f  \\[1.75ex]
        & \text{Kination} & f^{1/4} & f^{-1/2} & f \\[1.75ex]
  \end{array}
\]
\caption{Frequency dependence of the $m = 1$ mode amplitude, $\Omega^{(1)}(f)$, for loops that form and decay in a radiation, matter, or kination-dominated universe.}
\label{tab:frequencyDependence}
\end{table}
In the modified cosmology under consideration, the universe transitions from being dominated by radiation to matter at $T_{\rm RM}$, to kination at $T_{\rm MK}$, and back to radiation upon merging with the standard cosmology at $T_{\rm KR}$. From Table \ref{tab:frequencyDependence}, we may therefore expect for sufficiently long eras of radiation, matter, and kination that  $\Omega_{\rm GW}^{(1)} \propto f^0 \xrightarrow{T_{\rm RM}} f^{-1} \xrightarrow{T_{\rm MK}} f^{1} \xrightarrow{T_{\rm KR}} f^{0}$ as $f$ drops from high to low frequencies. That is, a triangular shaped peak in spectrum.

Although the first mode dominates the total power emitted by a string loop, the sum of the  contributions from all higher modes can appreciably change this power dependence of the total spectrum \cite{Blasi:2020mfx,Gouttenoire:2019kij}. The effect of the higher modes can be analytically estimated by noting that $\Omega_{\rm GW}^{(m)} = m^{-q} \Omega_{\rm GW}^{(1)}(f/m)$ \cite{Blasi:2020mfx}. For example, assuming that $\Omega_{\rm GW}^{(1)}$ is a broken power law proportional to $f^{\alpha}$ for $f < f_0$ and $f^{\beta}$ for $f \geq f_0$, we may write the total spectrum as
\begin{align}
    \label{eq:brokenPowerLawGW}
    \Omega_{\rm GW}(f) = \sum_{m = 1}^\infty m^{-q} \, \Omega_{\rm GW}^{(1)}(f/m) \approx \sum_{m = 1}^{f/f_0} m^{-q-\beta}\, \Omega_0 \left(\frac{f}{f_0}\right)^{\beta} + \sum_{m = f/f_0 + 1}^\infty m^{-q-\alpha}\,\Omega_0 \left(\frac{f}{f_0}\right)^{\alpha}
\end{align}
where $\Omega_0 = \Omega_{\rm GW}^{(1)}(f_0)$. In the limit $f \gg f_0$, \eqref{eq:brokenPowerLawGW} reduces to
\begin{align}
    \label{eq:highfreqLimit}
     \Omega_{\rm GW}(f) \xrightarrow[]{f \gg f_0} \Omega_0 \left[ \zeta(q+\beta)\left(\frac{f}{f_0}\right)^{\beta} + \left(\frac{1}{q + \alpha - 1} - \frac{1}{q + \beta - 1}\right) \left(\frac{f}{f_0}\right)^{-q + 1}\right].
\end{align}
Consequently, if $1-q > \beta$, the high frequency contribution from the sum over all string modes can make a steeply decaying spectrum shallower. Eq.~(\ref{eq:highfreqLimit}) shows that the $f^1$ power law during the kination era induced by axion rotations remains unchanged from summing all string modes, but the $f^{-1}$ power law during the preceding matter-dominated era becomes $f^{1-q} = f^{-1/3}$ for cusp dominated strings~\cite{Blasi:2020wpy}, $f^{-2/3}$ for kink dominated strings, and unchanged for kink-kink collision dominated strings. In this work, we focus on cusp dominated strings which are common on string trajectories \cite{vilenkin2000cosmic}. Interestingly, the  determination of the spectral slope during this early matter-dominated era can potentially indicate the value of $q$.

The peak amplitude and frequency of the stochastic string spectrum can be estimated analytically in terms of the key temperatures associated with axion kination, namely $T_{\rm KR}, T_{\rm MK}$, and $T_{\rm RM}$. From Table \ref{tab:frequencyDependence}, we see that loops forming in the matter dominated era and decaying in the late radiation dominated era enjoy an $f^{-1}$ growth, while loops that form in the kination era and decay in the late radiation dominated era experience an $f^1$ decay. Consequently, loops that form at time $t_k = t_{\rm MK}$ are responsible for the peak amplitude and frequency of the triangular peak spectrum when decaying. For example, the energy density of these loops immediately prior to decaying is $ \rho(t_\Gamma) \approx \mu \,l(t_{\rm MK}) n(t_{\rm MK}) a(t_{\rm MK}/a(t_{\Gamma}))^3$ where $l(t_{\rm MK}) = \alpha t_{\rm MK}$, $n(t_{\rm MK}) \approx \mathcal{F} C_{\rm eff}/3 \alpha t_{\rm MK}^3$, and the decay time $t_{\Gamma} = \mu l(t_{\rm MK})/\Gamma G \mu^2$. The resultant spectrum of gravitational waves is then given by
\begin{flalign}
  \left. \Omega_{\rm GW}h^2\right|_{\rm peak}  & \approx 10^{-8}
  \left(\frac{100 \, \rm MeV}{T_{\rm KR}}\right)^{ \scalebox{1.01}{$\frac{3}{2}$} }
  \left(\frac{T_{\rm MK}}{2 \, \rm GeV}\right)^{ \scalebox{1.01}{$\frac{3}{2}$} }
  \left(\frac{ G \mu}{6 \times 10^{-11}}\right)^{ \scalebox{1.01}{$\frac{1}{2}$} }
  \left(\frac{\alpha}{0.1}\right)^{ \scalebox{1.01}{$\frac{1}{2}$}
  }
  \left(\frac{50}{\Gamma}\right)^{ \scalebox{1.01}{$\frac{1}{2}$}
  }
  \\
  \label{eq:peakFreq}
  f_{\rm peak} & \approx 0.1 \, {\rm Hz}
  \left(\frac{100 \, \rm MeV}{T_{\rm KR}} \right)^{ \scalebox{1.01}{$\frac{1}{2}$} }
  \left(\frac{T_{\rm MK}}{2 \, \rm GeV}\right)^{ \scalebox{1.01}{$\frac{3}{2}$} }  
  \left(\frac{6 \times 10^{-11}}{ G \mu}\right)^{ \scalebox{1.01}{$\frac{1}{2}$} }
  \left(\frac{0.1}{\alpha}\right)^{ \scalebox{1.01}{$\frac{1}{2}$}
  }
  \left(\frac{50}{\Gamma}\right)^{ \scalebox{1.01}{$\frac{1}{2}$}
  }
  \\
  \label{eq:lowerFreq}
  f_{\rm KR} & \approx  1 \, {\rm mHz}
  \left(\frac{T_{\rm KR}}{100 \, \rm MeV}\right)
  \left(\frac{6 \times 10^{-11}}{ G \mu}\right)^{ \scalebox{1.01}{$\frac{1}{2}$}
  }
  \left(\frac{0.1}{\alpha}\right)^{ \scalebox{1.01}{$\frac{1}{2}$}
  }
  \left(\frac{50}{\Gamma}\right)^{ \scalebox{1.01}{$\frac{1}{2}$}
  }.
\end{flalign}
where the peak amplitude of the triangular spectrum, $\left. \Omega_{\rm GW}h^2 \right|_{\rm peak}$, and the peak frequency $f_{\rm peak}$, can be thought of as $\left. \Omega_{\rm GW}h^2 \right|_{\rm MK}$ and $f_{\rm MK}$, since the peak is associated with loops formed at $T_{\rm MK}$. 
The frequency of the peak, Eq.~(\ref{eq:peakFreq}), is set by the invariant size of the loop at the formation time $t_{\rm MK}$ with the emission frequency at decay $ 2/l(t_{\rm MK})$ redshifted to the present. 
Similarly, from Table \ref{tab:frequencyDependence}, we can see that the loops that form at the transition from matter to late era radiation, $t_{\rm KR}$, are responsible for the amplitude of the lower left vertex of the axion kination triangle. Again, the frequency of these loops is the emission frequency at decay, $2/l(t_{\rm KR})$ redshifted to the present as given by Eq.~(\ref{eq:lowerFreq}). Last, note that the fiducial values of $T_{\rm KR} = 100 \, \rm MeV$ and $T_{\rm MK} = 2 \, \rm GeV$, correspond to $T_{\rm RM} \approx 100 \, \rm GeV$, which corresponds to the dark purple curve of Fig.~\ref{fig:GW_stringSpectrum}. 

In general, for brief eras of kination and matter domination, the gravitational wave spectrum will not reach its asymptotic dependence, $\Omega_{\rm GW}^{\rm tot} \propto f^0 \xrightarrow{T_{\rm RM}} f^{-1/3} \xrightarrow{T_{\rm KM}} f^{1} \xrightarrow{T_{\rm KR}} f^{0}$; nor will the kination era peak be sharply defined. Consequently, we numerically evaluate Eq.~(\ref{eq:omegaGWSpec}) to precisely determine $\Omega_{\rm GW}$ over a wide range of $\{T_{\rm KR}, T_{\rm RM}, T_{\rm MK}\}$. In doing so, we numerically compute the time evolution of the scale factor from the Friedmann equation
\begin{align}
    \label{eq:hubbleScaleFactor}
    \frac{\dot{a}(t)}{a(t)} = H_0 \Bigl[\Omega_{\Lambda} + \Omega_r \left(\frac{a(t_0)}{a(t)}\right)^4 +  \Omega_m \left(\frac{a(t_0)}{a(t)}\right)^3 + \Omega_{k,\theta} \left(\frac{a_{\rm KR}}{a(t)}\right)^6 + \Omega_{m,\theta}\left(\frac{a_{\rm MK}}{a(t)}\right)^3 \Bigr]^{ \scalebox{1.01}{$\frac{1}{2}$} }
\end{align}
where $\Omega_{k,\theta} =\Omega_r \left(\frac{a(t_0)}{a_{\rm KR}}\right)^4 \, \Theta(a(t) - a_{\rm MK})$ and $\Omega_{m,\theta} =\Omega_{k,\theta}  \left(\frac{a_{\rm KR}}{a_{\rm MK}}\right)^6 \, \Theta(a(t) - a_{\rm RM})$ are the critical densities of the axion induced kination and matter dominated eras, respectively, while $\Omega_r =9.038 \times 10^{-5}$, $\Omega_m =0.315$, and $\Omega_\Lambda = 1 - \Omega_r - \Omega_m$ \cite{Planck:2018vyg} are the critical energy densities of radiation, matter and vacuum energy in the standard $\Lambda$CDM cosmology. $H_0 \simeq 67.4 \, {\rm km}\, {\rm s}^{-1}\, {\rm Mpc}^{-1} $ is the present-day Hubble constant \cite{Planck:2018vyg}.
\begin{figure}
    \centering
    \includegraphics[width=0.495\columnwidth]{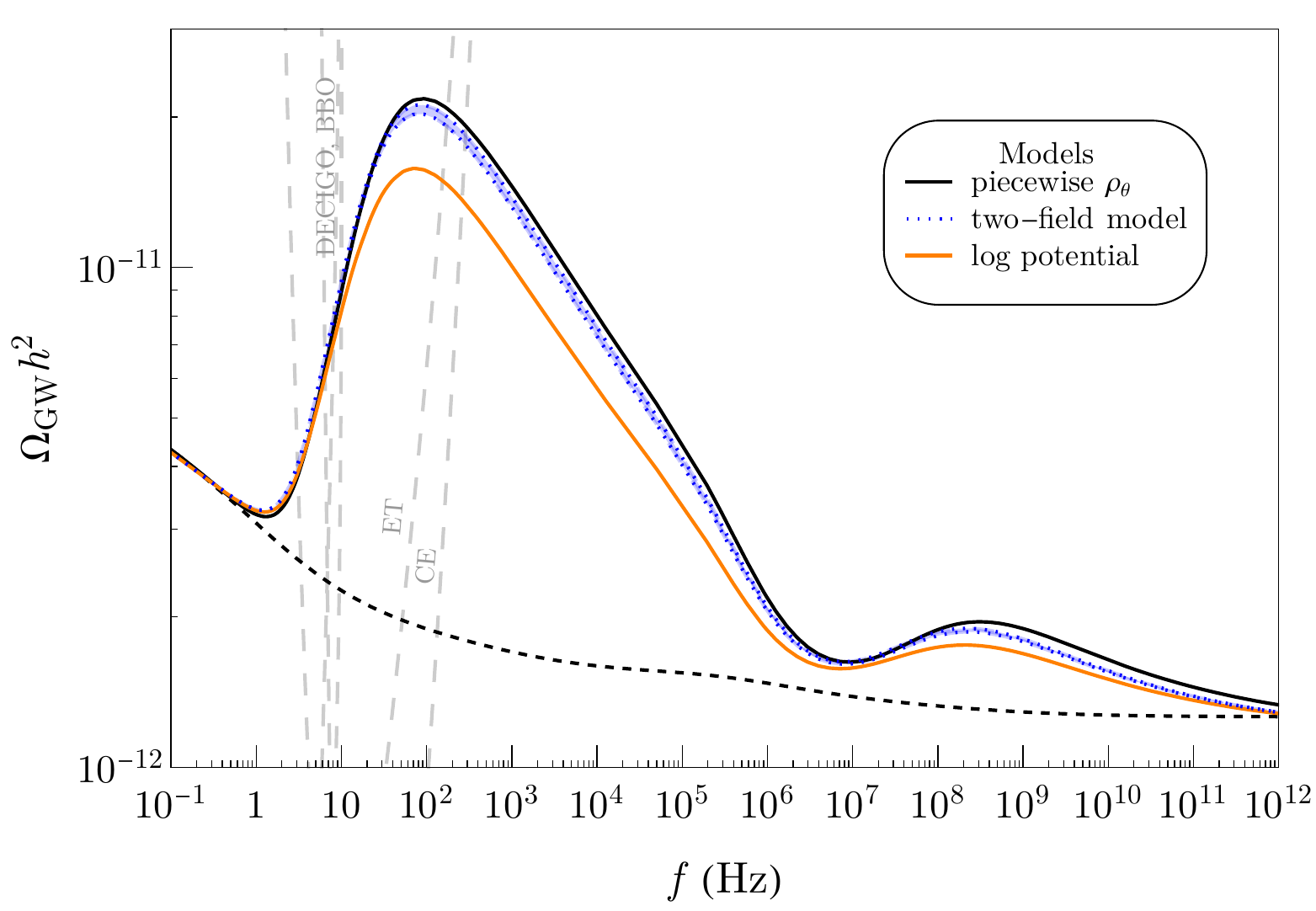}
    \includegraphics[width=0.495\columnwidth]{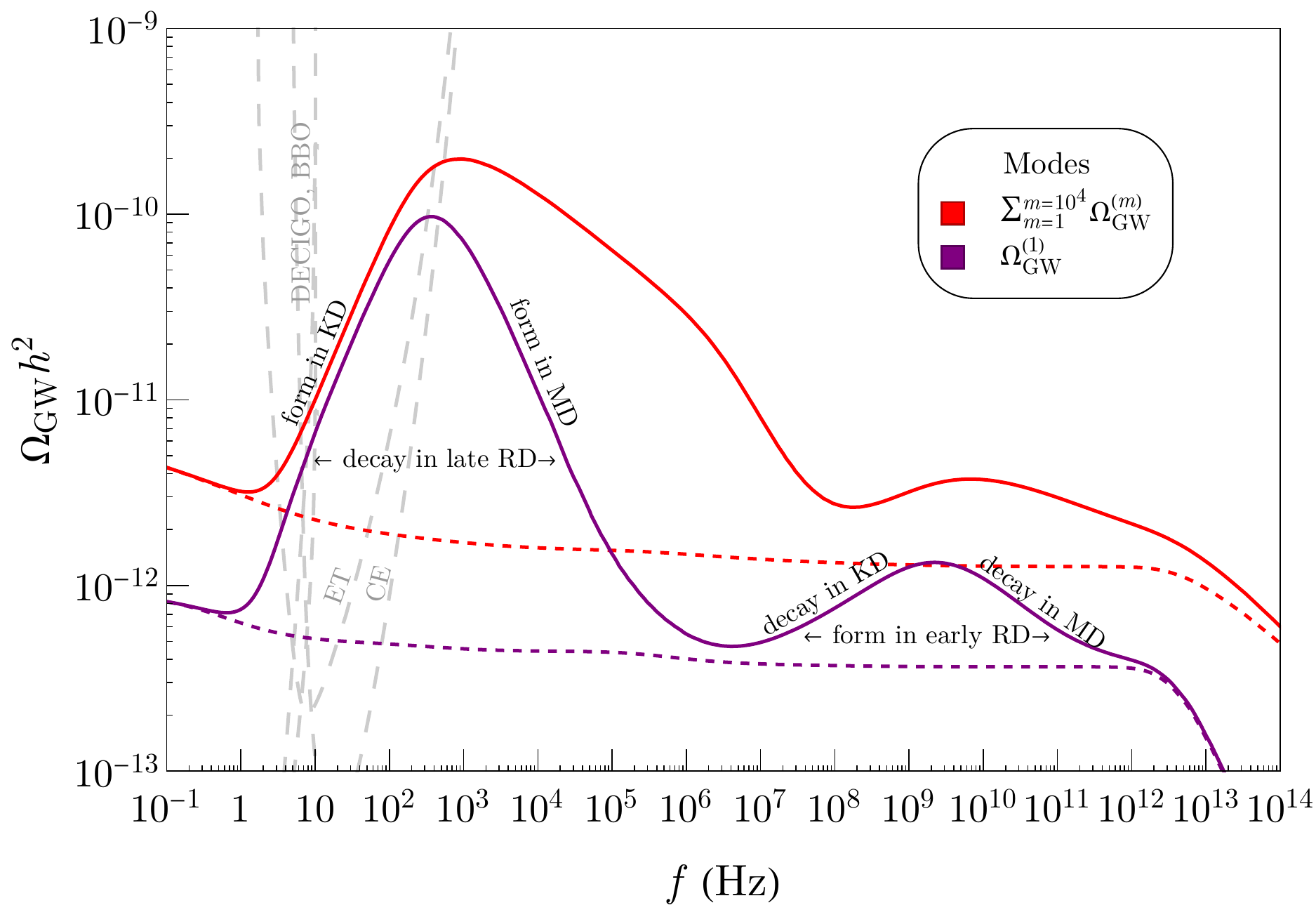}
    \caption{Left: An illustration of the model dependence in the stochastic string gravitational wave spectrum. The solid black line is the case where the rotation energy density $\rho_{\theta}$ follows a piecewise scaling $T \lessgtr T_{\rm MK}$ as shown in Fig.~\ref{fig:EoS}. The colored curves are for the two-field model (blue) and the logarithmic potential (orange) with evolution demonstrated in Fig.~\ref{fig:EoS}. For the two-field model, we show the blue dotted curves for different ratios of the soft masses of the two fields $\bar{P}$ and $P$, $m_{\bar P}/m_P = 1, 100$. The dashed black curve shows the standard string spectrum in a $\Lambda$CDM cosmology. We fix ($T_{\rm KR}, T_{\rm RM}) = (1 \, {\rm GeV}, 100 \, {\rm GeV})$. Right: An illustration of the difference between the $m = 1$ amplitude (purple) and the total amplitude summed over $10^4$ harmonics (red). The sum over high modes partially flattens the right side of the kination induced peak, shifting the spectral dependence from $f^{-1}$ to $f^{-1/3}$. We fix ($T_{\rm KR}, T_{\rm RM}) = (1 \GeV, 10 \TeV)$. In both panels, the second, smaller triangle at high frequencies is an additional fingerprint of axion kination and arises from loops that form in the early radiation dominated era and decay in the subsequent matter or kination dominated eras (see Table  \ref{tab:frequencyDependence}). Both panels assume $G\mu = 5 \times 10^{-15}$, and $\alpha = 0.1$. The drop in the spectrum above $f \sim 10^{12}$ Hz arises from only considering loops that form after the string network reaches scaling, $t_k > t_{\rm scl}$. We take scaling to be reached shortly after string formation, $t_k \sim 1/H(T = \sqrt{\mu})$. However, string friction with the thermal bath can delay scaling and shift this high frequency cutoff to lower frequencies \cite{Alford:1988sj,Vilenkin:1991zk,vilenkin2000cosmic,Gouttenoire:2019kij}. We do not include this model dependent effect in this work.
    }
    \label{fig:twinpeaks}
\end{figure}

The left panel of Fig.~\ref{fig:twinpeaks} shows the imprint of the saxion potential on the stochastic string gravitational wave background. The black curve corresponds to the piecewise approximation of the $\rho_\theta$ contribution to the Hubble rate as used in Eq.~\eqref{eq:hubbleScaleFactor}. The blue dotted and orange solid lines show the spectrum for the two-field model and the log potential, respectively. Similar to the gravitational wave spectrum of Fig.~\ref{fig:PGW_models} for inflation, the spectrum for the two-field model is close to that for the piecewise approximation, while that for the log potential deviates from them. In what follows, we use the piecewise approximation of $\rho_{\theta}$.

The right panel of Fig.~\ref{fig:twinpeaks} illustrates two key features that axion kination imparts to the stochastic string gravitational wave background. First, the purple curves show the $m=1$ contribution to the spectrum, $\Omega_{\rm GW}^{(1)}$, while the red curves shows $\Omega_{\rm GW}$ after summing over $10^4$ harmonics. For the $m =1$ amplitude, the triangular shaped peak approaches the expected $f^{-1}$ rise  and $f^1$ fall as shown in Table \ref{tab:frequencyDependence}. The amplitude summed over $10^4$ modes, however, demonstrates how the total amplitude deviates from the $m=1$ amplitude. Summing over higher harmonics increases the amplitude roughly by a factor of $\zeta(4/3)$, and most importantly, the contribution from the higher harmonics changes the $f^{-1}$ tail on the right side of the kination induced triangle into a much shallower $f^{-1/3}$ tail while leaving the $f^{1}$ decay on the left side the triangle the same. Such a long and shallow UV tail allows high frequency gravitational wave detectors to discern axion kination from the standard $\Lambda$CDM spectrum even when the triangular kination peak is at much lower frequencies. In addition, a second key feature of axion kination is shown in the second, smaller triangle at higher frequencies compared to the main triangle. The second triangular bump in the spectrum arises from loops that form in the early radiation dominated era prior to kination, and decay in the matter,  kination, or late radiation era. As seen from Table \ref{tab:frequencyDependence}, loops that form in the early radiation era and decay in these other eras are expected to exhibit a shallower rise and fall in amplitude akin to the main triangular shaped enhancement from loops that form in the matter or kination eras.  For sufficiently short eras of kination and matter domination, the smaller, second bump is visible even after summing over higher harmonics. For long eras of kination, the sum over higher harmonics can merge the main kination induced peak with this smaller second peak, as shown for instance, in the purple curves in the left panel of Fig.~\ref{fig:GW_stringSpectrum}. Nevertheless, the slightly broken power law near $10^3$ Hz for the solid purple and $10^5$ Hz for the lighter purple contour is a remnant left over from this second triangular peak. The observation of a broken decaying power law or the second triangular bump itself may provide a unique gravitational wave fingerprint for axion induced kination.

\begin{figure}
    \centering
    \includegraphics[width=0.48\columnwidth]{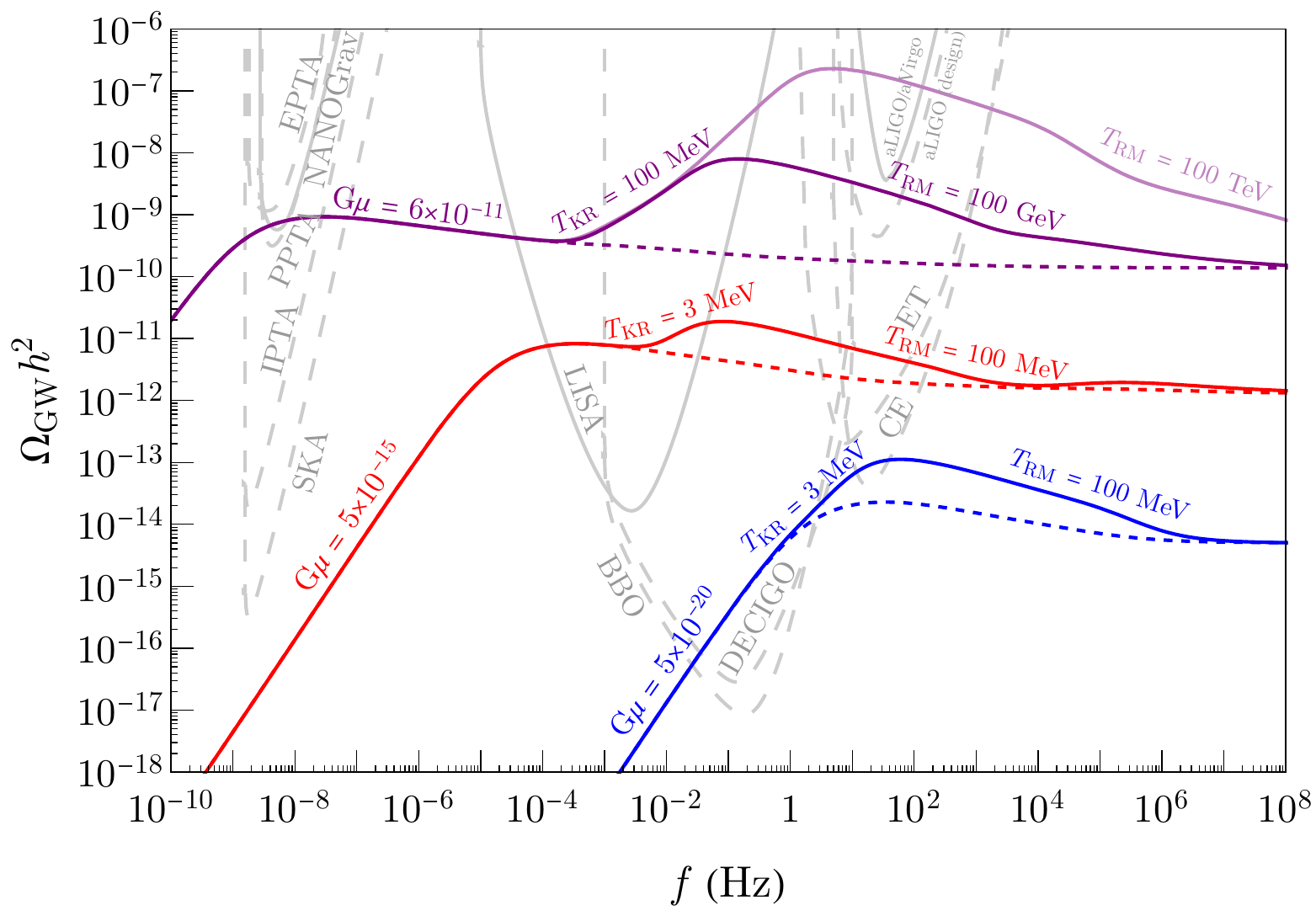}
    \includegraphics[width=0.48\columnwidth]{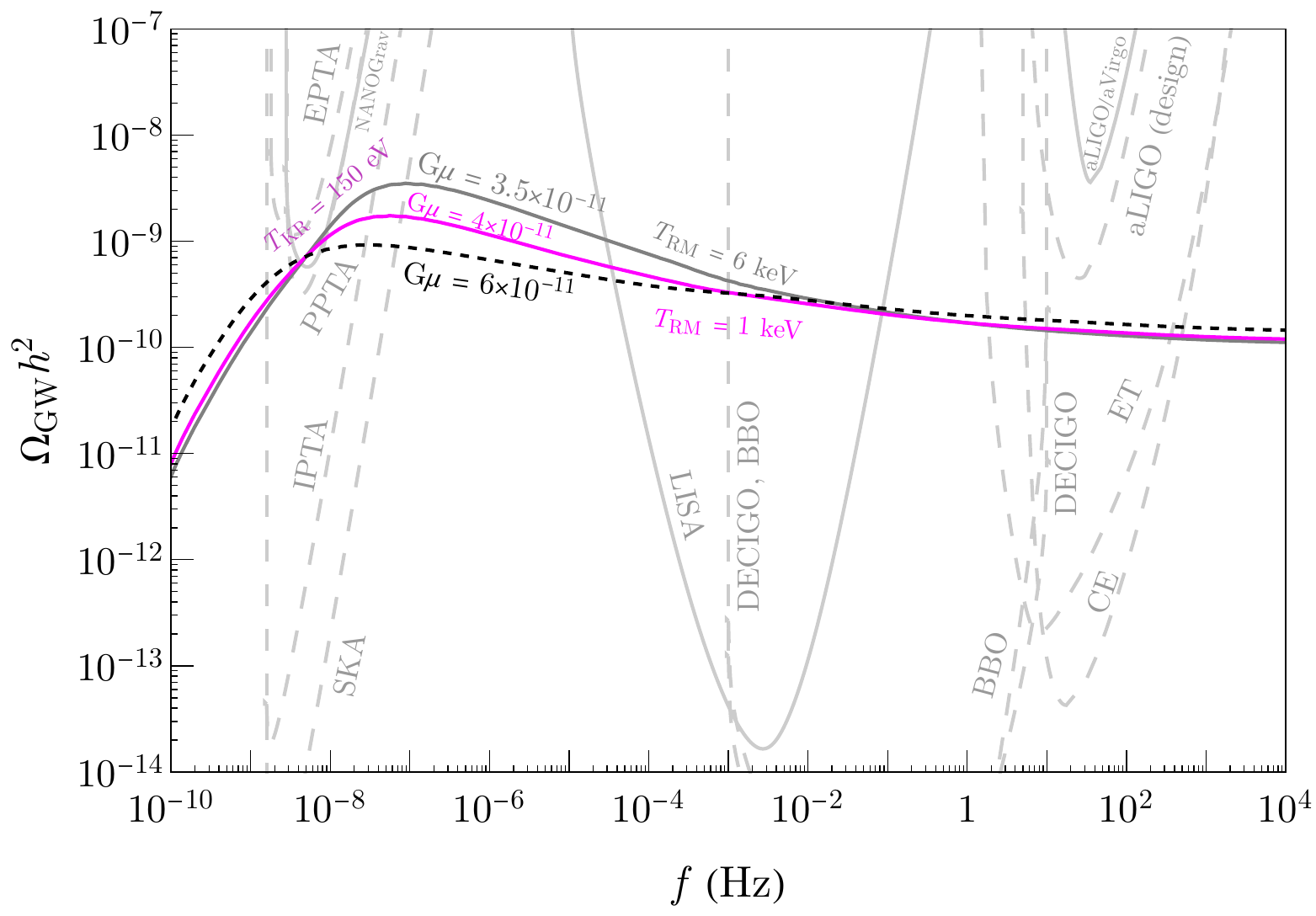}
    \caption{Representative spectra of primordial gravitational waves emitted from local cosmic strings experiencing axion kination (solid) and the standard $\Lambda$CDM cosmology (dashed).  Long eras of kination exhibit greater amplitudes in the triangular shaped peak of $\Omega_{\rm GW}h^2$, which is a key signature of axion kination. Of crucial importance is the slowly decaying high frequency tail arising from the sum over high mode numbers which enables detectors like BBO, DECIGO, and CE to detect deviations from the $\Lambda$CDM spectrum even when the kination peak is not located within their frequency domain. Left: Early axion kination cosmology where kination occurs before BBN. The top most contour shows the gravitational wave amplitude when $G\mu$ is fixed to pass through the NANOGrav signal. Right: Late axion kination cosmology where kination occurs in the epoch between CMB and BBN. For each contour, we plot the required $G\mu$ to pass through the NANOGrav signal.}
    \label{fig:GW_stringSpectrum}
\end{figure}

Fig.~\ref{fig:GW_stringSpectrum} shows the typical gravitational wave spectrum for a stochastic string background experiencing axion kination. Here, we numerically compute \eqref{eq:omegaGWSpec} up to $10^4$ modes and fix $\alpha = 0.1$ in accordance with simulations \cite{Blanco-Pillado:2017oxo,Blanco-Pillado:2013qja}. The solid contours show $\Omega_{\rm GW}h^2$ in the modified axion kination cosmology for a variety of $\{T_{\rm KR}, T_{\rm MK}, T_{\rm RM}\}$, while the dashed contours show the amplitude in the standard $\Lambda$CDM cosmology. The left and right panels of Fig.~\ref{fig:GW_stringSpectrum} represent the expected spectral shape for early and late axion kination eras, respectively. We define early (late) axion kination cosmologies as those that end before (after) BBN. To be consistent with BBN and CMB bounds, this entails that  $T_{\rm KR} \gtrsim \, 2.5 \, {\rm MeV}$ for early axion kination cosmologies and that $130 \, {\rm eV} \lesssim T_{\rm KR} \leq T_{\rm RM} \lesssim 6 \,{ \rm keV}$ for late kination cosmologies as discussed in Sec.~\ref{sec:constraints}.

The free parameters $\{G \mu, \alpha, T_{\rm KR}, T_{\rm MK}\}$ set the spectral shape of the stochastic background. Independent of the axion kination cosmology, larger $G\mu$ and $\alpha$ elevate the overall amplitude of the spectrum such that the base amplitude of the kination induced triangle scales as $\Omega_{\rm GW, \rm base} h^2 \simeq 2\times 10^{-4} \sqrt {G \mu \, \alpha}$ for $\alpha \gtrsim \Gamma G \mu$ \cite{Blasi:2020wpy}. On the other hand, the parameters $T_{\rm KR}$ and  $T_{\rm MK}$ determine the size and location of the triangular `bump' with a larger peak corresponding to longer duration of kination
\footnote{Equivalently, the greater the duration of the matter-dominated era. This follows from the temperature relationship 
$T_{\rm MK}^3 = T_{\rm RM} T_{\rm KR}^2 g_*(T_{\rm RM})/g_*(T_{\rm KR})$.}
and occurring at a higher frequency the greater $T_{\rm MK}$ is. For example, the solid and light purple contours on the left panel of  Fig.~\ref{fig:GW_stringSpectrum} illustrate the growth and blueshift of the kination induced peak when the duration of the kination era increases for fixed $G\mu$ and $T_{\rm{KR}}$. The red and blue contours in the same panel illustrate the overall decrease in amplitude and the blueshift of the spectrum when lowering $G\mu$ for a fixed duration of kination. In addition, the purple and red contours ($G\mu = 6 \times 10^{-11}$  and  $5 \times 10^{-15}$, respectively) illustrate that an era of axion induced kination provides future detectors such as LISA with an excellent opportunity to measure the significant deviation from the standard string spectrum (dashed) over a wide range of $G\mu$ and $(T_{\rm KR}, T_{\rm MK}$). Moreover, an axion induced kination cosmology may provide the only way to detect extremely small string tensions in future detectors like BBO and CE as shown by the blue contour ($G\mu = 5 \times 10^{-20}$). 

Similarly, the right panel of Fig.~\ref{fig:GW_stringSpectrum} shows the modified gravitational wave spectrum for late kination cosmologies. Here we show a collection of spectra that pass through the observed NANOGrav signal \cite{NANOGrav:2020bcs} that are consistent with CMB and BBN constraints. The dashed black contour shows $\Omega_{\rm GW}h^2$ for the standard $\Lambda$CDM cosmology \cite{Blasi:2020mfx, Ellis:2020ena} while the gray contour ($G\mu = 3.5 \times 10^{-11}$) shows $\Omega_{\rm GW}h^2$ for the maximum allowed $T_{\rm RM}$ ($6 \,{\rm keV})$ and near the minimum allowed $T_{\rm KR}$ ($130 \,{\rm eV})$, producing the largest kination peak passing through NANOGrav that is consistent with CMB and BBN. As $T_{\rm KR}$ and $T_{\rm RM}$ converge and the kination era decreases in duration, $\Omega_{\rm GW}h^2$ converges with the standard result, shown, for example, by the magenta contour ($G\mu = 4.0 \times 10^{-11}$). Fig.~\ref{fig:GW_stringSpectrum} demonstrates that a striking difference can exist between $\Omega_{\rm GW}h^2$ in the axion kination cosmologies and the standard $\Lambda$CDM  cosmology.

\begin{figure}
    \centering
    \includegraphics[width=0.495\columnwidth]{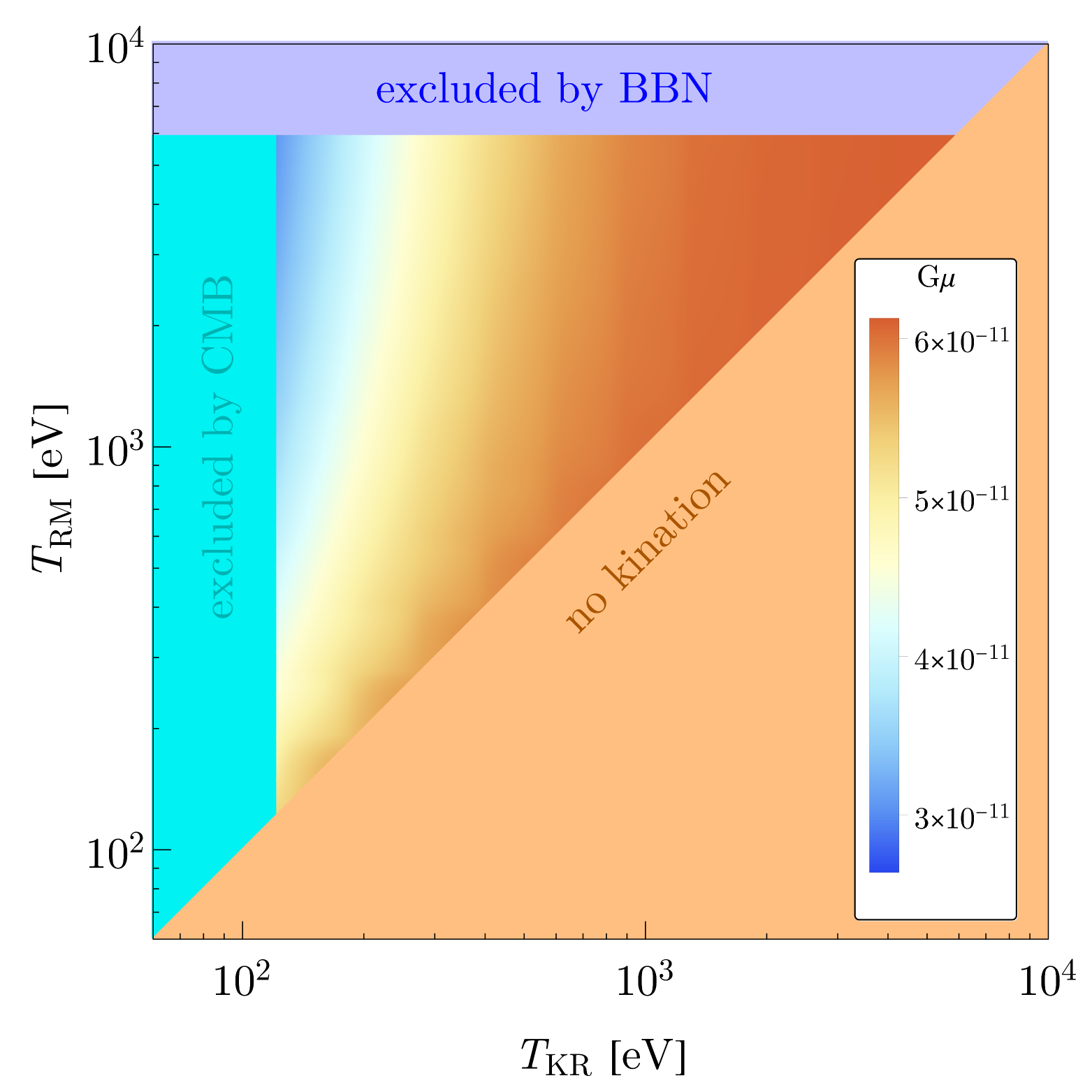}
    \includegraphics[width=0.495\columnwidth]{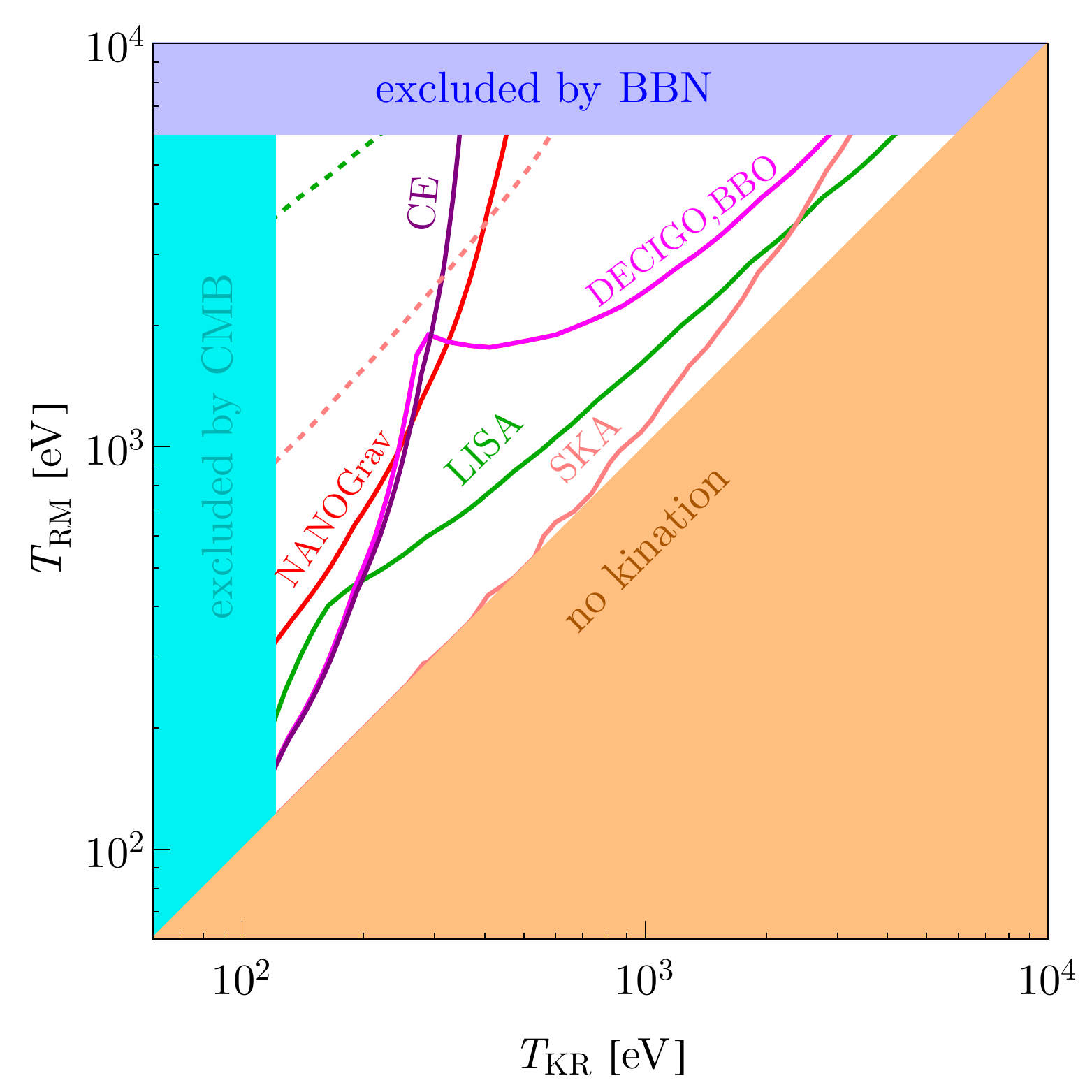}
    \caption{Left: Required $G\mu$ for $\Omega_{\rm GW}h^2$ to pass through the NANOGrav signal~\cite{NANOGrav:2020bcs,Blasi:2020mfx, Ellis:2020ena}. For long kination eras, which occur when $T_{\rm RM} \gg T_{\rm KR}$, $G\mu$ decreases with respect to the standard $\Lambda$CDM cosmology so that the kination peak does not exceed the NANOGrav signal.
    Right : The parameter region of axion kination whose imprints on the gravitational wave spectrum from cosmic strings can be detected. 
    For each ($T_{\rm RM},T_{\rm KR})$, we fix $G\mu$ according to the left panel so that spectrum passes through the NANOGrav signal. For the reference $\Lambda$CDM cosmology, we fix $G\mu$ and $\alpha$ to $6 \times 10^{-11}$ and $0.1$, respectively, to also fit NANOGrav. For a given $(T_{\rm KR}, T_{\rm MK})$, a detection is registered when the difference in amplitudes, $\Omega_{\rm GW} - \Omega_{\rm GW,0}$ is greater than $10\%$ (solid) or 100\% (dashed) of the standard cosmological amplitude, $\Omega_{\rm GW,0}$, within the sensitivity curve of the detector.}
    \label{fig:GmuLateKination}
\end{figure}

To understand the connection between the experimental detection of axion kination via string gravitational waves and axion kination parameters, we first reduce the four dimensional parameter space $\{G \mu, \alpha, T_{\rm KR}, T_{\rm MK}\}$ into a simpler two dimensional space of $T_{\rm KR}$ and $T_{\rm MK}$. This is achieved by fixing $G \mu$ so that the gravitational wave amplitude in the modified cosmology passes through the NANOGrav signal $(\Omega_{\rm GW}h^2, f) \simeq (7.5\times 10^{-10}, \, 5.2 \times 10^{-9} \,{\rm Hz})$)~\cite{NANOGrav:2020bcs}
\footnote{In this work, we do not fit the spectral index to NANOGrav. For early kination cosmologies, the nanohertz region of $\Omega_{\rm GW}h^2$ is effectively identical to the standard cosmology result and the best fit results of \cite{Blasi:2020mfx, Ellis:2020ena} apply. For the late kination cosmologies, the slope of the signal through NANOGrav can increase compared to the relatively flat slope of $\Lambda$CDM spectrum. It is possible a larger spectral index provides a better fit, but we leave that for future work.}
, and fixing $\alpha  = 0.1$ which best matches simulations. For early kination cosmologies (left panel of Fig.~\ref{fig:GW_stringSpectrum}), this requires $G\mu \simeq 6 \times 10^{-11}$. For late kination cosmologies (right panel Fig.~\ref{fig:GW_stringSpectrum}), which generally exhibit a `bump' in the spectrum at nanohertz frequencies, we decrease $G\mu$ for a given $(T_{\rm KR},T_{\rm MK})$ so that $\Omega_{\rm GW}h^2$ still crosses through the NANOGrav signal. The left panel of Fig.~\ref{fig:GmuLateKination} demonstrates this effect by showing the necessary $G \mu$ to match the NANOGrav signal in the ($T_{\rm KR}$ $T_{\rm RM}$) plane. For relative long durations of kination ($T_{\rm RM} \gg T_{\rm KR}$), the necessary $G\mu$ decreases by a factor of a few to near $G\mu = 3 \times 10^{-11}$ (blue region) whereas in the limit of no kination ($T_{\rm RM} \ll T_{\rm KR}$), the necessary $G\mu$ asymptotes to its $\Lambda$CDM value of $6 \times 10^{-11}$ (red region).

\begin{figure}
    \centering
     \includegraphics[width=0.475\columnwidth]{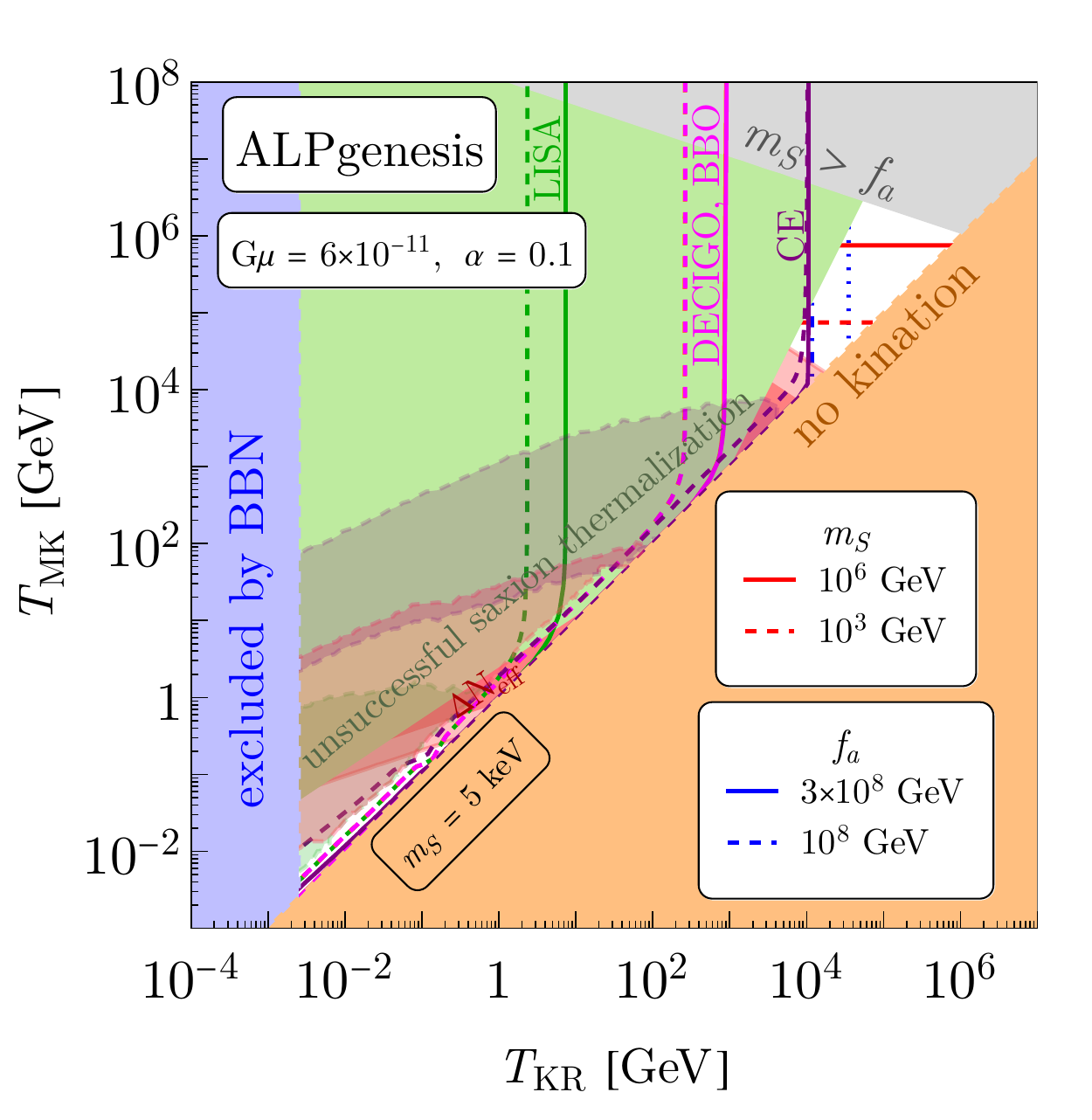}
     \includegraphics[width=0.475\columnwidth]{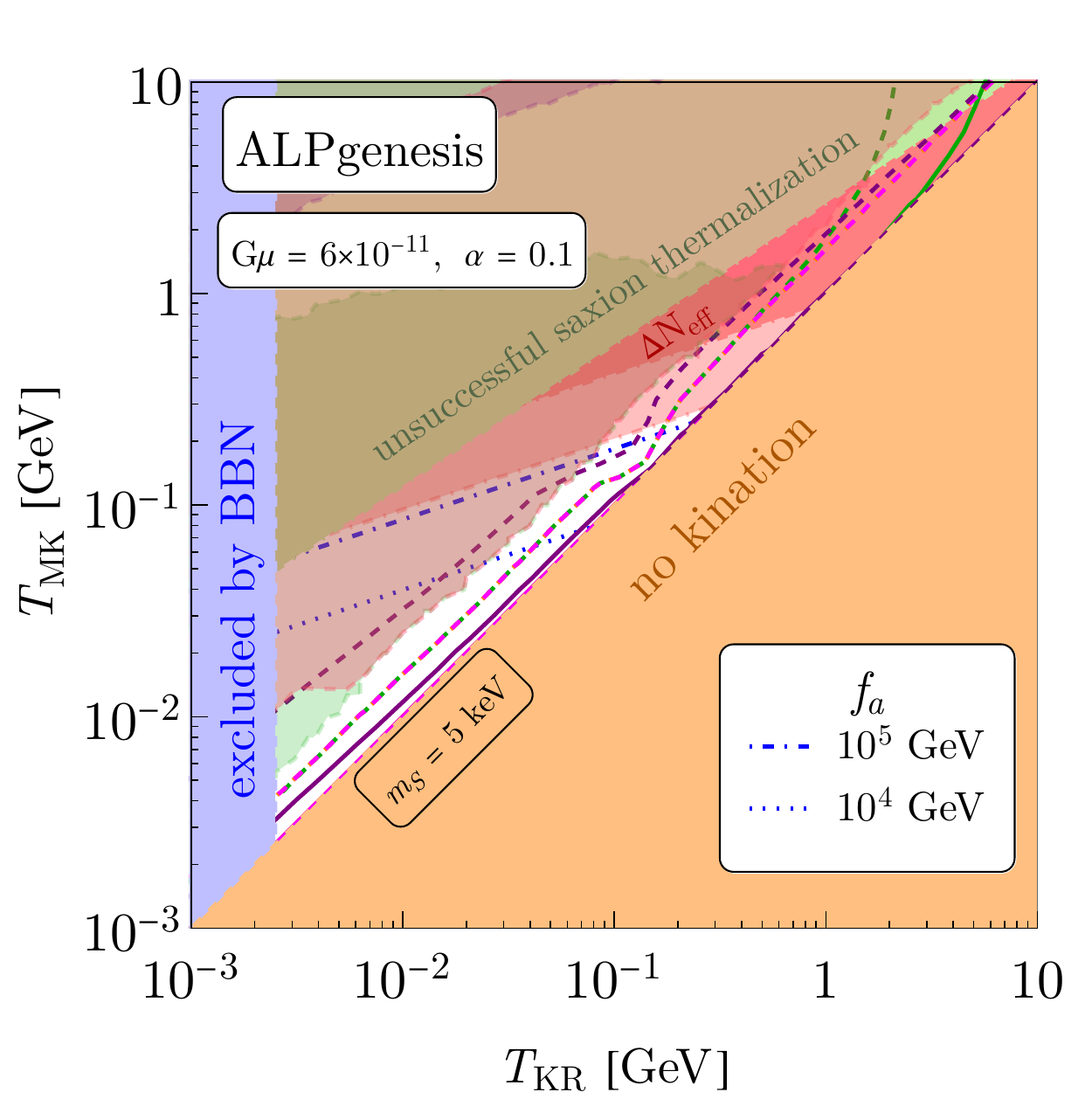}
    \includegraphics[width=0.475\columnwidth]{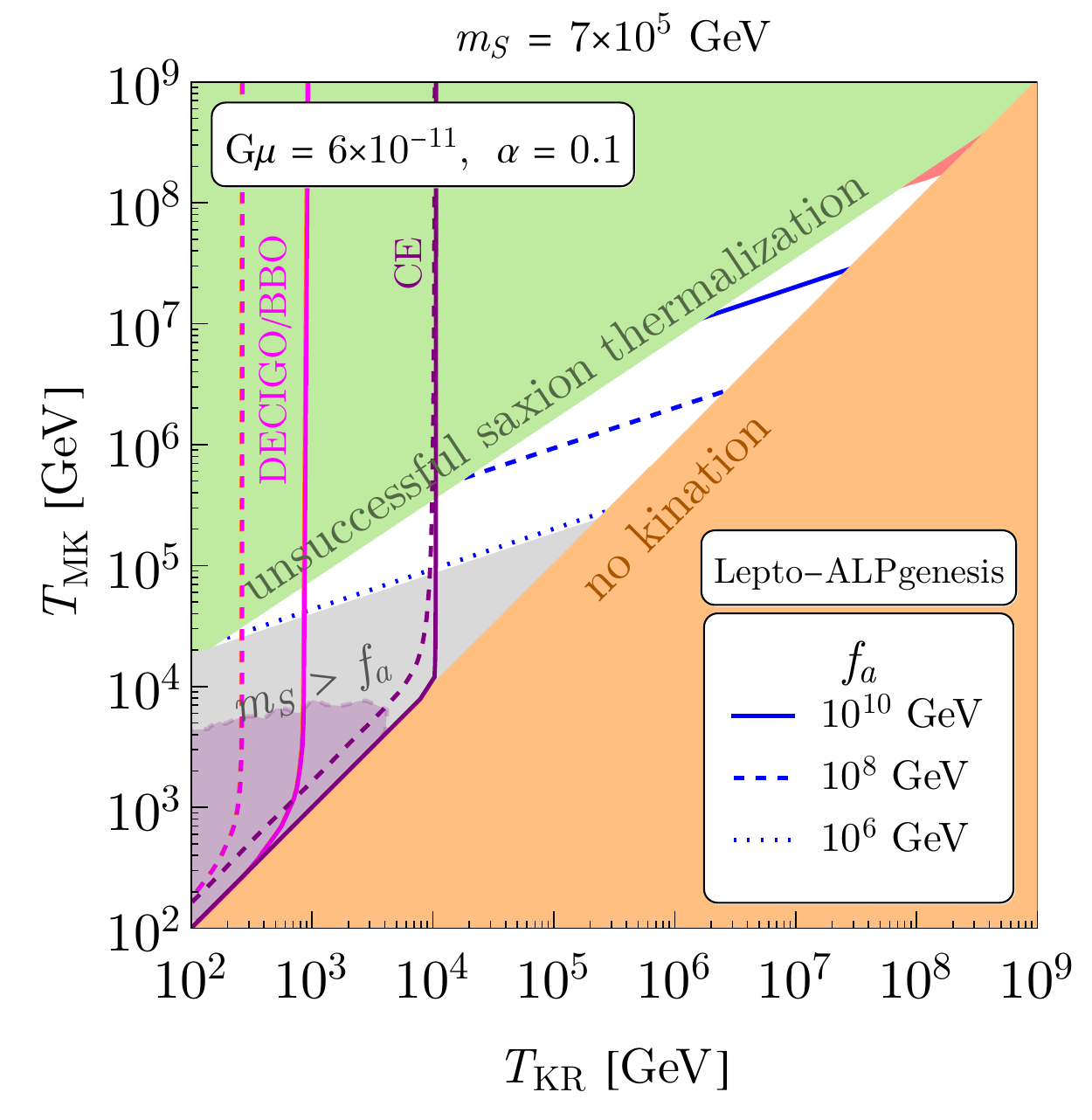}
    \includegraphics[width=0.475\columnwidth]{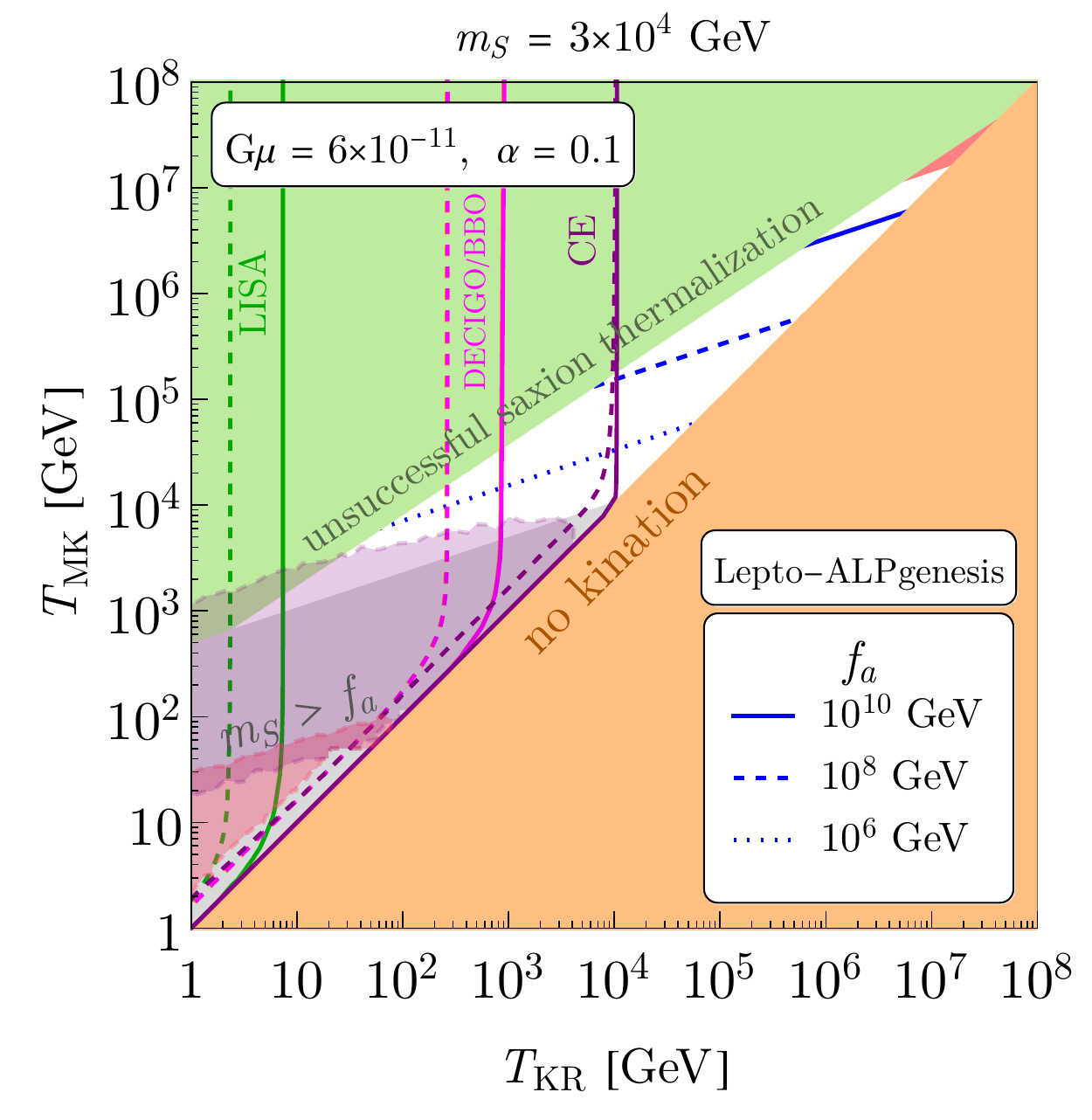}
    \caption{ Detector reach of the kination cosmic string gravitational wave spectrum for a range of $T_{\rm KR}$ and $T_{\rm MK}$ consistent with minimal ALPgenesis (top) and lepto-ALPgenesis (bottom). The top-right panel zooms in on the bottom-left part of the top-left panel. $G\mu$ and $\alpha$ are fixed at $6 \times 10^{-11}$ and $0.1$, respectively, to fit the NANOGrav data~\cite{NANOGrav:2020bcs}. For a given $(T_{\rm KR}, T_{\rm MK})$, a detection is registered when the difference in amplitudes, $\Omega_{\rm GW} - \Omega_{\rm GW,0}$ is greater than $10\%$ (solid) or 100\% (dashed) of the standard cosmological amplitude, $\Omega_{\rm GW,0}$, within the sensitivity curve the detector. In the transparent shared regions, the peak of the spectrum originated from axion kination can be detected.}
    \label{fig:GW_string_axiogenesis}
\end{figure}

For a given $(T_{\rm KR},T_{\rm MK})$, we register a detection of axion kination in a similar manner to the “turning-point” prescription of \cite{Gouttenoire:2019kij}: First, $\Omega_{\rm GW}h^2$ must be greater than the threshold for detection in a given experiment. Second, to actually distinguish between $\Omega_{\rm GW}h^2$ in the axion cosmology and the $\Lambda$CDM cosmology, we require that  their percent relative difference be greater than a certain threshold within the frequency domain of the experiment. Following~\cite{Gouttenoire:2019kij}, we take this threshold at a realistic 10\% and a more conservative 100\% relative difference. For more rigorous approaches in distinguishing similar gravitational wave spectra, see \cite{Kuroyanagi:2018csn,Caldwell:2018giq}.

The right panel of Fig.~\ref{fig:GmuLateKination} shows the parameter space in the $(T_{\rm KR},T_{\rm RM})$ plane where late axion kination can be detected and distinguished from the standard cosmology with difference of 10\% (solid) and 100\% (dashed) in $\Omega_{\rm GW}h^2$. Here we choose $G\mu$ for each point according to the left panel for axion kination cosmology and take $G\mu = 6\times 10^{-11}$ for the standard cosmology.
For most of the parameter space consistent with CMB and BBN, an era of kination can be detected and distinguished from the standard cosmological stochastic string background. 
In addition to the change of the required value of $G\mu$, remarkably, the slowly decaying $f^{-1/3}$ tail originating from the sum over high frequency harmonics, as shown for example by the right panel of Fig.~\ref{fig:GW_stringSpectrum}, allows detectors like LISA, BBO, DECIGO, and CE to detect late axion kination cosmology.
Future detectors like SKA can probe the nanohertz triangular bump. For sufficiently low $T_{\rm KR}$ and high $T_{\rm RM}$, a kination signal may already be observable or excluded at NANOGrav.

Early axion kination is consistent with axiogenesis above the electroweak scale, and can be probed by laser interferometers. We show the constraints on the parameter space of minimal ALPgenesis together with the detection prospects in the upper panels of Fig.~\ref{fig:GW_string_axiogenesis}. The top-right panel zooms in on the bottom-left part of the top-left panel.
The slowly decaying $f^{-1/3}$ tail allows detectors like LISA, BBO, DECIGO, and CE to distinguish an early era of axion cosmology for most $T_{\rm KR} \in (10^{-3} \, {\rm GeV}, 5 \times 10^{4} \, {\rm GeV})$. For $T_{\rm KR} \gtrsim 5 \times 10^4 \, {\rm GeV}$, the kination spectrum merges with the standard spectrum at frequencies above $f \gtrsim 10^3$ Hz, thereby evading detection. Still, a good portion of the parameter space with $f_a \lesssim 10^8$ GeV can imprint signals that are detectable by future observations. Future gravitational wave detectors that can observe super-kilohertz frequencies can potentially probe earlier eras of axion kination and hence larger $f_a$.
In the transparent shaded region, the peak of the spectrum produced by axion kination can be detected. As we argued, this is a smoking-gun signature of axion kination and the detailed shape of the peak contains information about the shape of the potential of the complex field that breaks the $U(1)$ symmetry.
The lower two panels of Fig.~\ref{fig:GW_string_axiogenesis} show the constraints and prospects for lepto-ALPgenesis, for values of $m_S$ used in Fig.~\ref{fig:lepto-ALP}. Future laser interferometers can probe much of the parameter region with low $f_a \lesssim 10^8$ GeV.

\section{Summary and discussion}
\label{sec:summary}

Axion fields, due to their lightness, may have rich dynamics in the early universe. In this paper, we considered rotations of an axion in field space that naturally provide kination domination preceded by matter domination, which we call axion kination.
This non-standard evolution affects the spectrum of possible gravitational waves produced in the early universe. To be concrete, we investigated gravitational waves from inflation and local cosmic strings, which have a nearly flat spectrum when they begin oscillations or are produced during radiation domination.
We found that kination domination preceded by matter domination induces a triangular peak in the gravitational wave spectrum.

We studied the theory for axion kination, which involves an approximately quadratic potential for the radial mode and has three parameters: the mass of the radial mode, the axion decay constant, and the comoving charge density. We derived constraints on this parameter space from successful thermalization of the radial mode, BBN, and the CMB.  We found large areas of fully realistic parameter space where the theory yields axion kination.  The allowed region splits into two pieces, one having early kination domination before BBN and the other having late kination after BBN but well before the CMB last scattering. 

Introducing a mass for the axion, we found that part of the axion kination parameter space is consistent with axion dark matter by the kinetic misalignment mechanism while part is not, due to the warmness constraint on dark matter.  Similarly, we showed that part of the axion kination parameter space is consistent with generating the baryon asymmetry by ALPgenesis. Furthermore, there are constrained regions with ALP cogenesis yielding both dark matter and the baryon asymmetry, and also regions with the baryon asymmetry successfully generated by lepto-ALPgenesis.

As demonstrated in Sec.~\ref{sec:GW}, axion kination modifies the spectrum of possible primordial gravitational waves through the modification of the expansion history of the universe. By analyzing the spectrum, we can in principle determine the product of the radial mode mass $m_S$ and the decay constant $f_a$ using the relations given in Eqs.~(\ref{eq:TRM}), (\ref{eq:TMK}), and (\ref{eq:TKR}). By further determining $f_a$ from axion searches, we may obtain $m_S$. In the simplest scenario of gravity mediation in supersymmetry, $m_S$ is as large as the masses of the gravitino and scalar partners of Standard Model particles; in other words, we can determine the scale of supersymmetry breaking.

We can further narrow down the parameter space by requiring that the baryon asymmetry of the universe be created from the axion rotation. As shown in Sec.~\ref{subsec:baryogenesis}, this imposes an extra relation on $(m_S, f_a)$, and in conjunction with the gravitational wave spectrum, we may make a prediction on both $f_a$ and $m_S$, which could be confirmed or excluded by measuring $f_a$ in axion experiments or $m_S$ in collider experiments assuming that $m_S$ is tied to the masses of the scalar partners of Standard Model particles.

If the inflation scale is not much below the current upper bound, future observation of gravitational waves can detect the spectrum modified by axion kination, or even the peak of the spectrum that contains information on the shape of the potential of the $U(1)$ symmetry breaking field.
In particular, if the QCD axion accounts for dark matter via the kinetic misalignment mechanism, a modification of the gravitational wave spectrum is predicted at high frequencies, $f \gtrsim 10^{-2}$ Hz, as shown in Fig.~\ref{fig:PGW}. In this case it is very interesting that this gravitational wave signal can be detected by DECIGO and BBO over most of the allowed parameter space, as shown in Fig.~\ref{fig:kination_QCD_axion}; in a significant fraction of the parameter space the gravitational wave peak will be probed. Furthermore, a signal may also be seen at CE if the inflation scale is very near the current upper bound or the sensitivity of CE is improved.

For gravitational waves from cosmic strings, for fixed axion kination cosmology parameters, the modification of the spectrum is predicted at higher frequency, so the QCD axion will not affect the spectrum observable by near future planned experiments. ALPs can affect the spectrum in an observable frequency range.

Gravitational waves from cosmic strings provide signals that can probe axion kination over a wide range of $G\mu$, $T_{\rm{RM}}$ and $T_{\rm{KR}}$, as illustrated in Fig.~\ref{fig:GW_stringSpectrum}.
We examined cosmic strings with a tension suggested by NANOGrav in detail. If axion kination occurs before BBN, the NANOGrav signal can be fitted by the same cosmic strings parameters as in standard cosmology. Importantly, axion kination enhances the spectrum at higher frequencies, allowing laser interferometers to probe the kination era. The enhancement can occur in the parameter region consistent with axiogenesis scenarios, as shown in Fig.~\ref{fig:GW_string_axiogenesis}. If axion kination occurs after BBN, the NANOGrav signal is fitted by a smaller string tension, as shown in the left panel of Fig.~\ref{fig:GmuLateKination}, and a detailed examination of the spectrum will determine if axion kination is involved. The spectrum at higher frequencies is suppressed, which can be detected by laser interferometers in the parameter region shown in the right panel of Fig.~\ref{fig:GmuLateKination}.

Our kination era is preceded by an epoch of matter domination that ends without creating entropy. Therefore, matter and kination domination can occur even after BBN. This allows for enhancements to the matter spectrum on small scales that may be probed by observations of Lyman-$\alpha$ and 21 cm lines. Evolving the enhanced matter power spectrum into the non-linear regime and understanding its effects on the Lyman-$\alpha$ flux spectrum as well as hierarchical galaxy formation, and constraints arising from corresponding observations will be discussed in future work.

In this paper, we concentrated on gravitational waves produced by inflation or local cosmic strings and modified by axion kination. At any temperature with early matter or kination domination, the Hubble scale is larger than with radiation domination, and hence, quite generally, primordial gravitational waves are enhanced by axion kination. Furthermore, a distinctive feature appears in the spectrum, a peak or bump depending on the field potential, containing information that probes in detail the era of kination and its origin. It will be interesting to investigate other sources of primordial gravitational waves.

\vspace{0.5cm}

{\it Note added.} Soon after the current manuscript was announced on the arXiv, Ref.~\cite{Gouttenoire:2021wzu} appeared as well. While our analyses have been conducted independently, Ref.~\cite{Gouttenoire:2021wzu} also discovered the triangular peak signature of axion kination in the spectrum of the primordial gravitational waves from inflation.

\section*{Acknowledgement}
R.C.~and K.H.~thank Kai Schmitz for fruitful discussion.
The work was supported in part by DoE grant DE-SC0011842 at the University of Minnesota (R.C.) and DE-AC02-05CH11231 (L.J.H.), the National Science Foundation under grant PHY-1915314 (L.J.H.), Friends of the Institute for Advanced Study (K.H.),  DE-SC0017840 (N.F., J.S.)  and DE–SC0007914 (A.G). The work of R.C., N.F., and L.J.H. was performed in part at Aspen Center for Physics, which is supported by National Science Foundation grant PHY-1607611. This work was partially supported by a grant from the Simons Foundation.

\appendix

\section{Evolution of the energy density of axion rotations}
\label{app:evol_axion}
In this appendix, we derive the evolution of the axion rotations for the nearly quadratic potentials.

\subsection{Logarithmic potential}
\label{app:log_pot}
We first consider the logarithmic potential in Eq.~(\ref{eq:dim_trans}). The evolution of the axion rotations for this potential is derived in Ref.~\cite{Co:2019wyp}.

We are interested in the rotation dynamics when the Hubble expansion is negligible, $m_S \gg H$. In this case, we may obtain short-time scale dynamics ignoring the Hubble expansion and include it when we derive the scaling in a long cosmic time scale. We are also interested in the circular motion after thermalization. Under these assumptions, we may put $\ddot{S} = \dot{S}=0$, and the equation of motion of $S$ requires that
\begin{align}
    \dot{\theta}^2 = \frac{V'(S)}{S} = \frac{m_S^2}{2} {\rm ln}\frac{S^2}{f_a^2}.
\end{align}
The equation of motion of $\theta$ gives a conservation law of the angular momentum in the field space up to cosmic expansion,
\begin{align}
\label{eq:eomtheta_log}
  n_\theta = \dot{\theta}S^2 = m_S S^2 \left( {\rm ln}\frac{S}{f_a}\right)^{1/2} \propto a^{-3}.
\end{align}
Using these two equations, we may obtain the dependence of $S$ and $\dot{\theta}$ on the scale factor $a$,
\begin{align}
\label{eq:S-theta-one}
    S^2 = S_i^2 \frac{a_i^3}{a^3} 2 \sqrt{\frac{{\rm ln} S_i/f_a}{W(\frac{4 a_i^6 S_i^4 {\rm ln}(S_i/f_a)}{a^6 f_a^4})}}, \nonumber \\
    \dot{\theta}^2 = \frac{1}{2}m_S^2 W(\frac{4 a_i^6 S_i^4 {\rm ln}(S_i/f_a)}{a^6 f_a^4}), 
\end{align}
where $W$ is the Lambert $W$ function and $S_i$ is an initial field value at a scale factor of $a_i$.
When $S_i \gg f_a$ and $a$ is not much above $a_i$, $S^2 \propto a^{-3}$ and $\dot{\theta}$ is nearly a constant. For $a \gg a_i$, $S\simeq f_a$ and $\dot{\theta} \propto a^{-3}$.

The dependence of the energy density
\begin{align}
    \rho_\theta = \frac{1}{2}\dot{\theta}^2 S^2 + \frac{1}{4}m_S^2 S^2 \left({\rm ln} \frac{S^2}{f_a^2}-1\right)+ \frac{1}{4}m_S^2 f_a^2
\end{align}
can be obtained by using Eq.~(\ref{eq:S-theta-one}). One can then show that the dependence of $\rho_\theta$ on the scale factor $a$ is
\begin{align}
\label{eq:rhologevol}
    \frac{d\ln\rho_\theta}{d\ln a} = \frac{ -6 \left(\frac{S}{f_a}\right)^2 \ln \left(\frac{S^2}{f_a^2}\right)}{1-\left(\frac{S}{f_a}\right)^2+2 \left(\frac{S}{f_a}\right)^2 \ln \left(\frac{S^2}{f_a^2}\right)} = 
    \begin{cases}
     -3 & : S \gg f_a \\
     -6 & : S \simeq f_a.
    \end{cases}
    ,
\end{align}
so at early times, $S \gg f_a$, the rotation behaves as matter $\rho_\theta \propto a^{-3}$, while at late times, $S \simeq f_a$, the rotation behaves as kination $\rho_\theta \propto a^{-6}$. This behavior is seen in the orange curve of Fig.~\ref{fig:EoS}.

\subsection{Two-field model}
\label{app:twofield}

We next consider the two-field model in Eq.~(\ref{eq:two_field}). We assume that the saxion field value is much larger than the soft masses, so that we may integrate out a linear combination of $P$ and $\bar{P}$ that is paired with $X$ and obtain a mass $\sim S$. Using the constraint $\bar{P} = v_P^2 / P $, from the kinetic and mass terms of $P$ and $\bar{P}$, we obtain an effective Lagrangian
\begin{align}
    {\cal L} = \left( 1 + \frac{v_P^4}{|P|^4} \right) |\partial P|^2 - m_P^2|P|^2 \left( 1 + r_P^2 \frac{v_P^4}{|P|^4} \right),~~r_P \equiv \frac{m_{\bar{P}}}{m_P}.
\end{align}
The potential has a minimum at $|P| = \sqrt{r_P} v_P$ when both $m_P^2$ and $m_{\bar{P}}^2$ are positive.

The equation of motion of $S \equiv \sqrt{2} |P|$ with $\ddot{S} = \dot{S}=0$ requires that
\begin{align}
\label{eq:eomS}
\dot{\theta}^2\left( 1 - \frac{4 v_P^4}{S^4}\right) - m_P^2 \left(1 - \frac{4 r_P^2 v_P^4}{S^4}\right) =0.
\end{align}
The equation of motion of $\theta$ gives a conservation of the angular momentum,
\begin{align}
\label{eq:eomtheta_two_field}
   n_\theta = \dot{\theta}S^2 \left( 1 + \frac{4 v_P^4}{S^4}\right) \propto a^{-3}.
\end{align}

Without loss of generality, we assume that $P \gg v_P$, i.e., $S \gg v_P$ initially. 
We first consider $r_P>1$ ($m_P < m_{\bar{P}}$). Eq.~(\ref{eq:eomS}) means
\begin{align}
\label{eq:theta_twofield}
    \dot{\theta}^2 = \frac{1 - 4 r_P^2 v_P^4/S^4}{1- 4 v_P^4 / S^4} m_P^2.
\end{align}
When $S \gg v_P$, we obtain $\dot{\theta}^2 \simeq m_P^2$. As $S$ approaches the minimum $\sqrt{2 r_P} v_P$, $\dot{\theta}$ approaches $0$. The scaling of $S$ can be derived from charge conservation,
\begin{align}
    n_\theta = \dot{\theta}S^2 \left( 1 + \frac{4 v_P^4}{S^4}\right) =  m_P S^2 \left( 1 + \frac{4v_P^4}{S^4} \right) \left( 1 - \frac{4v_P^4}{S^4} \right)^{-1/2}\left( 1 - \frac{4 r_P^2v_P^4}{S^4} \right)^{1/2} \propto a^{-3}.
\end{align}
The scaling of the energy density
\begin{align}
    \rho_\theta = \frac{1}{2}S^2 \dot{\theta}^2 \left( 1 + \frac{4 v_P^4}{S^4}\right) + \frac{1}{2}m_P^2 S^2 \left( 1 + \frac{4 r_P^2 v_P^4}{S^4}\right) - 2 r_P m_P^2 v_P^2
\end{align}
can be derived from these two equations. Here a constant term is subtracted from the energy density so that the energy density vanishes at the minimum. The dependence of $\rho_\theta$ on the scale factor $a$ is
\begin{align}
    \frac{d\ln\rho_\theta}{d\ln a} = \frac{-3 \left(1+\left(\frac{S}{\sqrt{2 r_P} v_P}\right)^2\right) \left(1 + r_P^2  \left(\frac{S}{\sqrt{2 r_P} v_P}\right)^4\right)}{1+r_P^2  \left(\frac{S}{\sqrt{2 r_P} v_P}\right)^6} = 
    \begin{cases}
     -3 & : S \gg \sqrt{2 r_P} v_P \\
     -6 & : S \simeq \sqrt{2 r_P} v_P
    \end{cases}
    ,
\end{align}
from which we again observe that at early times, $S \gg \sqrt{2 r_P} v_P$, the rotation behaves as matter $\rho_\theta \propto a^{-3}$, whereas at late times, $S \simeq \sqrt{2 r_P} v_P$, the rotation behaves as kination $\rho_\theta \propto a^{-6}$. This evolution is illustrated by the blue curves of Fig.~\ref{fig:EoS} for various values of $r_P$.
\begin{figure}[t!]
    \centering
    \includegraphics[width=\columnwidth]{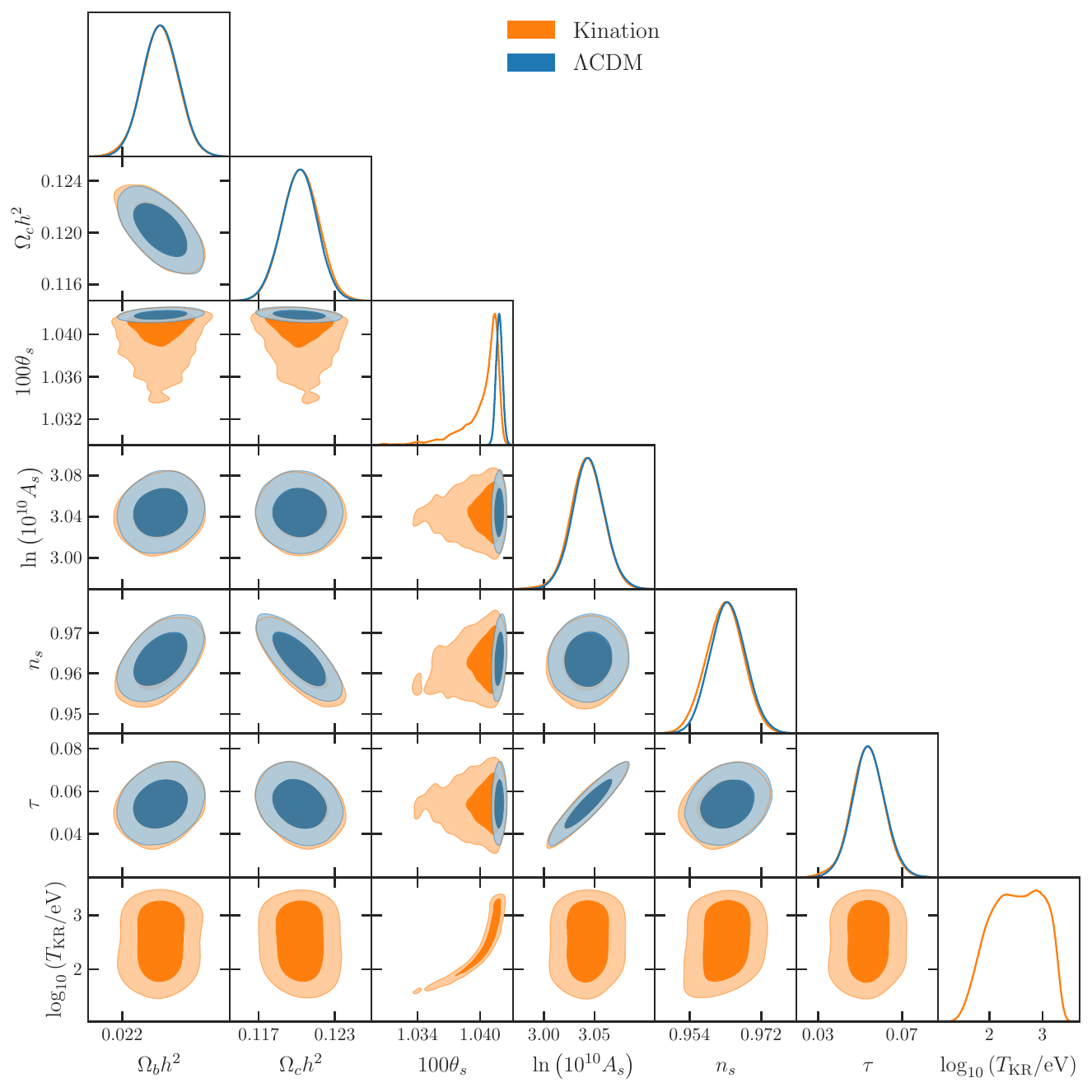}
    \caption{Corner plot for the posterior distributions for the $\Lambda$CDM independent parameters and for the axion kination model. We use the highTTTEEE+lowEE+lowTT likelihood combination from \textit{Planck} 2018. Contours contain 68\% and 95\% of the probability.}
    \label{fig:cornerplot}
\end{figure}

We next consider $r_P=1$. Eq.~(\ref{eq:eomS}) has two solutions, 1) $\dot{\theta}^2 = m_P^2$ with $S\neq \sqrt{2} v_P$ and 2) $S = \sqrt{2} v_P$ with unrestricted $\dot{\theta}$. For $S \gg v_P$, the solution is in the branch 1) and gives matter scaling. Charge conservation implies $S^2 ( 1 + 4 v_P^4 / S^4) \propto a^{-3}$. As $S$ decreases according to this scaling and reaches $\sqrt{2} v_P$, the branch 2) should be used and charge conservation implies $\dot{\theta} \propto a^{-3}$, giving kination scaling.

Finally, consider $r_P < 1$. When $S \gg v_P$, we again obtain $\dot{\theta}^2 \simeq m_P^2$. However, as $S$ approaches $\sqrt{2}v_P$ (before reaching the minimum at $\sqrt{2r_P} v_P$), $\dot{\theta}^2$ derived from Eq.~(\ref{eq:theta_twofield}) diverges. This is the point at which the solution becomes unstable.
Indeed, when $r_P<1$, for a fixed charge, it is energetically favored to have rotations in $\bar{P}$ rather than in $P$, so the rotation dominantly in $P$ is at the most a meta-stable solution. When $S$ reaches $\sqrt{2}v_P$, the solution becomes unstable. Quantum tunneling may occur before the instability is reached. We leave the investigation of this scenario to future work and assume $r_P \geq 1$.

\section{CMB cosmological constraints from \textit{Planck}}
\label{app:parameters}

\subsection{Perturbation equations of the $P$ field}

In order to consistently take into account the growth of perturbations in our axion kination cosmology, we need to account correctly for the perturbation equations of the complex scalar field. We focus on the model in Eq.~\eqref{eq:dim_trans}, where the kinetic term of $P$ is canonical. The complex scalar field can be written as
\begin{align}
    P = X + i Y .
\end{align}

We follow Ref.~\cite{Hu:1998kj} in deriving our perturbation equations and will separate the equations of motion for the zero modes of $X$ and $Y$ as
\begin{align}
\label{eq:backgroundeom}
    X''(\tau) + \frac{2 a'(\tau)}{a(\tau)}X'(\tau) + a^{2}(\tau)V_{X} &= 0 , \\
    Y''(\tau) + \frac{2 a'(\tau)}{a(\tau)}Y'(\tau) + a^{2}(\tau)V_{Y} &= 0 ,
\end{align}
where $V_X = \frac{\partial V}{\partial X}$ and $V_Y = \frac{\partial V}{\partial Y}$, and primes denote derivatives with respect to the conformal time $\tau$. In what follows, we will drop the explicit $\tau$ dependence for brevity.
The zeroth-order energy density and pressure can be written as
\begin{align}
    \label{eq:bgrhop}
    \rho &= \frac{1}{2 a^{2}}X'^{2} +\frac{1}{2 a^{2}}Y'^{2} + V , \\
    p &= \frac{1}{2 a^{2}}X'^{2} +\frac{1}{2 a^{2}}Y'^{2} - V .
\end{align}

The equations for the field perturbations $\delta X =x$ and $\delta Y = y$ are given by
\begin{align}
    \label{eq:perteom}
    x'' + \frac{2 a'}{a}x' + \left(k^{2} x + a^{2} x V_{X,X}  + a^{2} y V_{X,Y}\right) - \left(h_v' - 3 h_\delta'\right)X' + 2 a^{2} h_v V_{X}  &= 0 , \\
    y'' + \frac{2 a'}{a}y' + \left(k^{2} y + a^{2} y V_{Y,Y} + a^{2} x V_{X,Y}\right) - \left(h_v' - 3 h_\delta'\right)Y' + 2 a^{2} h_v V_{Y} &= 0 ,
\end{align}
where $h_v,h_\delta$ are gravitational potential perturbations. In synchronous gauge, which we will use for our end result, $h_\delta = \frac{h}{6}$  and $h_{v} = 0$ in notation of Ref.~\cite{Hu:1998kj}.
The perturbations of $\rho$ and $p$, and the fluid velocity $v$ are
\begin{align}
     \label{eq:pertrhop}
     \delta \rho &= \frac{1}{a^{2}}\left(X' x' + Y' y' -\left(X'^{2}+Y'^{2}\right)h_{v}\right) + x V_{X}+ y V_{Y} , \\
     \delta p &= \delta \rho - 2 \left(x V_{X} + y V_{Y}\right) , \\
     \left(\rho+p\right)v &= \frac{k}{a^{2}}\left(x X' + y Y'\right) .
\end{align}
Differentiating $\delta \rho$ with respect to the conformal time and using the above equations, we obtain
\begin{align}
    \label{eq:deltarhodot}
    \delta \rho' = -\left(\rho + p\right)\left(k v + 3 h_\delta'\right) - 6\frac{a'}{a}\delta \rho + 6 \frac{a'}{a}\left(x V_{X} +y V_{Y}\right),\\
    \left(\left(\rho + p\right)v/k \right)' = \delta \rho - 2 \left(x V_{X} + y V_{Y}\right) - 4 \frac{a'}{a}\left(\rho+p\right)v/k .
\end{align}

We can now transform from $X, Y$ to radial and angular degree of freedom. We can write
\begin{align}
    \label{eq:radial}
    X &= R \cos \theta , \hspace{- 1 in}
    && Y = R \sin \theta ,  \\
    x &= \delta R \cos \theta - R \sin \theta \delta \theta , \hspace{- 1 in}
    && y = \delta R \sin \theta + R \cos \theta \delta \theta . 
\end{align}
Assuming radial symmetry for our potential, the derivatives of the potential are
\begin{align}
  V_{X} = V_R\frac{X}{R},\qquad V_{Y} = V_{R}\frac{Y}{R},
\end{align}
leading to a simple expression
\begin{align}
    x V_X + y V_Y = V_{R} \delta R.
\end{align}
Also,
\begin{align}
\label{eq:aux}
     x X' + y Y' = \delta R R' + \delta \theta R^2 \theta', \nonumber \\
     x'X'+ y'Y' = \delta R' R' + \delta R R \theta'^{2} +  \delta \theta' \theta' R^2.
\end{align}

There generically exist two perturbation modes from $\delta R$ and $\delta \theta$. However, for $k/a \ll m_S$, we may integrate out the heavy degree of freedom (equivalently, the mode with a high frequency) that is nearly $\delta R$. The process goes as follows.
We can use Eq.~(5) in Ref.~\cite{Boyle:2001du} to write the corresponding equation for the evolution of $\delta R$ in synchronous gauge,
\begin{align}
    \label{eq:deltaReq}
    \delta R'' + 2 \frac{a'}{a}\delta R'+ (k^{2} + a^{2} V_{R,R}- \theta'^{2}) \delta R =  2 R \delta \theta ' \theta',
\end{align}
and we have used the fact that $R'$ is small to derive the above.
Note that for a logarithmic potential\footnote{The exact shape of the potential is crucial here. For a quadratic potential Eq.~(\ref{eq:vrr}) would be zero, prohibiting us from making simplifying  approximations in what follows. } in Eq.~(\ref{eq:dim_trans}),
 \begin{align}
 \label{eq:vrr}
     a^{2} V_{R,R} - \theta'^{2} = a^{2}\left(V_{R,R} - \frac{V_{R}}{R}\right) = a^{2} m_{S}^{2}.
 \end{align}
Since $m_{S} \gg H$, the coefficient of $\delta R$ in Eq.~\eqref{eq:deltaReq} dominates. For $k/a \ll  m_{S}$ we can write 
\begin{align}
    \label{eq:deltaR}
    \frac{\delta R}{R} =  \frac{2 \theta' \delta \theta'}{a^{2} m_S^{2}}.
\end{align}

Using Eqs.~\eqref{eq:pertrhop} and \eqref{eq:aux} and the fact that $R'$ is small and $\theta'^{2} = a^{2} V_{R}/R$  we can also write
\begin{align}
    \label{eq:deltarho}
    \delta \rho = 2  V_{R} \delta R + \frac{1}{a^{2}} R^{2} \theta' \delta \theta'.
\end{align}
Then using Eqs.~\eqref{eq:deltaR} and \eqref{eq:deltarho} we get
\begin{align}
    \label{eq:deltaRfinal}
    \delta R  = \frac{2 \delta \rho}{4 V_{R} + R m^{2}_{S}} .
\end{align}
The modified equations then become
\begin{align}
    \label{eq:pertfluidkin}
    \delta_{\phi}' +\left(1+w\right)\left(\Theta + \frac{h'}{2}\right) +\frac{3 a'}{a}\left(1- \frac{4 V_{R}}{4 V_{R} +  m^{2}_{S} R} -w \right)\delta_{\phi} &= 0 ,\\
    \label{eq:pertfluidkinTheta}
    \Theta ' + \frac{a'}{a}\Theta \left(1-3 c^{2}_{\phi} \right)-k^{2}\frac{\delta_{\phi}\left(1- \frac{4 V_{R}}{4 V_{R} +  m^{2}_{S} R}\right)}{1+w} &=0, 
\end{align}
where $c^{2}_{\phi} = w - \frac{w'}{3(1+w)\frac{a'}{a}}$ is the adiabatic speed of sound of the fluid.

Note that the above equations approach to matter perturbation equations as $w \rightarrow 0$ and kination perturbation equations as $w \rightarrow 1$.  This is consistent with the fact that radial perturbations do not grow in a logarithmic potential. The above equations are then implemented in CLASS.
\subsection{Implementation in CLASS}
\begin{figure}[t!]
    \centering
    \includegraphics[width=0.7\columnwidth]{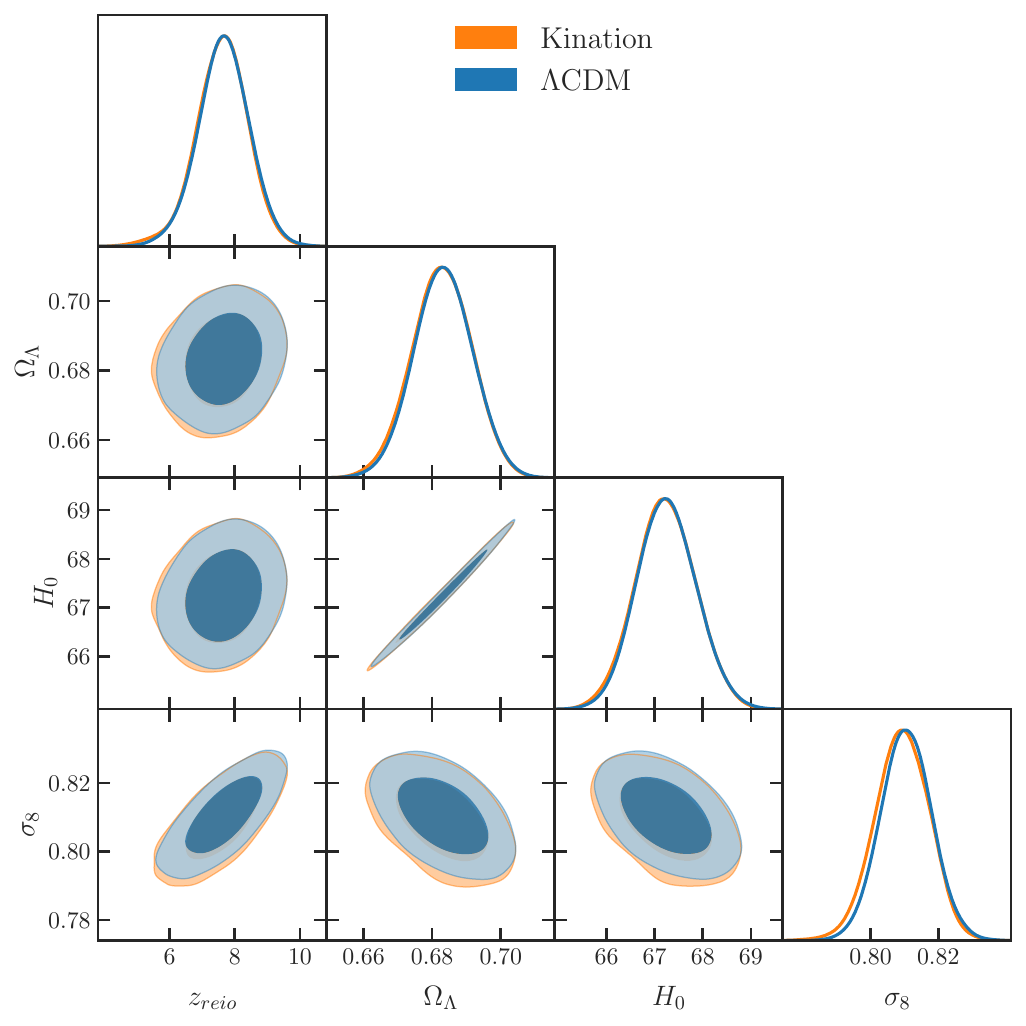}
    \caption{Corner plot for the posterior distributions for the calculated values $\Lambda$CDM parameters and axion kination cosmology. Contours contain 68\% and 95\% of the probability.}
    \label{fig:cornerplot_d}
\end{figure}

We use a modified version of CLASS \cite{Blas:2011rf} to solve the coupled Boltzmann equations and Monte Python \cite{Brinckmann:2018cvx}  and perform a parameter estimation with the \textit{Planck} 2018 likelihoods (TT,TE,EE+lowE) data \cite{Planck:2019nip}.

We take inspiration from the fluid module already present in CLASS to define our own ``kination'' module to take into account the perturbations of the $\phi$ fluid given in Eqs.~\eqref{eq:pertfluidkin} and \eqref{eq:pertfluidkinTheta}. Along with the perturbation equations listed above, we need to provide CLASS with the equation of state of the $\phi$ fluid. This can be derived by noting that $\frac{d \ln \rho_{\theta} }{d \log a}=-3(1+\omega(a))$ and using Eqs.~\eqref{eq:rhologevol} and \eqref{eq:S-theta-one}.

Fig.~\ref{fig:cornerplot} shows the posterior distributions of the six cosmological input parameters for $\Lambda$CDM and the axion kination model: baryon density $\Omega_{b} h^{2}$, cold DM density $\Omega_{c} h^{2}$, spectral index $n_{s}$, primordial amplitude $A_{s}$, optical depth at reionization $\tau$ and the temperature at kination-radiation equality $T_{\rm KR}$. We fix $N_{\rm eff} = 3.046$ and assume $\log$ flat prior for $T_{\rm KR}$ between $1 \eV \leq T_{\rm KR} \leq 5\keV$. We fix $T_{\rm RM} = \mathcal{O}(1)\keV$, obtaining $T_{\rm{KR}} > 50\, \rm{eV}$ at 95\%.

For completeness, Fig.~\ref{fig:cornerplot_d} shows the derived parameters: the effective redshift at reionization $z_{reio}$, dark energy density $\Omega_{\Lambda}$, Hubble expansion rate today $H_{0}$ in units of km~s$^{-1}$~Mpc$^{-1}$, and matter fluctuation amplitude $\sigma_{8}$.

\bibliography{kination} 

\end{document}